\documentclass[useAMS,usegraphicx,usenatbib]{mn2e}

\usepackage{amsmath}

\def\H2{{\mathrm{H}_2}}
\def\HI{{\mathrm{H}_{\rm I}}}
\def\SFR{\mathrm{SFR}}
\def\sSFR{\mathrm{sSFR}}

\usepackage{url}
\usepackage{times}

\usepackage{graphicx}

\usepackage{amssymb,amsmath}

\newcommand{\acknowledgements}{\begin{small}\section*{Acknowledgments}\end{small}}

\title[Low dust-to-metal ratios driven by outflows]{The equilibrium view on dust and metals in galaxies: Galactic outflows drive low dust-to-metal ratios in dwarf galaxies}

\author[R. Feldmann]{Robert Feldmann,$^{1,\ast}$\thanks{Hubble fellow}
\\
\normalsize{$^{1}$Department of Astronomy, University of California, Berkeley, CA 94720-3411, USA}\\
\\
\normalsize{$^\ast$E-mail: feldmann@berkeley.edu}
}

\begin{document} 

\maketitle 

\begin{abstract}
Most galaxy evolution simulations as well as a variety of observational methods assume a linear scaling between the (galaxy-averaged) dust-to-gas ratio $D$ and metallicity $Z$ of the interstellar medium (ISM). Indeed, nearby galaxies with solar or moderately sub-solar metallicities clearly follow this trend albeit with significant scatter. However, a growing number of observations show that the linear scaling breaks down for metal-poor galaxies ($Z\lesssim{}0.2$ $Z_\odot$), highlighting the need for a more sophisticated modeling of the dust-to-metal ratio of galaxies. 
Here we study the co-evolution of dust and metal abundances in galaxies with the help of a dynamical, one-zone model that incorporates dust formation and destruction processes in addition to gas inflows, outflows, and metal enrichment. The dynamical model is consistent with various observational constraints, including the stellar mass -- metallicity relation, the stellar mass -- halo mass relation, and the observed $Z$ -- $D$ relation for both metal-poor and metal-rich galaxies. 
The functional form of the $Z$ -- $D$ relation follows from a basic equilibrium ansatz, similar to the ideas used previously to model the stellar mass -- metallicity relation. 
Galactic outflows regulate the inflow rate of gas from the cosmic web for galaxies of a given star formation rate. The mass loading factor of outflows thus dictates the rate at which the dust and metal content of the ISM is diluted. The stellar mass dependence of the mass loading factor drives the evolution of metallicities, dust-to-gas ratios, and dust-to-metal ratios in galaxies.
\end{abstract}


\section{Introduction}
\label{sect:introduction}

The abundance of dust in the interstellar medium (ISM) is regulated by a complex network of dust formation and destruction channels (e.g., \citealt{2011piim.book.....D, 2011EP&S...63.1027I}). Even though many details of the involved physical and chemical processes are far from fully understood, the study of interstellar dust is crucial for a number of reasons. Dust grains are a main component of the ISM (e.g., \citealt{2003ARA&A..41..241D} and references therein) and affect star formation processes, the chemistry of the ISM, and emission properties of galaxies. For instance, dust grains are the primary catalysts that convert atomic ($\HI$) into molecular hydrogen ($\H2$) and, hence, are largely responsible for the formation of molecular gas in galaxies (e.g., \citealt{1971ApJ...163..155H, 2004ApJ...604..222C, 2014arXiv1402.0867K}). Dust grains also act as a working surface for radiation pressure exerted by ionizing and non-ionizing photons (e.g., \citealt{2003JKAS...36..167S, 2005ApJ...630..167T, 2014arXiv1406.5206T}). Furthermore, dust is an important shielding agent, protecting the molecular ISM from photo-dissociating and photo-ionizing radiation produced by nearby star clusters (e.g., \citealt{2007ApJS..169..239G, 2008ApJ...689..865K, 2009ApJ...697...55G}).

Radiation absorbed by dust grains is eventually re-emitted in mid- and far-infrared wavelengths enabling astronomers to infer rates of (obscured) star formation in galaxies (e.g., \citealt{1998ARA&A..36..189K, 2010ApJ...714.1256C}) and to estimate dust masses and temperatures (e.g., \citealt{2011A&A...536A..88G, 2014arXiv1406.6066G}). The mass of ISM can then be derived, provided the proper conversion factor is known. This technique is becoming increasingly popular (e.g., \citealt{2011ApJ...737...12L, 2011ApJ...740L..15M, 2012ApJ...760....6M, 2012ApJ...761..168E, 2014MNRAS.441.1017R, 2014A&A...562A..30S, 2014arXiv1411.2975G}) and may offer the most efficient way to characterize the ISM of large samples of high redshift galaxies \citep{2014ApJ...783...84S}. However, the utility of this method stands and falls with the accuracy of the employed dust-to-gas conversion factor. 

The measured dust-to-gas ratios of many nearby galaxies scale proportionally with their ISM metallicities, at least down to $Z\sim{}1/5\,Z_\odot$ \citep{2011ApJ...737...12L}. Hence, a common strategy is to adopt a dust-to-gas ratio that is simply proportional to the observationally more easily accessible ISM metallicity. However, recent measurements of the dust-to-gas ratio in nearby, metal-poor\footnote{We consider an ISM with an oxygen abundance of $12 + \log_{10}({\rm O}/{\rm H})\leq{}8$ as metal poor, i.e., a metallicity lower than $\sim{}1/5$ the metallicity of the Sun. In the remainder of the paper we will refer to both oxygen abundances and actual metal-to-gas ratios as ``metallicities''. We adopt the conversion $Z_\odot=0.014\equiv{}12 + \log_{10}({\rm O}/{\rm H})_\odot = 8.69$ \citep{2001ApJ...556L..63A}.}, and low mass
galaxies indicate that such galaxies have much less dust than expected based on the linear metallicity scaling (e.g., \citealt{2014A&A...563A..31R, 2014Natur.514..335S}). Hence, the use of dust masses to infer ISM masses in high redshift and/or metal-poor galaxies is potentially questionable unless we understand in quantitative detail which (and how) physical processes drive the dust-to-gas ratio of galaxies in the first place.

Clearly, the evolution of the dust, metal, and gas masses in galaxies is linked to a variety of physical processes, including dust and metal enrichment, star formation, stellar feedback, gas inflows, and feedback driven outflows. Hence, the $Z$ -- $D$ relation encodes not only invaluable information about dust formation and destruction in the ISM but, as we discuss in detail in this paper, it is also intimately connected to the overall baryonic cycle of gas inflows, outflows, and star formation that drives galaxy evolution.

Various aspects of the physics of interstellar dust grains have been modeled over the last 30 yr. The modeled processes include the condensation of dust grains in stellar outflows of asymptotic giant branch (AGB) stars (e.g., \citealt{1997A&A...326..305W, 1999A&A...347..594G, 2006A&A...447..553F, 2008A&A...479..453Z, 2008A&A...491L...1H, 2013MNRAS.434.2390N}), the formation of dust grains in type II supernovae (SNe II; e.g. \citealt{1978ApJ...219..230L, 1991A&A...249..474K, 2003ApJ...598..785N, 2007ApJ...666..955N}), the destruction of dust in interstellar shocks (especially supernova remnants; e.g., \citealt{1987ApJ...318..674M, 1989IAUS..135..431M, 1996ApJ...469..740J, 2011ApJ...735...44Y}), and the accretion of metal atoms out of the ISM gas phase on to pre-existing dust grains (e.g., \citealt{1990ASPC...12..193D, 1998ApJ...501..643D, 2000PASJ...52..585H, 2003PASJ...55..901I, 2011EP&S...63.1027I}).

Chemical models combine dust formation and destruction with other processes occurring in the ISM such as gas inflows, metal enrichment, and star formation. A simple, and often used approach, is to follow the evolution of the \emph{integrated} dust, metal, gas, and stellar masses of a galaxy using a system of ordinary differential equations (see e.g., \citealt{1980ApJ...239..193D, 1991ApJ...374..456W, 1998ApJ...501..643D, 2002MNRAS.337..921H, 2008ApJ...672..214G, 2011EP&S...63.1027I, 2011A&A...528A..13G, 2013EP&S...65..213A, 2014A&A...562A..76Z}). The solution depends on the specified initial conditions (e.g., the initial gas mass), on the choice of the boundary conditions (e.g., the gas inflow rate), and on the adopted ISM processes (e.g., whether dust formation via SNe is included). The functional form of the $D$ -- $Z$ relation can then be analyzed by varying the adopted initial conditions, boundary conditions, and ISM processes and by comparing the model predictions with observations. 

An alternative numerical ansatz is to include dust chemistry in galaxy-scale hydrodynamical simulations \citep{2013MNRAS.432.2298B}. This approach has the benefit of resolving the spatial distribution of dust, metal, and gas masses in galaxies although this comes at an increased computational cost. In this work we focus on galaxy-integrated quantities and a traditional one-zone approach is a sufficient and efficient choice.

It has long been recognized that, at least in galaxies like the Milky Way, dust destruction in the ISM proceeds at a much faster pace than dust formation via stellar outflows (e.g., \citealt{1978MNRAS.183..367B, 1979ApJ...231..438D, 1980ApJ...239..193D, 1987ApJ...318..674M, 1989IAUS..135..431M, 1994ApJ...433..797J, 1996ApJ...469..740J}). The predicted low dust abundances stand in stark conflict, however, with the observed high levels of metal depletion (e.g., \citealt{1984MNRAS.208..481R, 1996ARA&A..34..279S, 2004oee..symp..336J}). A popular solution to this puzzle is that in metal-rich galaxies dust formation outpaces dust destruction and, hence, the majority of the gas-phase metals deplete on to grains (e.g. \citealt{2011EP&S...63.1027I, 2013EP&S...65..213A}). A prime candidate for fast dust formation is the accretion of metal atoms on to dust grains in the cold ISM (e.g., \citealt{1989IAUS..135..431M}). In contrast, \cite{1998ApJ...501..643D} argue that the linear scaling between $D$ and $Z$ in a metal-rich ISM is a consequence of dust formation and destruction processes operating (coincidentally) on similar time scales as star formation (and, hence, metal enrichment) in galaxies. Alternatively, a linear relation between the dust-to-gas ratio $D$ and the metallicity $Z$ (the metal-to-gas ratio) of the ISM may point to a common origin of dust formation and metal production, namely via SNe II (e.g., \citealt{2008ApJ...672..214G}).

The predictions and interpretations also differ for galaxies with a metal-poor ISM. Some models predict that an approximately linear $Z$ -- $D$ relation holds down to very low metallicities (e.g., \citealt{1998ApJ...501..643D, 2008ApJ...672..214G}). Instead, other models predict a substantial lower $D$/$Z$ ratio in metal-poor than in metal-rich galaxies (e.g., \citealt{2011MNRAS.416.1340H, 2013EP&S...65..213A, 2013MNRAS.436.1238K, 2014A&A...562A..76Z}) in qualitative agreement with recent observations (e.g., \citealt{2014A&A...563A..31R}). The theoretical interpretation is that dust growth in the ISM is ineffective at low ISM metallicities. Instead, the relatively slow dust production in stellar outflows ($\sim{}1-2$ Gyr; \citealt{1989IAUS..135..431M}) is the main dust formation channel. The short timescales for grain destruction in the ISM ($\sim{}0.2-0.4$ Gyr; \citealt{1994ApJ...433..797J}) then imply very low dust abundances.
 
The sometimes conflicting theoretical interpretations of the $Z$ -- $D$ relation can be attributed to differences in the model assumptions. They thus highlight intrinsic degeneracies in the modeling. In this paper, we mitigate the degeneracy problem by comparing our model predictions against a large number of essential observational constraints not just the $Z$ -- $D$ relation. These additional tests include: (i) the stellar mass -- metallicity relation, its evolution with redshift, and its dependency on star formation rate (SFR), (ii) the stellar mass -- halo mass relation as function of stellar mass and redshift, and (iii) the gas mass -- stellar mass relation of nearby galaxies. 

Models that adopt a quasi-steady state, or ``equilibrium'', ansatz are frequently used to explore various aspects of galaxy evolution, e.g., the evolution of SFRs (e.g., \citealt{2010ApJ...718.1001B, 2013MNRAS.433.1910F,2014MNRAS.444.2071D}), ISM metallicities (e.g., \citealt{1972NPhS..236....7L}), or the stellar mass -- metallicity relation (e.g., \citealt{2012MNRAS.421...98D, 2013ApJ...772..119L}). In contrast to dynamical models, the predictions of equilibrium models are insensitive to the chosen initial conditions. Furthermore, correlations between model variables,  e.g., between stellar mass and gas phase metallicity or between $D$ and $Z$, directly highlight the regulatory and competing aspects of the underlying physical processes. As we will demonstrate in this paper the functional form of the $Z$ -- $D$ relation can be accurately understood within the context of an equilibrium model.

The outline of this paper is as follows. We introduce the basic analytical framework and the equilibrium predictions for the $Z$ -- $D$ relation in section \ref{sect:equilibriumDust}. We subsequently present a simple dynamical model that integrates the evolution of dust, metals, gas, and stars in galaxies (section \ref{sect:equilibriumTest}). This model reproduces a variety of observations and, furthermore, matches the $Z$ -- $D$ relation as predicted by the equilibrium approach. We highlight the connection between the $Z$ -- $D$ relation and galactic outflows in section \ref{sect:RoleOutflows} and discuss potential caveats in section \ref{sect:Caveats}. We summarize our results and conclude in section \ref{sect:Conclusions}.

\section{The basic equilibrium picture of dust and metal abundances in galaxies}
\label{sect:equilibriumDust}

We follow the integrated gaseous, stellar, metal, and dust components of a galaxy via a set of continuity equations with specified sources and sinks. The full (time-dependent) solution of this problem can be complex and we defer its discussion to section \ref{sect:equilibriumTest}. However, as we show in this section, the general problem has simple and sufficiently accurate (quasi-)equilibrium solutions that allow us to understand quantitatively the functional form of the $Z$ -- $D$ relation. In the following subsections we introduce the general analytical framework of the equilibrium approach, briefly remind the reader of the concept of an equilibrium metallicity, and then derive an analogous (although somewhat more complicated) expression for the equilibrium dust-to-gas ratio.

\subsection{Equilibrium metallicity}
\label{sect:EquivZ}

The gas mass $M_{\rm g}$ of a galaxy changes as a result of gas inflows, outflows, star formation, and stellar mass loss (e.g. \citealt{2010ApJ...718.1001B, 2012MNRAS.421...98D, 2012ApJ...753...16K, 2013MNRAS.433.1910F, 2013ApJ...772..119L}):
\begin{equation}
\dot{M}_{\rm g} = \dot{M}_{\rm g, in} - (1 - R + \epsilon_{\rm out})\,\SFR.
\label{eq:dotMg}
\end{equation}
Here,  $\dot{M}_{\rm g, in}$ is the instantaneous gas inflow rate, $\SFR$ is the instantaneous SFR, and $R$ is the return fraction of gaseous material from the formed stars in the instantaneous recycling approximation (IRA). The dimensionless quantity $\epsilon_{\rm out}$ is the mass loading factor of gas outflows.
Star formation results in the growth of the stellar mass $M_*$
\begin{equation}
\dot{M}_* = (1 - R)\,\SFR.
\label{eq:dotMs}
\end{equation}
The SFR can re-written, without loss of generality, as a function of the gas mass and gas depletion time $t_{\rm dep} \equiv \SFR/M_{\rm g}$ or as a function of molecular gas fraction $f_\H2 = M_\H2 / M_{\rm g}$, gas mass, and molecular gas depletion time $t_{\rm dep, \H2}$. The latter form is particularly useful because observations indicate an approximately constant depletion time $t_{\rm dep,\H2} \sim{} 1-2$ Gyr of the molecular ($\H2$) gas component (e.g., \citealt{2008AJ....136.2846B, 2014arXiv1409.1171G}).
\begin{equation}
 \SFR = \frac{M_{\rm g}}{t_{\rm dep}} = f_\H2 \frac{M_{\rm g}}{t_{\rm dep,\H2}}.
 \label{eq:SFR}
\end{equation}
The metal mass of the ISM changes because of metal injection from supernovae with yield\footnote{We follow \cite{2012ApJ...753...16K} and define the yield as the amount of metals per units stellar mass of a single stellar population relative to $1-R$, see their equation (A2).} $y$, astration (a metal mass per unit time of $Z(1-R)\,\SFR{}$ is permanently locked up in stellar remnants), inflows with a metallicity $Z_{\rm in} = r_{\rm Z} Z$, and outflows (we discuss metal-loading in section \ref{sect:Caveats}; here we assume the outflow metallicity to equal the metallicity $Z$ of the ISM). Hence,
\begin{equation}
\dot{M}_{\rm Z} = \left[(y - Z)(1-R) - Z\epsilon_{\rm out}\right]\SFR + r_{Z} Z \dot{M}_{\rm g, in}
\label{eq:dotMZ}
\end{equation}
Dynamical models integrate this set of equations based on a specified initial state and a chosen gas infall rate as function of time. The solution $M_{\rm g}(t)$, $M_*(t)$, $M_{\rm Z}(t)$ represents the evolution of these galaxy components between the initial state and some final time. Equilibrium models instead search for quantities that approach an equilibrium value, i.e., are time-independent after some transitory period\footnote{The (quasi-)equilibrium may be reached after some finite time, i.e., we do \emph{not} restrict our analysis to the asymptotic state at time $t\rightarrow{}\infty$.}.

When we plug (\ref{eq:dotMg}, \ref{eq:SFR}, \ref{eq:dotMZ}) into $M_{\rm g}\dot{Z} = \dot{M}_{\rm Z} - Z\dot{M}_{\rm g}$, we obtain a general solution for $Z$ in terms of $\dot{Z}$, $r$ and the model parameters:
\begin{equation}
Z = \frac{y(1-R)-t_{\rm dep}\dot{Z}}{1-r_{\rm Z}}\,r,\textrm{ with }r=\frac{\SFR}{\dot{M}_{\rm g, in}}.
\label{eq:Zgen}
\end{equation}

In a steady state ($\dot{Z}\equiv{}0$) the ISM metallicity equals the equilibrium metallicity
\begin{equation}
\langle{}Z\rangle{} = \frac{y(1-R)}{1-r_{\rm Z}}\,r,
    \label{eq:Zeq}
\end{equation}
see also \cite{2012MNRAS.421...98D}, \cite{2013MNRAS.433.1910F}, and \cite{2013ApJ...772..119L}.

The equilibrium metallicity is proportional to the stellar metal yield $y$ and to the ratio between the SFR and the \emph{effective} inflow rate of pristine gas $(1-r_{\rm Z})\dot{M}_{\rm g, in}$. The effective inflow rate is zero if $Z_{\rm in}=Z$ or if $\dot{M}_{\rm g, in}=0$. 

If the ISM mass of the galaxy is unchanging ($\dot{M}_{\rm g}=0$), $r$ is simply $1/(1-R+\epsilon_{\rm out})$, see equation (\ref{eq:dotMg}). Hence, in this case the equilibrium metallicity of the ISM is set by the mass loading factor $\epsilon_{\rm out}$ of stellar outflows and by $r_{\rm Z}$. If there are no outflows ($\epsilon_{\rm out}=0$) and $r_{\rm Z}\ll{}1$, then $\langle{}Z\rangle{}\sim{}y$ (\citealt{1972NPhS..236....7L, 2012MNRAS.421...98D}).

However, the assumption of a constant ISM mass is neither necessary, nor actually correct (\citealt{2013MNRAS.433.1910F, 2013ApJ...772..119L}). Instead, using equations (\ref{eq:dotMg}-\ref{eq:SFR}) we can rewrite $r$ as \citep{2013ApJ...772..119L}: 
\begin{flalign}
r  &= \frac{1}{1 - R + \epsilon_{\rm out} + \dot{M}_{\rm g}/\SFR}\nonumber \\
    &= \frac{1}{1 - R + \epsilon_{\rm out} + t_{\rm dep} \frac{d\ln\SFR}{dt} + \frac{dt_{\rm dep}}{dt}  } \nonumber,\\
    &=\frac{1}{1 - R + \epsilon_{\rm out} + t_{\rm dep}  \left[ (1-R)\sSFR + \frac{d\ln\sSFR}{dt}  + \frac{d\ln{}t_{\rm dep}}{dt} \right]  }.
    \label{eq:r}
\end{flalign}
Provided it is known how the various quantities in equation (\ref{eq:r}) scale with stellar mass, equations (\ref{eq:Zeq}, \ref{eq:r}) lead to a prediction for the stellar mass -- metallicity relation (e.g., \citealt{1979A&A....80..155L, 2004ApJ...613..898T, 2005ApJ...634..849M, 2006ApJ...644..813E, 2008A&A...488..463M}). For instance, the term in brackets vanishes for a galaxy with a constant ISM mass. In this case, the stellar mass -- metallicity relation is driven solely by the functional form of the mass loading factor $\epsilon_{\rm out}$ on, e.g., the stellar mass of galaxies.

In general, however, the term in brackets is non-negligible and introduces an additional dependence of the equilibrium metallicity of a galaxy on its specific SFR (sSFR), i.e., the SFR per unit stellar mass of a galaxy \citep{2013ApJ...772..119L}. In fact, everything else being equal, equation (\ref{eq:r}) predicts an anti-correlation between SFRs and metallicities of galaxies with the same stellar mass, or more generally, a relation between stellar mass, metallicity and SFR (\citealt{2010MNRAS.408.2115M, 2013ApJ...765..140A, 2013MNRAS.434..451L, 2014arXiv1408.2521S}). The second and third terms in the brackets are typically much smaller than the sSFR and, hence, an approximate equation for $r$ is $1/((1-R)(1+\mu)+\epsilon_{\rm out})$, where $\mu=M_{\rm g}/M_{*}$. In this form, equation (\ref{eq:Zeq}) represents a relation between stellar mass, metallicity, and gas fraction of galaxies, in qualitative agreement with observations \citep{2013MNRAS.433.1425B}.

In the remainder of this paper we will assume that the actual, instantaneous metallicity of a given galaxy can be approximated with the equilibrium metallicity (\ref{eq:Zeq}). A linear stability analysis shows that $Z\rightarrow{}\langle{}Z\rangle{}$ over the timescale of a few $\sim{}M_{\rm g}/\dot{M}_{\rm g, in} = r\,t_{\rm dep}$ if $r$ does not change strongly with time \citep{2013MNRAS.433.1910F, 2013ApJ...772..119L}. The gas depletion time is short compared with the age of the Universe at low $z$ (e.g., \citealt{2014arXiv1409.1171G}), while $r\ll{}1$ at high $z$ \citep{2013MNRAS.433.1910F}. We thus expect $r\,t_{\rm dep}$ to be shorter than the age of the Universe at any given redshift and, hence, $Z\approx{}\langle{}Z\rangle{}$ to be a reasonably good approximation.

A different way to arrive at the same conclusion is to compare the two terms $y(1-R)$ and $t_{\rm dep}\dot{Z}$ in equation (\ref{eq:Zgen}). We can estimate $\dot{Z}$ from the time derivative of equation (\ref{eq:Zgen}). For illustrative purposes we assume that all model parameters are time independent. If we then drop terms proportional to $\ddot{Z}$ we obtain 
\[
\dot{Z}=\frac{y(1-R)}{1-r_{\rm Z}+t_{\rm dep}\dot{r}}\dot{r}.
\]
Hence, $t_{\rm dep}\dot{Z}/(y(1-R))\ll{}1$ if $\dot{r}\,t_{\rm dep}\sim{}t_{\rm dep}r/t_{\rm H} \ll{} 1$.

We can derive a first order correction to equation (\ref{eq:Zeq}) by inserting the above approximation for $\dot{Z}$ into equation (\ref{eq:Zgen}):
\begin{equation}
\langle{}Z\rangle{}^{(1)}  = \langle{}Z\rangle{}\frac{1-r_{\rm Z}}{1-r_{\rm Z}+t_{\rm dep}\dot{r}}.
\end{equation}
This equation is somewhat more accurate than equation (\ref{eq:Zeq}) in predicting the actual ISM metallicity of a galaxy. However, computing $\langle{}Z\rangle{}^{(1)}$ requires the knowledge of both $r$ \emph{and its time derivative}.

\subsection{Equilibrium dust-to-gas ratio}
\label{sect:eqD}

We can extend the idea of an equilibrium metallicity to that of an equilibrium dust-to-gas ratio $\langle{}D\rangle{}$. First, however, we need to specify the continuity equation for the dust mass. Our simple, one-zone model includes the following dust-related processes: 
\begin{itemize}
\item Dust is produced in the late stages of intermediate-mass and massive stars. We assume that the produced dust mass per unit time is proportional to $y_{\rm D}\,\SFR{}(1-R)$. The dust yield $y_{\rm D}$ quantifies the combined dust injection via AGB stars and type II SN of a stellar population. It is defined analogously to the metal yield $y$, see section \ref{sect:EquivZ}. Given the large uncertainties of the actual dust production efficiencies of AGN stars and SN, we explore a range of dust yields in the later sections of this paper. Note that $y_{\rm D}$ cannot exceed $y$.
\item Dust in the ISM is consumed by star formation. This astration process amounts to a net removal $\propto{}D\,\SFR$.
\item Dust is further destroyed by SN shockwaves in the ISM. We can write the dust destruction rate as $D\epsilon_{\rm SN}\,\SFR{}$ with the dimensionless parameter $\epsilon_{\rm SN}$. Following \cite{1989IAUS..135..431M}, we assume $\epsilon_{\rm SN} \sim{} 1000 M_\odot E_{51} {\rm SNR} / \SFR$, where $E_{51}$ is the initial kinetic energy of a typical supernova in units of $10^{51}$ erg and SNR is the supernova rate. By choosing canonical values of $E_{51}\sim{}1$ and ${\rm SNR} / \SFR\sim{}0.01 M_\odot^{-1}$ we obtain the result $\epsilon_{\rm SN}\sim{}10$. 
\item Galactic outflows expel gas and dust from the ISM. We assume that the dust-to-gas ratio of outflows equals that of the ISM (but see section \ref{sect:Caveats}).
\item Dusty galactic inflows increase the dust and gas mass of the ISM. We parametrize the dust-to-gas ratio of the inflows as $D_{\rm in}=r_{\rm D}\,D$.
\item Dust grains in the ISM can grow by accreting metal atoms out of the gas phase. Our modeling of this process mimics earlier work where $\dot{M}_{\rm D}\propto{}D (Z-D) M_{\rm cold}$ (e.g., \citealt{2000PASJ...52..585H}), but differs in two important aspects. First, we identify the cold gas mass with the molecular mass of the ISM and, using equation (\ref{eq:SFR}), link the dust growth directly to the SFR. Secondly, we limit the maximum depletion fraction of gas phase metals to $f^{\rm dep}=70\%$ to account for the incomplete depletion of some metal species (e.g., oxygen, nitrogen, and neon) even in the dense ISM \citep{2011piim.book.....D}. The growth rate of the dust mass is thus
\[
\left.\dot{M}_{\rm D}\right\vert_{\rm acc}=\frac{t_{\rm dep, \H2}}{t_{\rm ISM}}D(f^{\rm dep}Z-D)\SFR.
\]
The growth time scale $t_{\rm ISM}$ can be computed from basic collision theory \citep{1999ApJ...517..292W} and equals $Z_\odot\,t_{\rm grow}(Z_\odot)$, where $t_{\rm grow}(Z_\odot)\sim{}10^7$ yr is the exponential growth time of dust in solar metallicity, cold clouds (e.g., \citealt{2000PASJ...52..585H}). In this paper we treat the ratio $t_{\rm dep, \H2}/t_{\rm ISM}$ as a free parameter, however.
\end{itemize}
By combining the various dust formation and destruction processes we arrive at the continuity equation for the dust mass
\begin{flalign}
\dot{M}_{\rm D} &= \left[y_{\rm D}(1-R) - D(1 + \epsilon_{\rm SN} +\epsilon_{\rm out})\right]\SFR  \nonumber \\
                          &+ r_{\rm D} D \dot{M}_{\rm g, in} +\frac{t_{\rm dep, \H2}}{t_{\rm ISM}}D(f^{\rm dep}Z-D)\SFR. \label{eq:dotMd}
\end{flalign}
We can now make the equilibrium ansatz $\dot{D}\equiv{}0$ \& $\dot{Z}\equiv{}0$ to solve for the equilibrium dust-to-gas ratio $\langle{}D\rangle{}$. This approach results in the following  quadratic equation
\begin{equation}
\alpha + \beta{}\langle{}D\rangle{} - \gamma{}\langle{}D\rangle{}^2 = 0.
\label{eq:DeqImplicit}
\end{equation}
The coefficients are
\begin{flalign}
\alpha &= y_{\rm D}(1-R), \label{eq:alpha}  \\
\beta &= \gamma{}f^{\rm dep}\langle{}Z\rangle{} - \left[ \epsilon_{\rm SN} + R + \frac{1-r_{\rm D}}{r}\right], \label{eq:beta} \\
\gamma &= \frac{t_{\rm dep, \H2}}{t_{\rm ISM}} \label{eq:gamma}.
\end{flalign}
Here $r=\frac{\SFR}{\dot{M}_{\rm g, in}}$, as before. Note that $\alpha$ and $\gamma$ are positive, while $\beta$ can be positive, negative, or zero. The quadratic equation is therefore guaranteed to have a unique non-negative solution:
\begin{equation}
\langle{}D\rangle{} = \frac{\beta}{2\gamma} + \left[ \left(\frac{\beta}{2\gamma}\right)^2 + \frac{\alpha}{\gamma} \right]^{1/2}.
\label{eq:Deq}
\end{equation}

It is instructive to analyze solution (\ref{eq:Deq}) in the three regimes: (A) $\beta<-2\left(\alpha{}\gamma\right)^{1/2}$ (B) $\beta>2\left(\alpha{}\gamma\right)^{1/2}$, and (C) $\beta^2<4\alpha{}\gamma$.
\begin{itemize}

\item In the first regime 
\begin{equation}
\langle{}D\rangle{}\approx{}-\alpha/\beta. \label{eq:DeqA}
\end{equation}
Here, the dust-to-gas ratio is determined by the competition between stellar dust production, dust destruction in shockwaves, and dilution of dust via dust-poor inflows. The latter process is regulated by the $(1-r_{\rm D})/r$ term and dominates over the destructive effect of SN shockwaves at sufficiently low $r$. Typically, $\alpha \ll{} \vert{}\beta\vert{}$ and, hence, the dust-to-gas ratio in this regime is quite small. Specifically, the dust-to-gas ratio approaches $\frac{y_{\rm D}}{y}\frac{1-r_{\rm Z}}{1-r_{\rm D}}\langle{}Z\rangle{}$ as $r/(1-r_{\rm D})\rightarrow{}0$.

\item In the second regime, 
\begin{equation}
\langle{}D\rangle{}\approx{}\beta/\gamma = f^{\rm dep}\langle{}Z\rangle{} - \frac{\epsilon_{\rm SN}+R}{\gamma} - \frac{1-r_{\rm D}}{r\gamma}. \label{eq:DeqB}
\end{equation}
The condition $\beta>0$ implies that $r$ cannot be very small. Hence, the dust-to-gas ratio in this regime is close to the depletion limit, i.e, $\approx{}f^{\rm dep}\langle{}Z\rangle{}$, because $\gamma{}\gg{}1$ under typical ISM conditions.

\item The third regime corresponds to $\beta\sim{}0$, i.e., when the dust growth in the cold ISM is balanced by dust destruction via SN shocks and the dilution of dust via dust-poor inflows. Here,
\begin{equation} 
\langle{}D\rangle{}\approx{}(\alpha/\gamma)^{1/2}. \label{eq:DeqC}
\end{equation}
Given that  $t_{\rm ISM}$  scales inversely proportional to the ISM density, such a balance likely only occurs over a very narrow range of ISM properties. However, this regimes becomes more important with increasing $\alpha$, i.e. for higher dust yields, and with increasing $\gamma$, i.e., for shorter $t_{\rm ISM}$ or longer $t_{\rm dep,\H2}$.
\end{itemize}

\begin{figure*}
\begin{tabular}{cc}
\includegraphics[width=80mm]{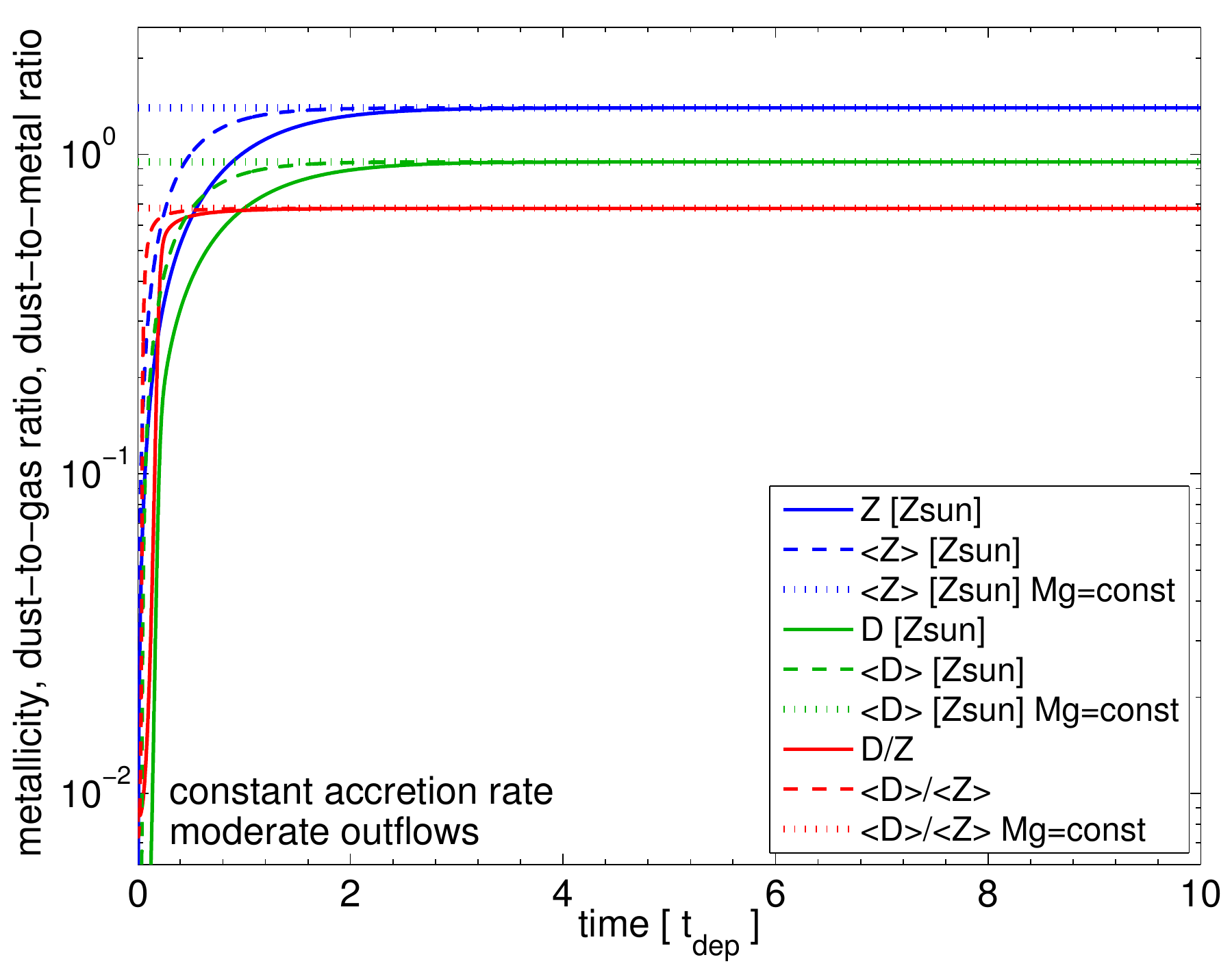} &
\includegraphics[width=80mm]{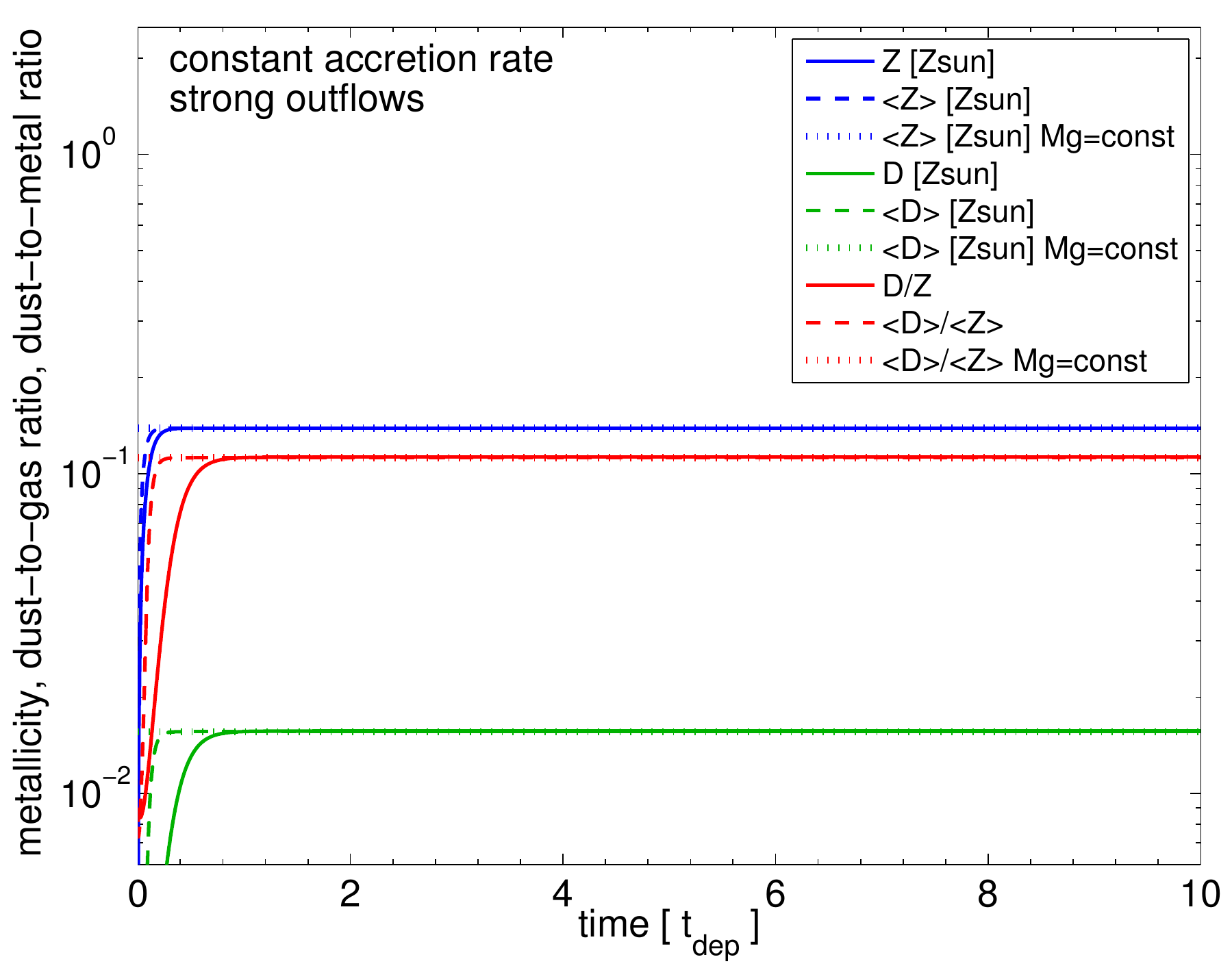} \\
\includegraphics[width=80mm]{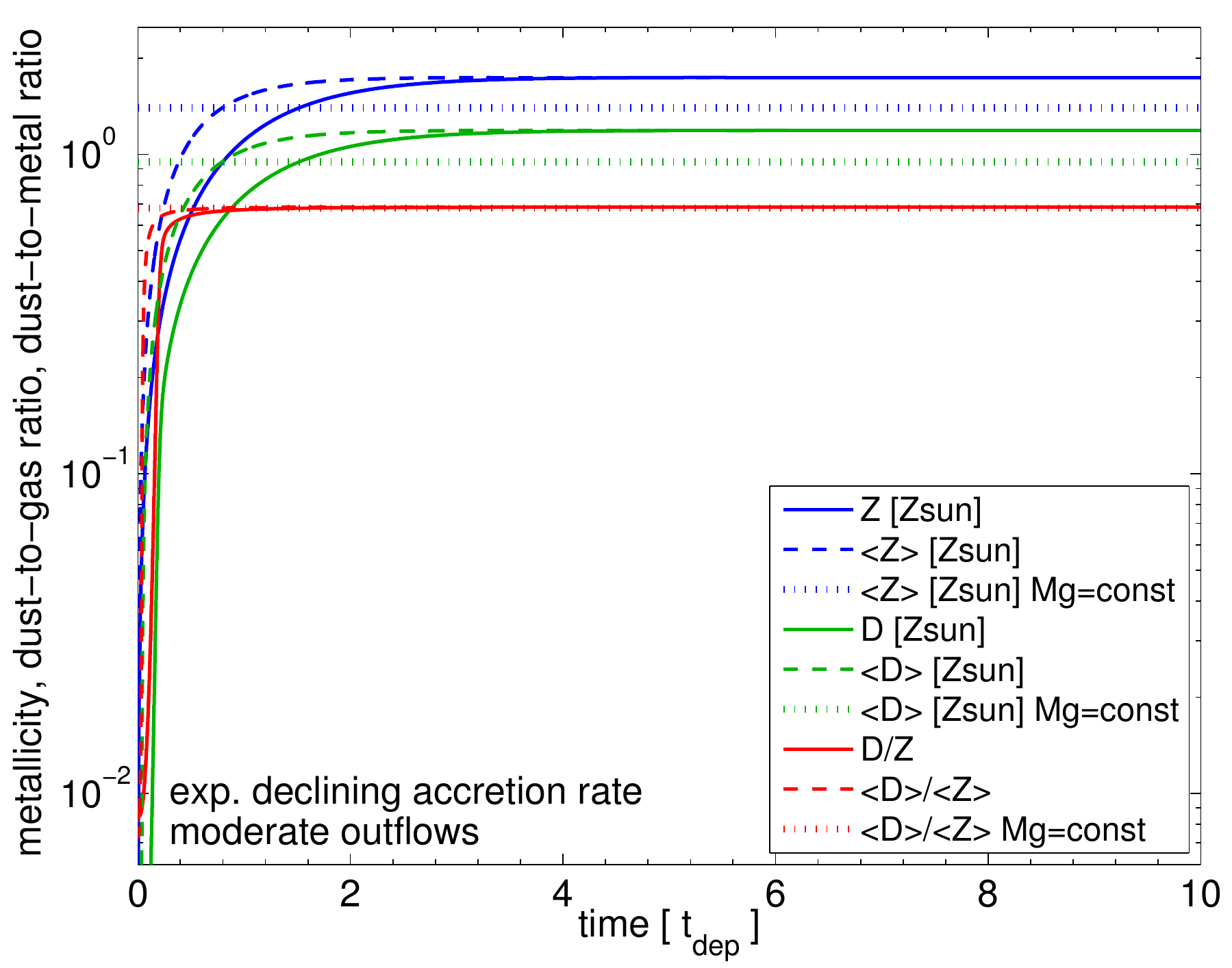} &
\includegraphics[width=80mm]{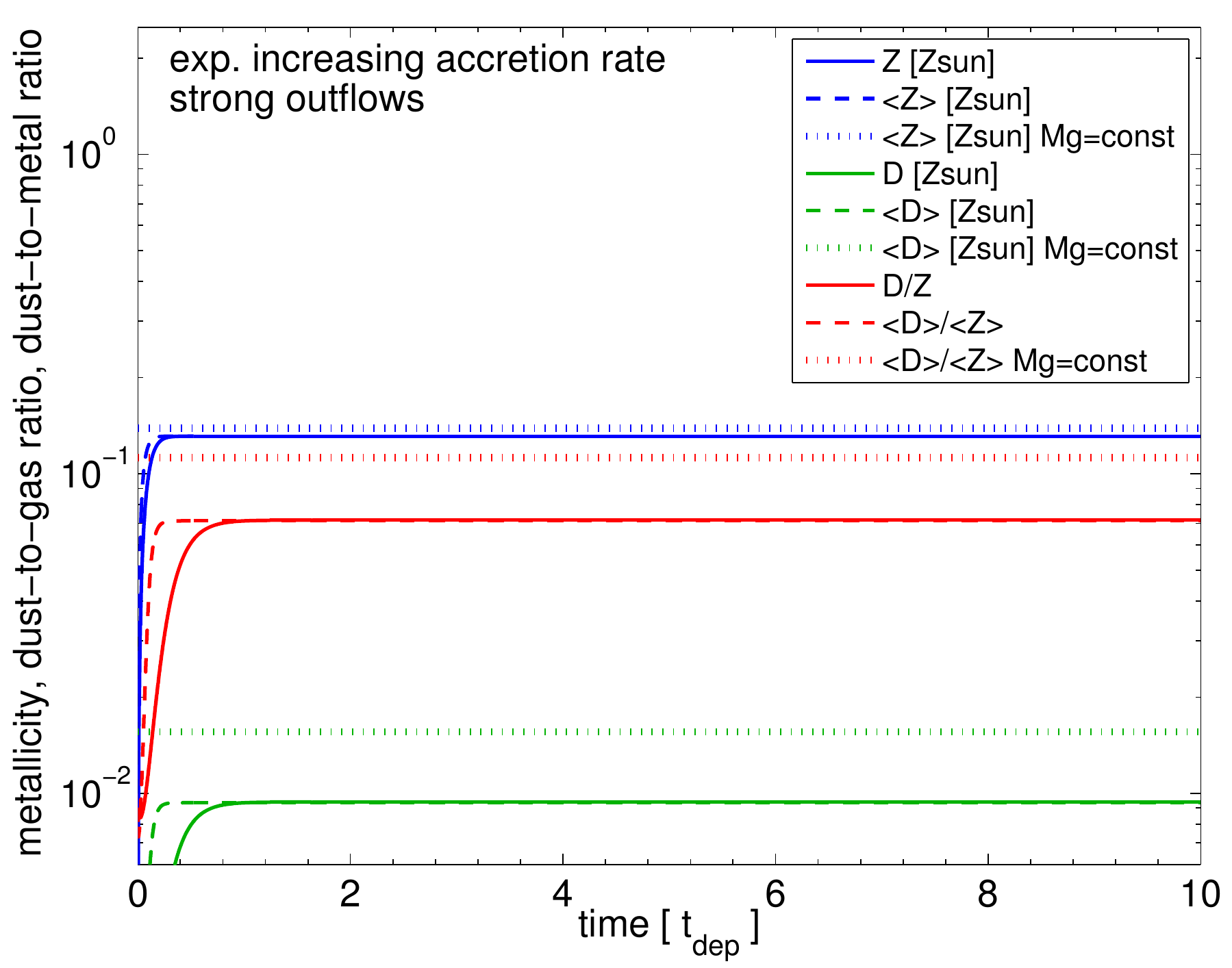}
\end{tabular}
\caption{ISM metallicity, dust-to-gas ratio, and dust-to-metal ratio as predicted by a simple dynamical model (see the text for details) and by the equilibrium ansatz. ISM metallicities, dust-to-gas ratios, and dust-to-metal ratios are shown in blue, green, and red, respectively. 
(Top left) for a constant gas accretion rate of 1 $M_\odot$ yr$^{-1}$ and a moderate mass loading factor $\epsilon_{\rm out}=2$. (Top right) same as top left but for a larger mass loading factor $\epsilon_{\rm out}=25$. (Bottom left) Same as top left but for an exponentially declining gas infall rate $\dot{M}_{\rm g, in}=\exp(-t / 3\times{}10^9 {\rm yr})$ $M_\odot$ yr$^{-1}$. (Bottom right) Same as top right but for an exponentially increasing gas infall rate $\dot{M}_{\rm g, in}=\exp(t / 10^9 {\rm yr})$ $M_\odot$ yr$^{-1}$. 
In each of the panels we plot the results of the explicit integration of equations (\ref{eq:dotMg}--\ref{eq:dotMZ}, \ref{eq:dotMd}) using solid lines. Dashed lines show the corresponding predictions of the equilibrium model, equations (\ref{eq:Zeq}, \ref{eq:Deq}). The (time-dependent) ratio $r$ between SFR and gas inflow rate is known from the explicit integration. Dotted lines show to the equilibrium predictions under the assumption that the ISM mass is constant, i.e., for the special case $r = 1/(1-R+\epsilon_{\rm out}) ={\rm const}$.
The systems evolves quickly (within $\lesssim{}1-2\,t_{\rm dep}$) to the equilibrium solution given by equations (\ref{eq:Zeq}) and (\ref{eq:Deq}).}
\label{fig:TestEqM}
\end{figure*}

\begin{figure*}
\begin{tabular}{cc}
\includegraphics[width=85mm]{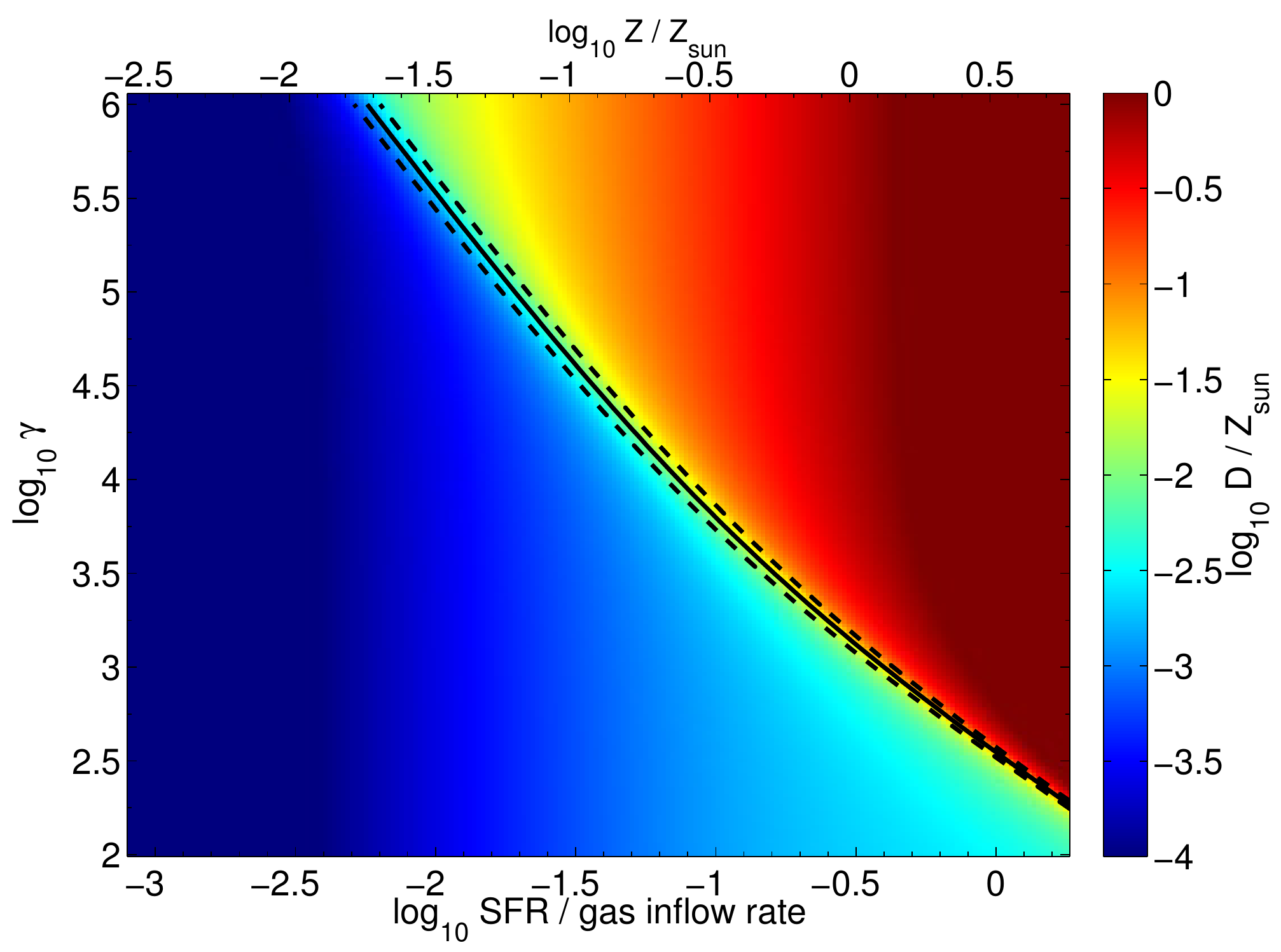} &
\includegraphics[width=85mm]{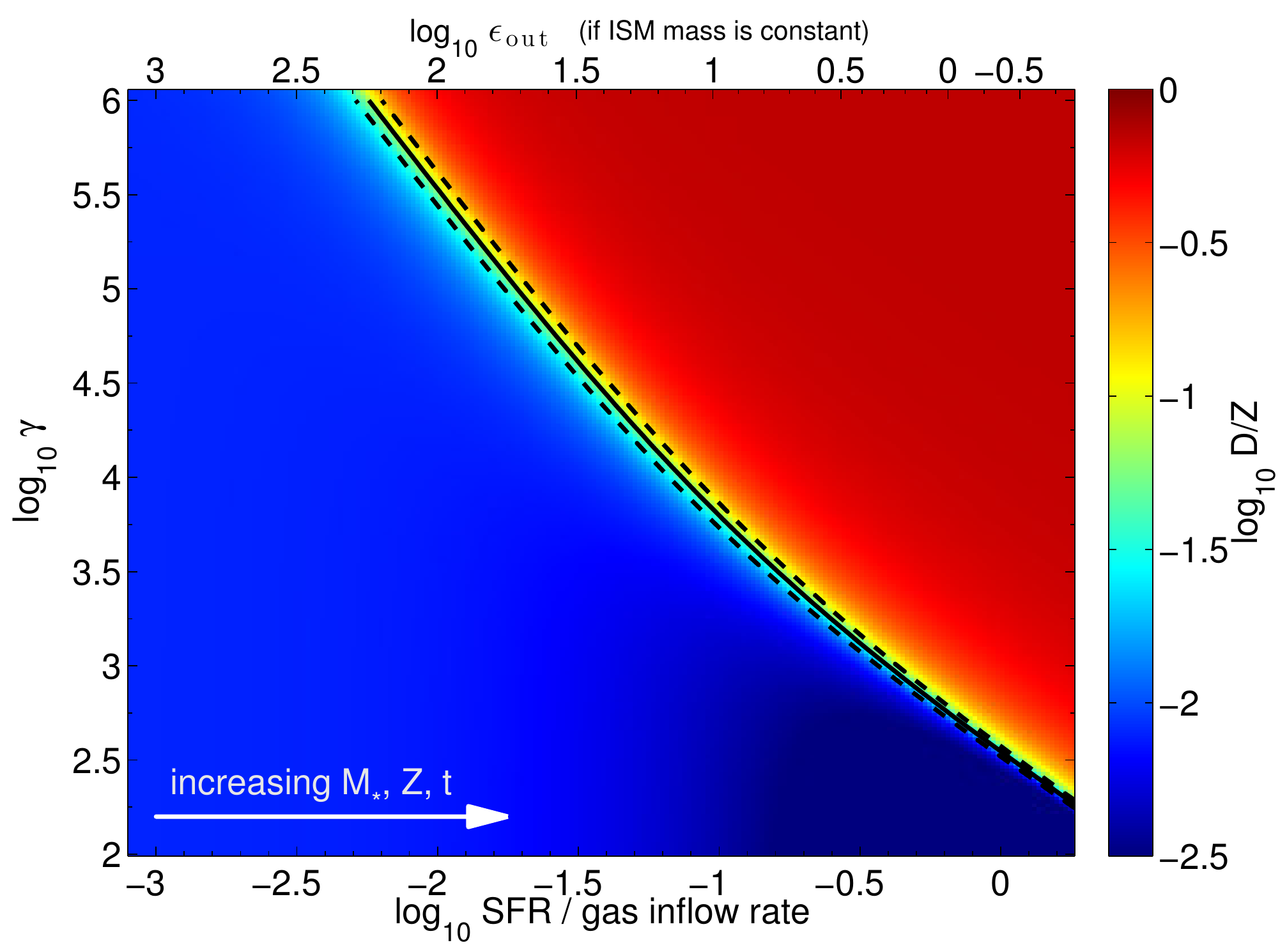} \\
\end{tabular}
\caption{Equilibrium dust-to-gas and dust-to-metal ratios as function of the ratio between SFR and gas inflow rate (horizontal axis) and the ratio between gas depletion time and dust growth time in the ISM (vertical axis). (Left-hand panel) equilibrium dust-to-gas ratio, equation (\ref{eq:Deq}), in units of $Z_\odot$. The x-axis is labelled with the ratio between SFR and gas inflow rate (bottom) and the corresponding equilibrium metallicity (top), respectively, see (\ref{eq:Zeq}). (Right-hand panel) dust-to-metal ratio as predicted by the equilibrium model, i.e., equation (\ref{eq:Deq}) divided by equation (\ref{eq:Zeq}). The $x$-axis is labelled with the ratio between SFR and gas inflow rate (bottom) and the corresponding mass loading factor of outflows (top), respectively, see equation (\ref{eq:r}). The solid line in both panels shows $\beta=0$. This line separates a parameter region of low $D$ and low $D/Z$ (to the left of the line) from a region of high $D$ and high $D/Z$ (to the right of the line). Dashed lines show where $\beta=\pm{}2\sqrt{\alpha\gamma}$ and indicate the width of the transition region. The dust-to-gas and dust-to-metal ratios increase with increasing gas-phase metallicity, with decreasing mass loading factor, and with a decreasing dust formation time in the ISM.}
\label{fig:DZeq}
\end{figure*}

Fig.~\ref{fig:TestEqM} compares the predictions of the equilibrium model, equations (\ref{eq:Zeq}, \ref{eq:Deq}), with those of a one-zone dynamical model with a specified gas inflow rates. Specifically, we compare the dust-to-gas ratio $D$, the ISM metallicity $Z$, and the ratio of the two, the dust-to-metal ratio $D/Z$. The dynamical model is adequate to test the equilibrium ansatz for a range of inflow rates (constant, exponentially increasing, exponentially declining) and mass loading factors ($0-30$). We will introduce a more advanced dynamical model in section \ref{sect:DynModel} in order to cross-check the equilibrium predictions under more realistic circumstances and to provide a connection with observations.

The dynamical model starts from negligible initial gas mass, metallicity, and dust-to-gas ratio. Stars are formed from the gas based on equation (\ref{eq:SFR}) and a depletion time  $t_{\rm dep}=1.5$ Gyr \citep{2008AJ....136.2846B}. The return fraction is $R=0.46$ \citep{2012ApJ...753...16K}. Metal and dust yields are $y=6.9\times{}10^{-2}$ \citep{2012ApJ...753...16K}, $y_{\rm D}=6.9\times{}10^{-4}$ (a free parameter). Inflows are assumed to have a relative metal and dust enrichment of $r_{\rm Z}=0.25$, and $r_{\rm D}=0$. The maximal dust depletion is $f^{\rm dep}=0.7$ and the ratio $\gamma$ between gas depletion and dust accumulation time in the ISM is $3\times{}10^4$. The dust destruction efficiency of SNe is $\epsilon_{\rm SN}=10$.

We find that the system approaches quickly (within $\lesssim{}1-2\,t_{\rm dep}$) the equilibrium solutions (\ref{eq:Zeq}, \ref{eq:Deq}) for $D$, $Z$, and $D/Z$. Galaxies with a declining gas infall rate have a somewhat longer equilibration time than galaxies with constant or increasing gas inflow rates. However, the equilibration time is significantly shorter than $t_{\rm dep}$ if galaxies experience strong outflows. We verified that this statement holds even for galaxies with slowly declining gas accretion rates. Hence, unless galaxies have \emph{both} quickly declining gas inflow rates (on time scales $\lesssim{}t_{\rm dep}$) \emph{and} weak/no outflows, the actual $Z$, $D$, and $D/Z$ of their ISM should be close to the equilibrium predictions.

Fig.~\ref{fig:TestEqM} also shows that $D$, $Z$, and $D/Z$ are lower in galaxies that have stronger outflows (top right panel) \emph{or} that have more quickly growing gas accretion rates (bottom right panel). This result is a consequence of $r$ decreasing with both increasing $\epsilon_{\rm out}$ and increasing $\dot{M}_{\rm g}/\SFR$, see equation (\ref{eq:r}).

It is sometimes possible to obtain reasonably accurate equilibrium solutions by replacing $r$ from equation (\ref{eq:r}) with the approximation $r = 1/(1-R+\epsilon_{\rm out})$. The accuracy of this approximation depends on the ratio between $\epsilon_{\rm out}$ and $\dot{M}_{\rm g}/\SFR$. The approximation is exact if the galaxy has a constant ISM mass (i.e., $\dot{M}_{\rm g}=0$). If the ISM mass is increasing (decreasing), the approximation overestimates (underestimates) $Z$, $D$, $D/Z$, see Fig.~\ref{fig:TestEqM}. 

As we will argue later in this paper (section \ref{sect:RoleOutflows}) $\epsilon_{\rm out} \gtrsim{} \dot{M}_{\rm g}/\SFR$ for galaxies near the main sequence of star formation. Hence, the stellar mass dependence of outflows is (at least qualitatively) the primary driver that determines the ratio $r$ between the SFR and gas inflow rate into a galaxy.

\subsection{Critical metallicity and critical ratio between SFR and gas inflow rate}
\label{sect:Zcrit}

The transition between the regimes of low and high dust-to-gas ratios takes place when $\beta=0$, i.e., for 
\begin{equation}
Z_{\rm crit}(r)=\frac{1}{\gamma}\left[\epsilon_{\rm SN} + R + \frac{1-r_{\rm D}}{r}\right]\frac{1}{f^{\rm dep}}.
\label{eq:Zcrit}
\end{equation}
In other words, the equilibrium model predicts that galaxies with $Z\lesssim{}Z_{\rm crit}(r)$ have very small dust-to-gas ratios, $\langle{}D\rangle{}\approx{}y_{\rm D}/(\epsilon_{\rm SN}+(1-r_{\rm D})/r)$, while galaxies with $Z\gtrsim{}Z_{\rm crit}(r)$ have a dust-to-gas ratio close to the depletion limit $f^{\rm dep}\langle{}Z\rangle{}$. Galaxies with $Z\sim{}Z_{\rm crit}(r)$ have $D\sim{}D_{\rm crit}=y_{\rm D}(1-R)/\gamma$. The smaller the product $\alpha\gamma$ the sharper is the transition.

Small values of $r$ result in low equilibrium metallicities $\langle{}Z\rangle{}$, see equation (\ref{eq:Zeq}), but large critical metallicities $Z_{\rm crit}(r)$. The opposite happens for large values of $r$. We can remove the explicit dependence of $Z_{\rm crit}(r)$ on $r$ by using equation (\ref{eq:Zeq}), i.e., by replacing $r$ with the value corresponding to $\langle{}Z\rangle{}=Z_{\rm crit}(r)$. This results in
\begin{flalign}
Z_{\rm crit} & = \frac{p}{2} + \left[ \left(\frac{p}{2}\right)^2 + q \right]^{1/2},\textrm{ with} \label{eq:Zcrit2} \\
p &= \frac{\epsilon_{\rm SN} + R}{\gamma f^{\rm dep}} \nonumber \\
q &= \frac{1-r_{\rm D}}{1-r_{\rm Z}}\frac{y(1-R)}{\gamma f^{\rm dep}}. \nonumber
\end{flalign}
Inserting typical values for the parameters (see previous section), we find that $q\gg{}(p/2)^2$ for $\gamma\gtrsim{}10^4$ and thus $Z_{\rm crit} \approx{} q^{1/2}$. The product $y(1-R)$ is theoretically/observationally well constrained and 
$(1-r_{\rm D})/(1-r_{\rm Z})$ as well as $f^{\rm dep}$ have values close to unity. Hence, it may be possible to extract the (uncertain) value of $\gamma$ from a measurement of $Z_{\rm crit}$. The dust yield $y_{\rm D}$ can then be derived from the corresponding measurement of 
\begin{equation}
D_{\rm crit}=y_{\rm D}(1-R)/\gamma.
\end{equation}

Critical metallicities have also been introduced by \cite{2011EP&S...63.1027I} and by \cite{2013EP&S...65..213A}. These previous works agree with this paper in that the dust-to-gas ratio increases substantially above a metallicity threshold. They differ, however, in methodology and in their findings. 

First, our approach is based on a steady-state picture in which inflows, outflows, and star formation balance each other. In contrast, \cite{2011EP&S...63.1027I} use a dynamical model with a slowly declining gas inflow rate, but no outflows, while \cite{2013EP&S...65..213A} adopt a dynamical, closed-box model. The metallicity in the latter model increases without bounds with time and, hence, does not allow for an equilibrium metallicity or dust-to-gas ratio.

Secondly, we define the critical metallicity based on the transition point ($\beta=0$) between low (for $\beta\ll{}0)$ and high (for $\beta\gg{}0$) dust-to-gas ratios. In physical terms, the regime $\beta\sim{}0$ corresponds to the balance between dust growth in the ISM, dust destruction by SNe, \emph{and dust dilution}, as we showed in section \ref{sect:eqD}. The definition of \cite{2011EP&S...63.1027I} is similar to ours ($\beta\equiv{}0$) but neglects the dilution effect produced by dust-poor inflows. Hence, their critical density does not depend on the ratio $r$ between SFR and gas inflow rate and corresponds to equation (\ref{eq:Zcrit}) with $(1-r_{\rm D})/r<{\epsilon_{\rm SN} + R}$, i.e., $Z_{\rm crit}\approx{}p$. If this were true, the critical metallicity would be set by the competition between dust growth in the ISM and dust destruction via SN shockwaves. 

In contrast, we argued above that $Z_{\rm crit}\approx{}q^{1/2}$ is a better approximation of our general solution (\ref{eq:Zcrit2}). This approximation corresponds to (\ref{eq:Zcrit}) with $\epsilon_{\rm SN} + R<\frac{1-r_{\rm D}}{r}$. For instance, for $Z_{\rm crit}=0.1 Z_{\rm sun}$, $1/r\sim{}35$ according to (\ref{eq:Zeq}), i.e., $r(Z_{\rm crit})$ is much larger than $\epsilon_{\rm SN} + R$. \emph{Hence, the critical metallicity of the ISM is set by the competition between dust growth in ISM and the dilution of dust via dust-poor gas inflows. The destruction of dust in supernova shock-waves plays a sub-dominant role unless $\gamma{}\ll{}10^4$.}

\cite{2013EP&S...65..213A} adopt as critical metallicity the metallicity at which stellar dust production and dust accretion in the ISM proceed at the same pace. Their definition thus refers to the relative strength of the two dust production channels and not to a balance between dust growth and dust dilution/destruction.

Our definition of a critical metallicity is designed to accurately pin-point the metallicity at which the ISM of a galaxy switches from low to high dust-to-gas ratios. The critical metallicity of \cite{2011EP&S...63.1027I} typically underestimates the metallicity at which the transition from low to high dust-to-gas ratios occurs. Their result, $Z_{\rm crit}\approx{}p$, neglects the contribution given by $q$, which dominates for typical values of $\gamma$. The definition adopted by \cite{2013EP&S...65..213A} also underestimates the transition metallicity by some factor (see their fig. 3).

According to equation (\ref{eq:Zeq}) the equilibrium metallicity depends primarily on the ratio 
\begin{equation}
\tilde{r}\equiv{}\frac{r}{1-r_{\rm Z}}.
\label{eq:tilder}
\end{equation}
The criticality condition $\beta=0$ is satisfied for $\tilde{r}_{\rm crit} = Z_{\rm crit}/[y(1-R)]$, i.e., we can replace the concept of a critical metallicity with that of a critical ratio between the SFR and the (effectively pristine) gas inflow rate. Similar to $Z_{\rm crit}$, $\tilde{r}_{\rm crit}$ separates the regimes of low (for $\tilde{r}\ll{}\tilde{r}_{\rm crit}$) and high (for $\tilde{r}\gg{}\tilde{r}_{\rm crit}$) dust-to-gas ratios.  The transition regime holds for $\vert\beta\vert<2\sqrt{\alpha\gamma}$, see section \ref{sect:eqD}. The corresponding values of $\tilde{r}$ can be obtained from equation (\ref{eq:Zcrit2}) but with $\epsilon_{\rm SN}$ replaced by $\epsilon_{\rm SN}\pm{}2\sqrt{\alpha\gamma}$.

We see from equations (\ref{eq:r}, \ref{eq:tilder}) that galactic outflows, increasing ISM masses, and metal-poor inflows affect $\tilde{r}$. These physical processes are thus the prime candidates for ultimately driving the metallicity, the dust-to-gas ratio, and the dust-to-metal ratios of galaxies. We will continue this discussion in section \ref{sect:Driver}.

\subsection{The $Z$ -- $D$ relation in the equilibrium picture}
\label{sect:ZDrel}

Both the equilibrium metallicity and the equilibrium dust-to-gas ratio of a given galaxy depend\footnote{In this subsection we adopt the view that other important model parameters, e.g., the metal yield, the dust yield, or the inflow enrichment factors $r_{\rm Z}$ and $r_{\rm D}$ are either constant or only evolve on sufficiently long ($\gg{}r\,t_{\rm dep}$) timescales.} on a dimensionless parameter: $r$, the ratio between SFR and gas inflow rate. The dust-to-gas and dust-to-metal ratios further depend on the ratio $\gamma$ between the gas depletion time and the dust growth time in the ISM, see equation (\ref{eq:gamma}).

We visualize the equilibrium predictions for $D$ and $D/Z$  in Fig.~\ref{fig:DZeq}. For a given value of $\gamma$, galaxies with small enough $r$ (or small enough $Z$) have low dust-to-gas and dust-to-metal ratios. In contrast, galaxies with a large enough $r$ (or large enough $Z$) have much larger dust-to-gas and dust-to-metal ratios. The transition is sharp, especially so for the dust-to-metal ratio. 

\begin{figure}
\begin{tabular}{c}
\includegraphics[width=80mm]{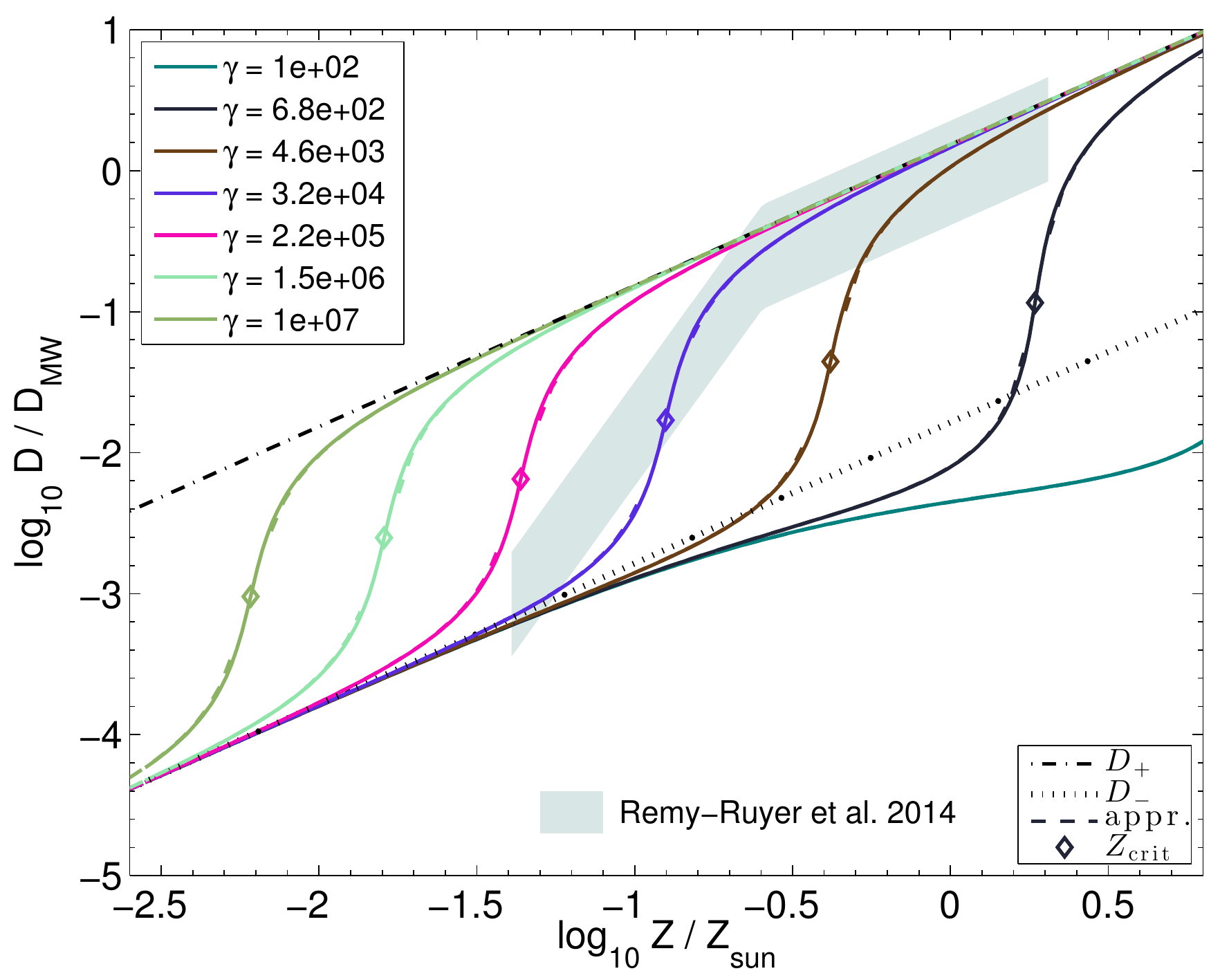}
\end{tabular}
\caption{Relation between metallicity (normalized to $Z_\odot=0.014$) and dust-to-gas ratio (normalized to $D_{\rm MW}=0.0064$; \citealt{2004ApJS..152..211Z}) as predicted by the equilibrium model, equations (\ref{eq:Zeq},\ref{eq:Deq}). Diamond symbols show the position of $(Z_{\rm crit}, D_{\rm crit})$. The dust-to-gas ratio is small ($\approx{}D_-\equiv{}\frac{y_{\rm D}}{y}\frac{1-r_{\rm Z}}{1-r_{\rm D}}Z$; dotted line) at $Z\ll{}Z_{\rm crit}$, and close to the depletion limit ($\approx{}D_+\equiv{}f^{\rm dep}Z$; dot-dashed line) at $Z\gg{}Z_{\rm crit}$. The transition shifts to higher metallicities, i.e., $Z_{\rm crit}$ increases, with decreasing $\gamma$ (see legend). Dashed lines (almost on top of the solid lines) are the approximations (equations \ref{eq:DeqA}, \ref{eq:DeqB}) for regimes A and B, see section \ref{sect:eqD}. The shaded area is a double-power law fit to the dust-to-gas ratio of nearby galaxies by \protect\cite{2014A&A...563A..31R}. The vertical width of the shaded region indicates the dispersion around the mean of 0.37 dex \protect\citep{2014A&A...563A..31R}. The predicted and observed $Z$ -- $D$ relations are in good agreement for $\gamma\sim{}3\times{}10^4$.}
\label{fig:DZrelation}
\end{figure}

In Fig.~\ref{fig:DZrelation}, we show the predictions of our equilibrium model for the parameters specified in section \ref{sect:eqD}. We obtain a $Z$ -- $D$ relation by varying $r$ and keeping all other model parameters (including $\gamma$) fixed. Hence, the $Z$ -- $D$ relation emerges as a consequence of the dependence of $Z$ and $D$ on $r$ (at fixed $\gamma$). Galaxies with sufficiently large $r$ have dust-to-gas ratios close to the depletion limit, i.e., here $D\approx{}D_+\equiv{}f^{\rm dep}Z$. Galaxies with sufficiently small $r$ (and thus low $Z$) have low dust-to-gas ratios $D\approx{}D_-\equiv{}\frac{y_{\rm D}}{y}\frac{1-r_{\rm Z}}{1-r_{\rm D}}Z$. The transition between low and high dust-to-gas ratios takes place at $Z\sim{}Z_{\rm crit}$, or equivalently, for $\tilde{r}\sim{}\tilde{r}_{\rm crit}$. $Z_{\rm crit}$ and $\tilde{r}_{\rm crit}$ both increase with decreasing $\gamma$.

Variations of $\gamma$ lead to scatter in the $Z$ -- $D$ relation. For instance, at a fixed metallicity, galaxies with longer $\H2$ depletion times or with shorter dust growth times in the ISM will have a larger dust-to-gas ratio. It remains to be seen whether this mechanism causes most of the scatter in the observed $Z$ -- $D$ relation. Short term fluctuations in the gas accretion or outflow rates are plausible alternatives (e.g., \citealt{2014A&A...562A..76Z}).

In Fig.~\ref{fig:DZrelation}, we also compare the predictions of the equilibrium model with the observational study by \cite{2014A&A...563A..31R}. The authors fit a broken-power law to the $Z$ -- $D$ relation based on an observational sample of 126 nearby galaxies (their table 1, right-hand column). With the choice $\gamma\sim{}3\times{}10^4$, the equilibrium model reproduces quite well the observed drop of the average dust-to-gas ratio for galaxies with $Z\lesssim{}1/5 Z_\odot$. This value corresponds to $t_{\rm grow}=t_{\rm dep,\H2}/t_{\rm ISM}/Z_\odot\sim{}4\times{}10^6$ yr, which is consistent with previous estimates ($\sim{}10^7$ yr) for the time scale of accretion-based dust growth in the ISM (e.g., \citealt{2000PASJ...52..585H}).

Strong outflows as well as a highly inefficient conversion of inflowing gas into stars result in small values of $r$ and, hence, in low equilibrium metallicities, low dust-to-gas ratios, and low dust-to-metal ratios. The equilibrium model by itself does not tell us which of these processes is/are actually responsible for keeping $r$ small. However, we can gain additional insights by combining the equilibrium model with empirical constraints, as we show in the next section.

\section{Testing the Equilibrium predictions}
\label{sect:equilibriumTest}

The dynamical model of the previous section, equations (\ref{eq:dotMg} - \ref{eq:dotMZ}, \ref{eq:dotMd}), can be integrated easily for a given gas inflow rate. So far, we considered various simple choices, ranging from constant rates to those that exponentially increase or decrease with time. In each case we found the equilibrium metallicity, equation (\ref{eq:Zeq}), to be an accurate predictor of the actual, time-dependent ISM metallicity. 

In this section we want to go one step further, by constructing a dynamical model that uses a more realistic, time-dependent accretion rate constrained by observations. In addition, we will allow model parameters, such as the molecular gas depletion time and the mass loading factor, to change over the evolution of a particular galaxy. We will show that such a dynamical model matches various observational constraints and, furthermore, that the equilibrium model predicts metallicities and dust-to-gas ratios in good agreement with those obtained by a direct integration of the dynamical model.

\subsection{An empirical dynamical model}
\label{sect:DynModel}

Starting with a galaxy of a given stellar mass $M_*$ at $z=0$ we reconstruct the evolution of its stellar mass, SFR, $Z$, and $D$ as follows:
\begin{itemize}
\item We assume that the SFR of the galaxy is proportional to the SFR of the main sequence of star formation. By integrating the observed evolution of the main sequence backward in time, we obtain the stellar mass and the SFR of this galaxy as function of redshift. A similar procedure has been used by \cite{2011ApJ...734...48L} and \cite{2012ApJ...745..149L} to constrain the star formation history of galaxies and to estimate the importance of stellar mass loss for galactic SFRs. 
\item We convert the SFR history into an ISM mass history based on observations of the (redshift and SFR dependent) gas depletion time.
\item We adopt a theoretically motivated functional form of the (stellar-mass dependent) mass loading factor.
\item By inserting the SFR, the mass loading factor, and the change of the ISM mass into equation (\ref{eq:dotMg}) we derive the gas inflow rate as function of redshift. 
\item Finally, we integrate equations (\ref{eq:dotMg} -- \ref{eq:dotMZ}, \ref{eq:dotMd}) forward in time to obtain time-dependent ISM metallicities and dust-to-gas ratios.
\end{itemize}

To implement our model, we need to specify functional forms for (i) the evolution of the main sequence of star formation, (ii) the gas depletion time, (iii) the $\H2$ fraction of galaxies, and (iv) the mass loading factor.

We adopt the parametrization of the main sequence of star formation by \cite{2013ApJ...772..119L}:
\begin{flalign}
\SFR_{\rm MS}(M*, z)= & \frac{M_*\,\text{Gyr}^{-1}}{1-R} \left(\frac{M_*}{10^{10.5}\,M_\odot}\right)^{-0.1} \nonumber \\ 
                                    \times  &  \begin{cases} 
                                     0.07(1+z)^3  & \text{if $z\leq{}2$},\\ 
                                     0.303(1+z)^{5/3}& \text{if $z>2$}. \end{cases} 
                                     \label{eq:SFRLilly}
\end{flalign}
We also tested the parametrization by \cite{2012ApJ...754L..29W} and found some quantitative, but no qualitative, differences.

Galaxies that lie on the main sequence of star formation (e.g., Noeske+2007) have a molecular gas depletion time of $t_{\rm dep,\H2}\sim{}1-2$ Gyr (e.g., \citealt{2008AJ....136.2846B, 2010MNRAS.407.2091G, 2014arXiv1409.1171G}). The large scatter of $t_{\rm dep,\H2}$ at fixed stellar mass arises to a large degree from the anti-correlation between depletion time and the offset from the main sequence of star formation \citep{2011MNRAS.415...61S, 2012ApJ...758...73S}. We model $t_{\rm dep,\H2}$ following \cite{2014arXiv1409.1171G}
\begin{equation}
t_{\rm dep,\H2} = 10^{9.08}\,{\rm yr}\,(1+z)^{-0.31}\,\mathcal{Q}_{\rm MS}^{-0.5}.
\label{eq:tdepH2}
\end{equation}
The term 
\[
\mathcal{Q}_{\rm MS} = \frac{\SFR}{\SFR_{\rm MS}(M_*, z)}
\]
in the above equation denotes the (multiplicative) offset of a given galaxy from the main sequence. We will assume that $\mathcal{Q}_{\rm MS}$ of a given galaxy does not change with time. 

The molecular gas depletion time enables us to convert SFRs into molecular ISM masses. Conversion to $M_{\rm g}$, however, requires knowledge of the $\H2$ fraction in the part of the ISM that is involved in both creation/destruction of dust and in the mixing of both dust and metals. For $z=0$ we adopt the stellar mass dependent scaling 
\begin{equation}
f_{\H2}=R_{\rm mol}/(1+R_{\rm mol})\text{ with }R_{\rm mol}=\xi{}(z)\,0.41\left(\frac{M_*}{10^{10.7} M_\odot}\right)^{0.425}
\label{eq:fH2}
\end{equation}
by \cite{2011MNRAS.415...32S}, based on observations of nearby galaxies over a broad range of stellar masses $M_*\sim{}10^{10}-10^{11.5}$ $M_\odot$. We added a fudge factor $\xi(z)$ to take into account that (i) the $\H2$ fraction is higher in the centers of galaxies and smaller at larger radii (e.g. \citealt{2008AJ....136.2782L}) and (ii) higher redshift galaxies are expected to have higher $\H2$ fractions at a fixed stellar mass given their larger gas surface densities \citep{2012MNRAS.424.2701F, 2013ApJ...778....2S}. We fit the prediction by \cite{2012MNRAS.424.2701F} (their ``Krumholz-$\H2$ prescription 1'') as function of stellar mass and redshift
\begin{equation}
\xi(z) = 1.2\times[1+z]^{1.6}.
\end{equation}
The shape of the $Z$ -- $D$ relation at $z=0$ is insensitive to the adopted scaling of $\xi(z)$ with redshift as long as $\xi(0)$ remains fixed. This result is a consequence of dust-to-gas ratios and metallicities of galaxies generally being close to the equilibrium prediction (equations \ref{eq:Zeq}, \ref{eq:Deq}), see section \ref{sect:ValidateEquilibriumAnsatz}. In the equilibrium limit, $Z$ and $D$ depend only on the current and not on the past values of galaxy properties.

\begin{figure}
\begin{tabular}{c}
\includegraphics[width=80mm]{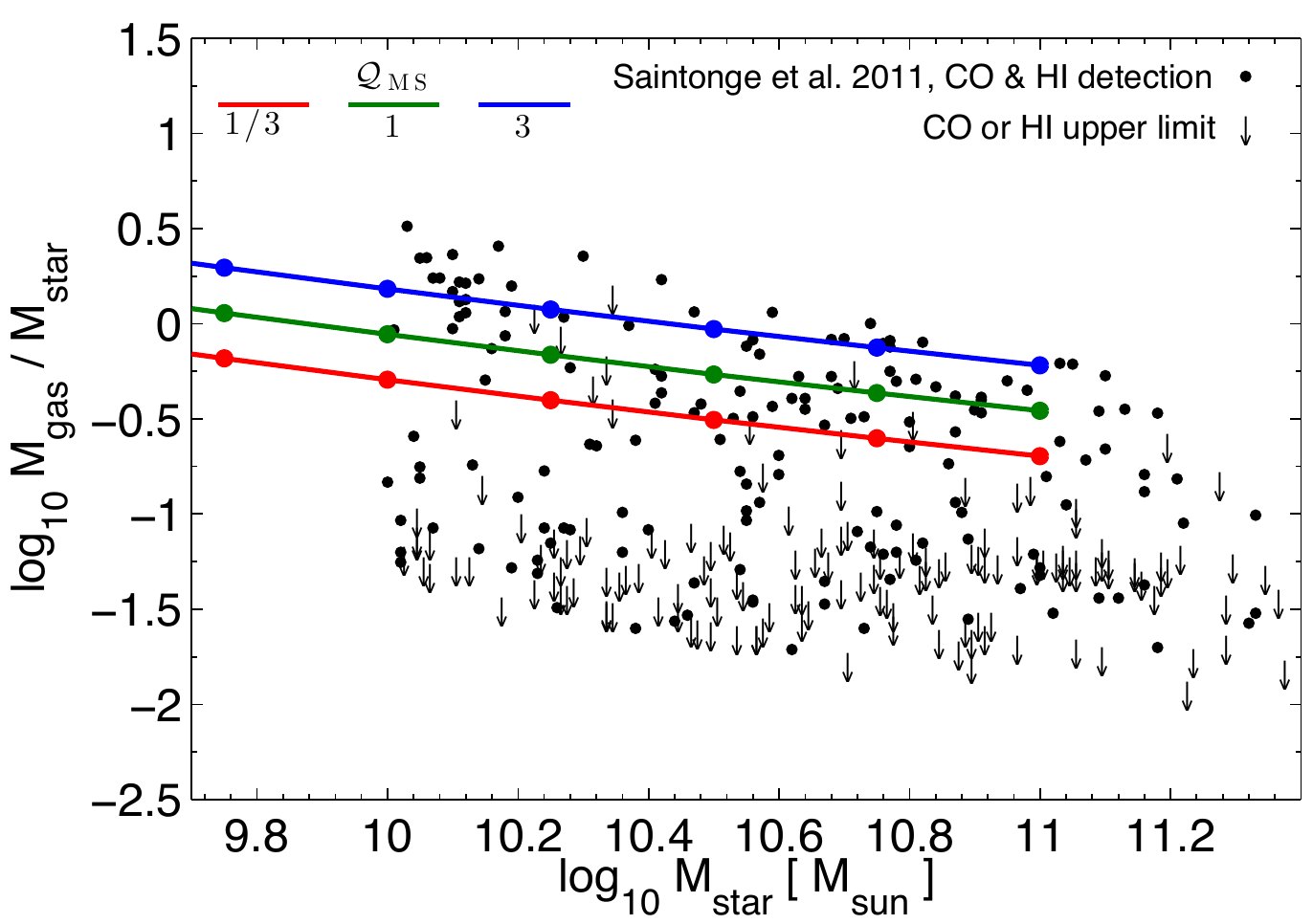}
\end{tabular}
\caption{ISM mass to stellar mass ratio $\mu$ as function of stellar mass at $z=0$. We combine measurements of SFRs (equation \ref{eq:SFRLilly}), molecular gas depletion times (equation \ref{eq:tdepH2}), and $\H2$ fractions (equation \ref{eq:fH2}) of galaxies to compute $\mu$ as function of stellar mass (solid lines), see text. Lines are colored according to the adopted offset from the main sequence of star formation (from top to bottom: $\mathcal{Q}_{\rm MS}=3,\,1,\,1/3$). Black circles show $\H2+\HI$ mass to stellar mass ratios based on CO and $\HI$ detections from the GASS \protect\citep{2010MNRAS.403..683C} and COLD GASS surveys \protect\citep{2011MNRAS.415...32S}. Arrows indicate upper limits on the $\H2+\HI$ mass to stellar mass ratios for galaxies without CO or $\HI$ detections.}
\label{fig:MgasMstar}
\end{figure}

The ratio between molecular gas mass and stellar mass of a galaxy, $M_\H2/M_*= t_{\rm dep,\H2}\,\sSFR$, is determined by equations (\ref{eq:SFRLilly}) and (\ref{eq:tdepH2}):
\begin{flalign}
  M_\H2/M_* =& \left(\frac{M_*}{10^{10}\,M_\odot}\right)^{-0.1} \, \mathcal{Q}_{\rm MS}^{0.5} \nonumber \\ 
                                    \times  &  \begin{cases} 
                                     0.175(1+z)^{2.69}  & \text{if $z\leq{}2$},\\ 
                                     0.757(1+z)^{1.357}& \text{if $z>2$}. \end{cases}
\label{eq:MH2Ms}                                     
\end{flalign}
The ISM mass to stellar mass ratio $\mu=M_{\rm g}/M_*$ of a galaxy is obtained by dividing $M_\H2/M_*$  by the $\H2$ fraction, equation (\ref{eq:fH2}). The so-computed $\mu$ compares well with the $\H2+\HI$ mass to stellar mass ratio measured by \cite{2011MNRAS.415...32S} for a sample of galaxies in the local Universe, see Fig.~\ref{fig:MgasMstar}.

The mass loading factor of galactic outflows has to increase with decreasing stellar mass of a galaxy in order to reproduce the empirically determined stellar mass -- halo mass relation (e.g. \citealt{2006MNRAS.373.1265O, 2009MNRAS.396..141D, 2010MNRAS.406.2325O}), and the mass -- metallicity relation (e.g., \citealt{2011MNRAS.417.2962P, 2012MNRAS.421...98D}). Furthermore, a dependence of the mass loading factor on the depth of the potential well of a galaxy (which is correlated with stellar mass) is expected on theoretical grounds resulting in specific power-law scaling for momentum or energy driven outflows, e.g., \cite{2005ApJ...618..569M}. Here we adopt a scaling of the mass loading factor that is in agreement with predictions based on self-consistent hydrodynamical simulations of galactic winds \citep{2012MNRAS.421.3522H}:
\begin{equation}
\epsilon_{\rm out}=2\,f_{\rm comb}\left(1, \left[ \frac{M_*}{10^{10} M_\odot} \right]^{-0.59}\right).
\end{equation}
$f_{\rm comb}(x,y) = x+y - (x^{-1} + y^{-1})^{-1}$ provides a smooth transition between high mass loading factors for low $M_*$ and our adopted lower limit $\epsilon_{\rm out}=2$ for large $M_*$. A significant fraction of the expelled material is expected to return, especially for more massive galaxies. We discuss some aspects of this ``wind recycling'' in section \ref{sect:PredDynModel}.

The relative metal enrichment $r_{\rm Z}$ of the inflowing gas is likely smaller at higher redshift when material that had been expelled in galactic fountains has not yet fallen back. We thus adopt a linear scaling with redshift similar to the one proposed by \cite{2012MNRAS.421...98D} based on hydrodynamical simulations (but without an explicit stellar mass dependence):
\begin{equation}
r_{\rm Z} = \max\left( 0, 0.25 - 0.05\,z \right). 
\end{equation}
While our choice is to some extent ad hoc, the precise value of $r_{\rm Z}$ plays little role as long as $r_{\rm Z}\ll{}1$, see equation (\ref{eq:Zeq}).

In order to numerically integrate equations (\ref{eq:dotMg} - \ref{eq:dotMZ}, \ref{eq:dotMd}) we need to specify some additional model parameters. We mostly stick to our choices from section \ref{sect:ZDrel}: $R = 0.46$, $y = 6.9\times{}10^{-2}$, $y_{\rm D} = 6.9\times{}10^{-4}$, $f^{\rm dep} = 0.7$, $r_D=0$, and $\epsilon_{\rm SN}=10$. To convert between redshift and time we adopt a \emph{Wilkinson Microwave Anisotropy Probe}-9 cosmology \citep{2013ApJS..208...19H} with the following parameters: $\Omega_{\rm m}=0.2821$, $\Omega_{\rm b}=0.0461$, $\Omega_\Lambda=1-\Omega_{\rm m}$, $\sigma_8=0.817$, $n=0.965$, $H_0=69.7$ km s$^{-1}$ Mpc$^{-1}$, and $f_{\rm bar}=0.163$.

\begin{figure*}
\begin{tabular}{cc}
\includegraphics[width=85mm]{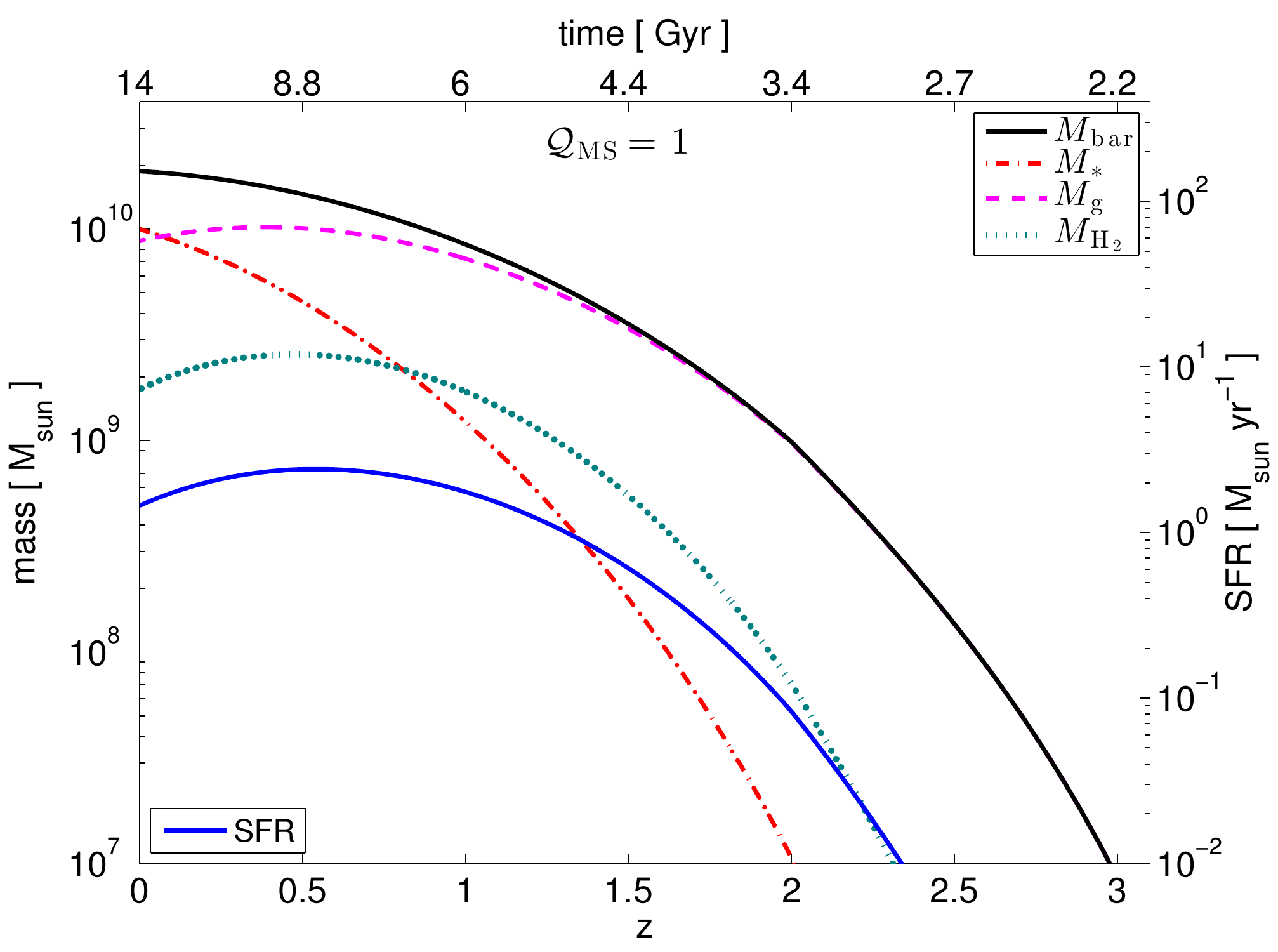} &
\includegraphics[width=85mm]{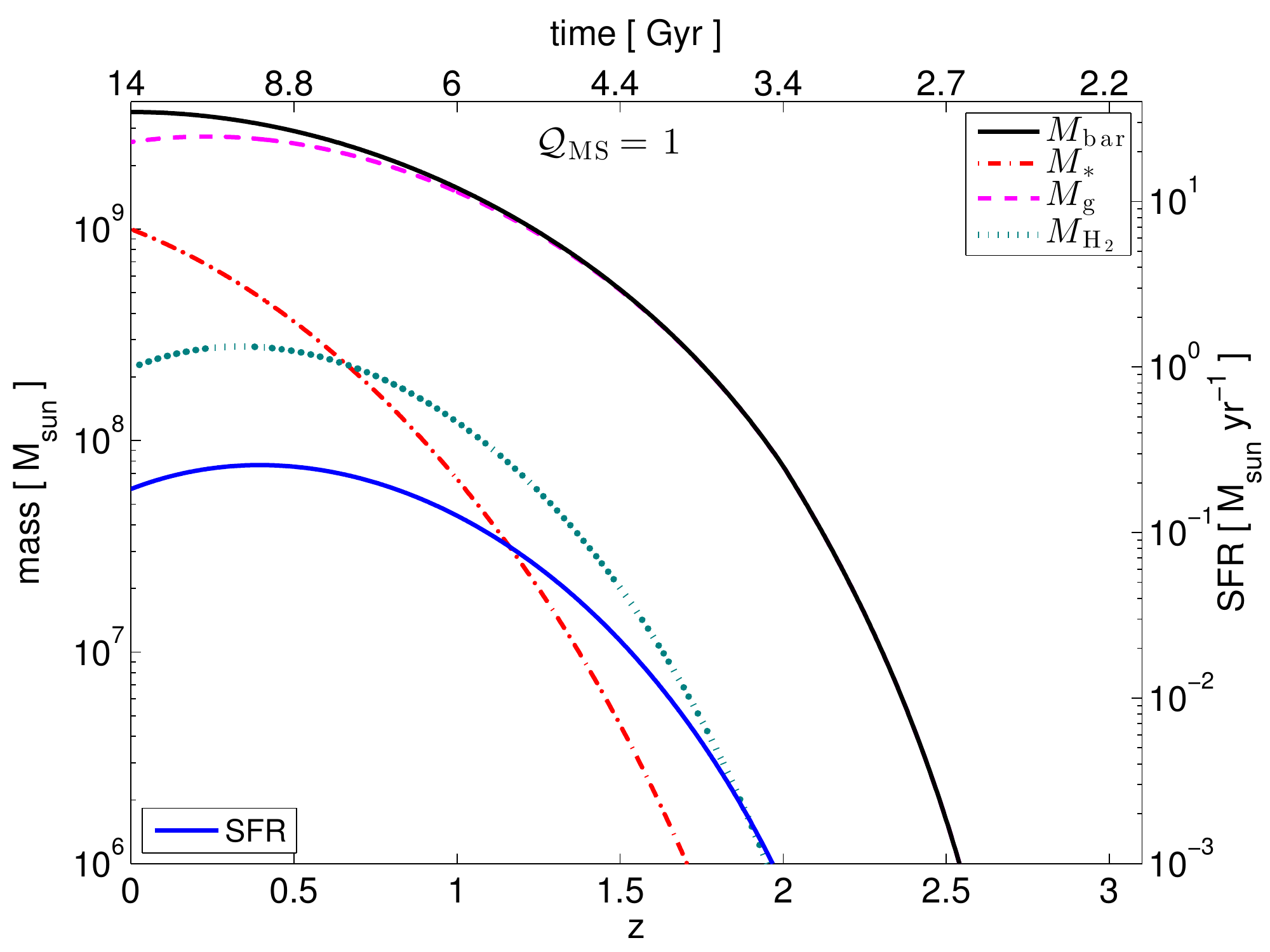} \\
\includegraphics[width=85mm]{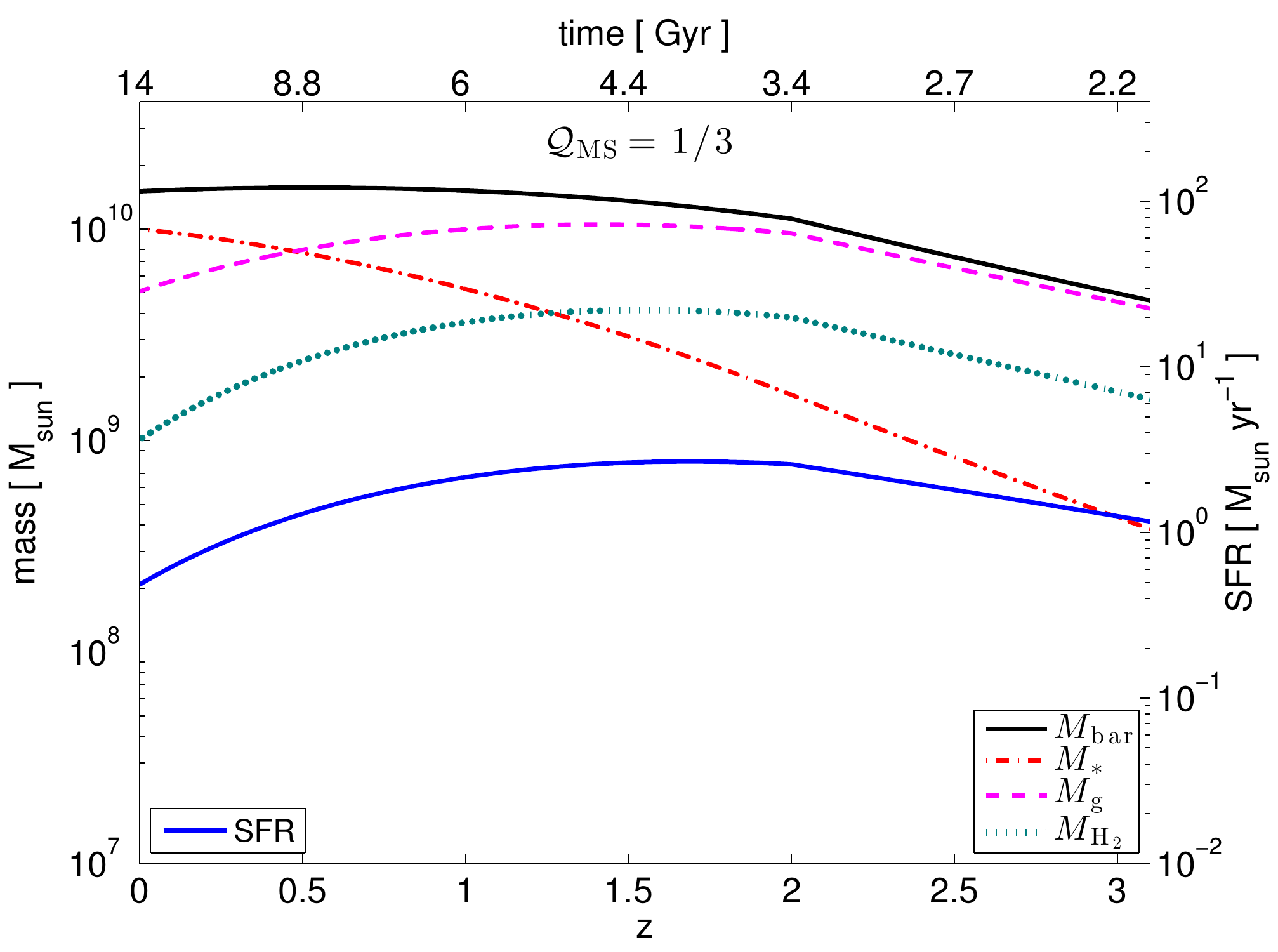} &
\includegraphics[width=85mm]{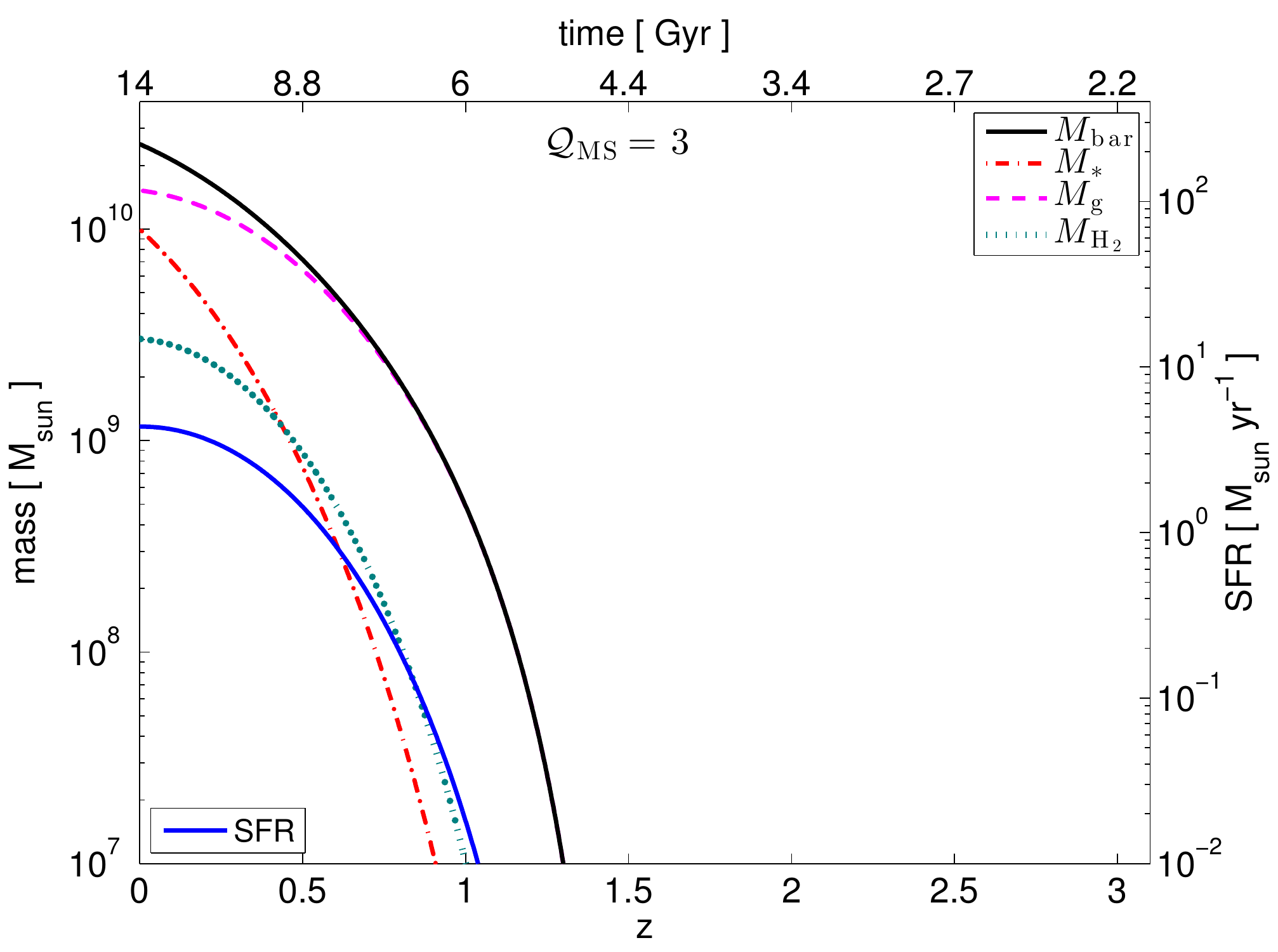} \\
\end{tabular}
\caption{Evolution of global galaxy properties as predicted by the dynamical model of section \ref{sect:DynModel} for our fiducial set of parameters. Each panel shows the baryonic mass of the galaxy (black solid lines on the top), the stellar mass (red dot-dashed lines), the gas mass (magenta dashed lines), the $\H2$ mass (green dotted lines), and the SFR (blue solid line, see axis on the right) as function of redshift (bottom axis) and time (top axis). (Top left) Masses and SFRs for a galaxy with a stellar mass of $10^{10}$ $M_\odot$ at $z=0$ that evolves on top of the main sequence of star formation, $\mathcal{Q}_{\rm MS}=1$. (Top right) same as top left, but for a galaxy with a stellar mass of $10^{9}$ $M_\odot$ at $z=0$. The $y$-axes are rescaled by a factor 1/10. Stellar masses and SFRs evolve in an approximately proportional way in high and low mass galaxies at fixed $\mathcal{Q}_{\rm MS}=1$. However, the gas-to-stellar ratio increases significantly with decreasing stellar mass. (Bottom left) Same as top left, but for a galaxy that evolves parallel to, but below, the main sequence of star formation, $\mathcal{Q}_{\rm MS}=1/3$. (Bottom right) Same as bottom left, but for a galaxy that lies above the main sequence of star formation, $\mathcal{Q}_{\rm MS}=1/3$. Galaxies that evolve below (above) the main sequence of star formation acquire much of their stellar mass at higher (lower) redshift and are more gas-poor (gas-rich) compared with galaxies on the main sequence.}
\label{fig:DynModelMassEvolution}
\end{figure*}

In Fig.~\ref{fig:DynModelMassEvolution}, we illustrate the evolution of various galaxy properties based on our one-zone dynamical model. Specifically, we plot how baryonic masses, stellar masses, gas masses, molecular gas masses, and SFRs change with time for galaxies with different stellar masses and different $\mathcal{Q}_{\rm MS}$. We find that galaxies lower sSFRs acquire their masses much earlier than galaxies with larger sSFRs. For instance, a galaxy with a $z=0$ mass of $10^{10}$ $M_\odot$ acquires 1/10-th (one half) of its final stellar mass (i) by $z\sim{}1.1$ ($z\sim{}0.45$) if it lives on the main sequence, (ii) by $z\sim{}0.46$ ($z\sim{}0.18$) if it lives a factor 3 above the main sequence, and (iii) by $z\sim{}2.4$ ($z\sim{}1.0$) if it lives a factor 3 below the main sequence. Hence, even a small (but persistent) offset from the main sequence of star formation has a large impact on the growth history of galaxies. The mass assembly also depends on the final stellar mass, although only weakly. For instance, a galaxy with a $z=0$ mass of $10^{9}$ $M_\odot$ acquires 1/10-th (one half) of its final stellar mass by $z\sim{}0.90$ ($z\sim{}0.38$), i.e., at a slightly later time than a galaxy with a $z=0$ mass of $10^{10}$ $M_\odot$.

\subsection{Predictions of the dynamical model}
\label{sect:PredDynModel}

The dynamical model fits a number of observations by construction, e.g., the SFR - stellar mass relation or the evolution of the gas depletion time. In this section, we test two genuine predictions of the model against observational data: the stellar mass -- ISM metallicity relation and the stellar mass -- halo mass relation. We discuss the $Z$ -- $D$ relation in section \ref{sect:ValidateEquilibriumAnsatz}.

\begin{figure}
\begin{tabular}{c}
\includegraphics[width=85mm]{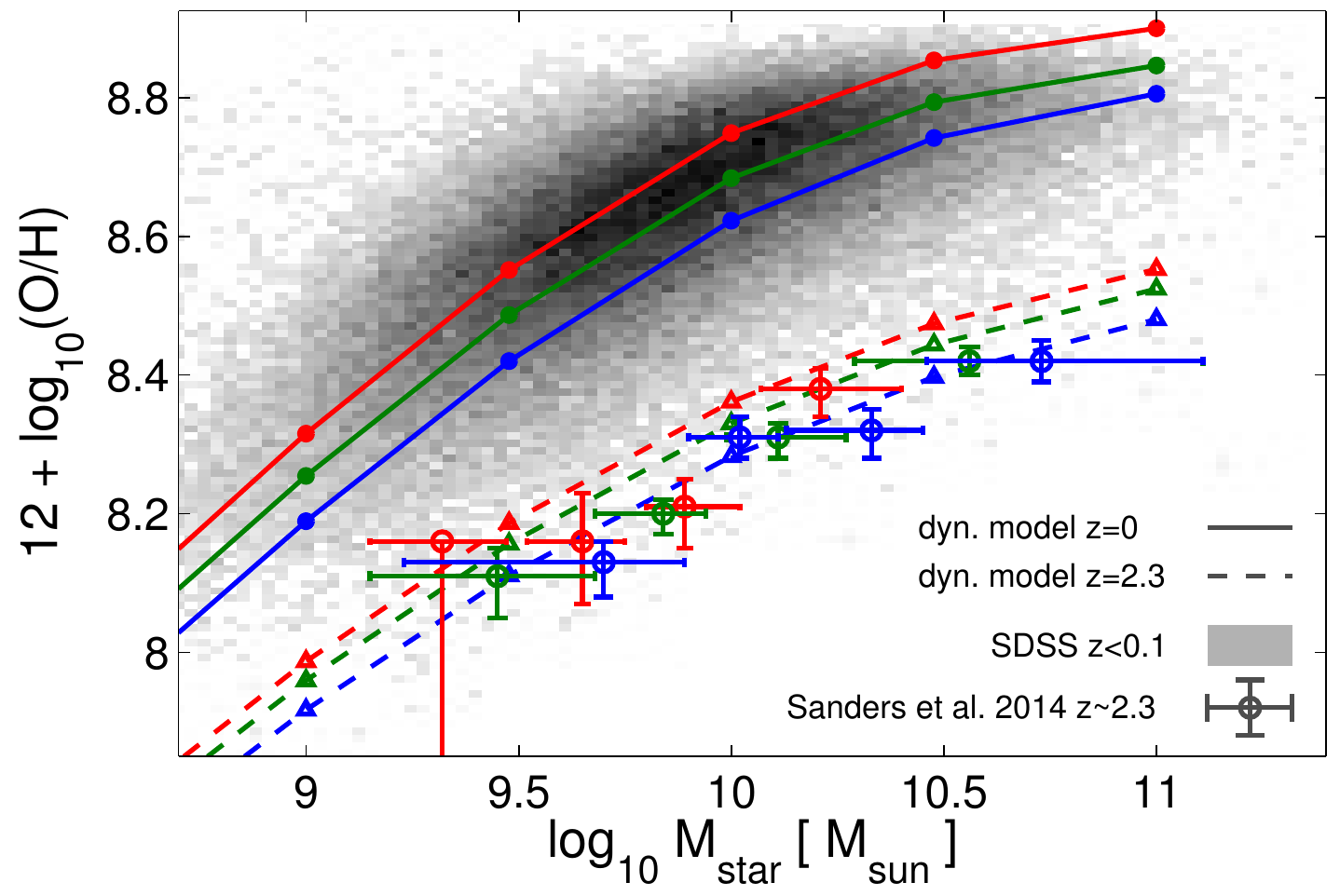}
\end{tabular}
\caption{Stellar mass - ISM metallicity relation as predicted by the dynamical model of section \ref{sect:DynModel} for our fiducial set of parameters. We compare the model predictions for the $z=0$ (solid lines) and $z=2.3$ (dashed lines) mass -- metallicity relation with measurements for local galaxies (SDSS; from \protect\citealt{2014arXiv1408.2521S}) and star forming galaxies at $z\sim{}2.3$ \protect\citep{2014arXiv1408.2521S}. The observational data uses the O3N2 metallicity indicator for both low and high redshift galaxies to minimize biases that arise from the metallicity calibration \citep{2008ApJ...681.1183K}. Blue, green, and red lines show samples binned according to their sSFR: $\mathcal{Q}_{\rm MS}=3$, 1, 1/3 (at $z=0$) and $\mathcal{Q}_{\rm MS}=0.75$, 1, 1.5 (at $z=2.3$). The lower range of $\mathcal{Q}_{\rm MS}$ at $z=2.3$ was chosen to roughly mimic the spread in SFRs in the observational data set by \protect\cite{2014arXiv1408.2521S}. The dynamical model reproduces the observed stellar-mass metallicity relation in shape, normalization, and redshift evolution. Furthermore, it predicts a dependence on the offset from the main sequence of star formation in (at least qualitative) agreement with observations \protect\citep{2010MNRAS.408.2115M, 2013ApJ...765..140A, 2013MNRAS.434..451L, 2014arXiv1408.2521S}.}
\label{fig:MZ}
\end{figure}

We show in Fig.~\ref{fig:MZ} the stellar mass -- ISM metallicity relation as predicted by the dynamical model. In particular, we plot $12 + \log_{10}({\rm O}/{\rm H}) = 8.69 + \log_{10}(Z/0.014)$ at $z=0$ and $z=2.3$ for a range of stellar masses and sSFRs. The ISM metallicities increase with decreasing redshift in good agreement with observational data (Sanders et al. 2014). Furthermore, we reproduce the observed trend of galaxies with larger sSFR having lower metallicities (e.g., \citealt{2010MNRAS.408.2115M, 2013ApJ...765..140A, 2013MNRAS.434..451L, 2014arXiv1408.2521S}).

The dynamical model predicts that offsets from the main sequence of star formation affect the ISM metallicity more strongly at higher redshift. In Fig.~\ref{fig:MZ}, we vary $\mathcal{Q}_{\rm MS}$ by a factor 3 at $z=0$ and by $\leq{}50\%$ at $z=2.3$, yet the change in metallicity is similar ($\sim{}0.04-0.05$ dex). The equilibrium framework offers an explanation for this result via equation (\ref{eq:r}). The normalization of the main sequence of star formation increases with redshift. Hence, the ratio between SFR and gas inflow rate, $r$, becomes more sensitive to the first term in the brackets in the equation, i.e., the term proportional to the sSFR, at higher $z$. The ISM metallicity is controlled by the value of $r$, or more precisely by $r/(1-r_{\rm Z})$, via equation (\ref{eq:Zeq}).

The stellar mass -- halo mass relation provides a further test of our dynamical model. Observationally, this relation can be derived from, e.g., galaxy clustering (e.g., \citealt{2003MNRAS.339.1057Y, 2004ApJ...608...16Z, 2009MNRAS.392.1080S}), weak lensing (e.g., \citealt{2006MNRAS.368..715M}), satellite kinematics (e.g., \citealt{1996ApJ...462...32C, 2011MNRAS.410..210M}), or abundance matching (e.g., \citealt{1999ApJ...520..437K, 2006ApJ...647..201C, 2010MNRAS.404.1111G, 2013MNRAS.428.3121M, 2013ApJ...770...57B}). Conciliating the predictions of galaxy evolution models with the observed stellar mass -- halo mass relation proved challenging for a long time. Even now models struggle at reproducing the evolution of this relation (e.g., \citealt{2011MNRAS.413..101G, 2012MNRAS.423.1992S, 2013MNRAS.428..129S, 2014ApJ...795..123L}). However, recent progress in the modeling of feedback processes (especially galactic winds) and star formation physics lead to an improved match with observations \citep{2013ApJ...766...56M, 2013MNRAS.431.3373H, 2014MNRAS.445..581H, 2014MNRAS.444.1518V, 2015MNRAS.446..521S}.

Our dynamical model tracks explicitly stellar, metal, dust, and gas masses of galaxies, but not the masses of their parent dark matter halos, $M_{\rm halo}$. We estimate the latter from the gas accretion rate on to the galaxy, $\dot{M}_{\rm g, in}$, and the gas outflow rate $\dot{M}_{\rm g, out}=\epsilon_{\rm out}\SFR$. We assume that a fraction $\chi{}_{\rm rec}$ of the currently outflowing mass is falling back on to the galaxy and contributes to $\dot{M}_{\rm g, in}$. Hence, the gas accretion rate \emph{on to the halo} is $\dot{M}_{\rm g, in} - \chi{}_{\rm rec}  \dot{M}_{\rm g, out}$ and the halo mass is
\begin{equation}
M_{\rm halo}(t) =\frac{1}{f_{\rm bar}} \int_0^t \left[ \dot{M}_{\rm g, in}(t') - \chi{}_{\rm rec}  \dot{M}_{\rm g, out}(t') \right] dt'.
\label{eq:Mhalo}
\end{equation}

The wind recycling fraction  $\chi{}_{\rm rec}$ at $z=0$ is not well constrained observationally. Hydrodynamical simulations and semi-analytical models \citep{2008MNRAS.387..577O, 2010MNRAS.406.2325O, 2013MNRAS.431.3373H} predict that wind recycling plays a larger role at lower redshift, i.e., that $\chi{}_{\rm rec}$ increases with decreasing $z$. These models also suggest that more massive halos have shorter return times because of their deeper potential wells and the larger ram-pressure forces resulting from higher densities of the circum-galactic medium. 

We adopt $\chi{}_{\rm rec}(z=0)=0.5$ and $\chi{}_{\rm rec}(z=2)=0.35$. We decided not to model the halo mass dependence of $\chi{}_{\rm rec}$ to avoid introducing additional free parameters into our model. Furthermore, we found that the precise value of $\chi{}_{\rm rec}$ has only a slight impact on $M_{\rm halo}$. For instance, changing $\chi{}_{\rm rec}(z=2)$ from 0.35 to 0.45 increases halo masses typically by $\sim{}0.06$ dex.

\begin{figure*}
\begin{tabular}{cc}
\includegraphics[width=85mm]{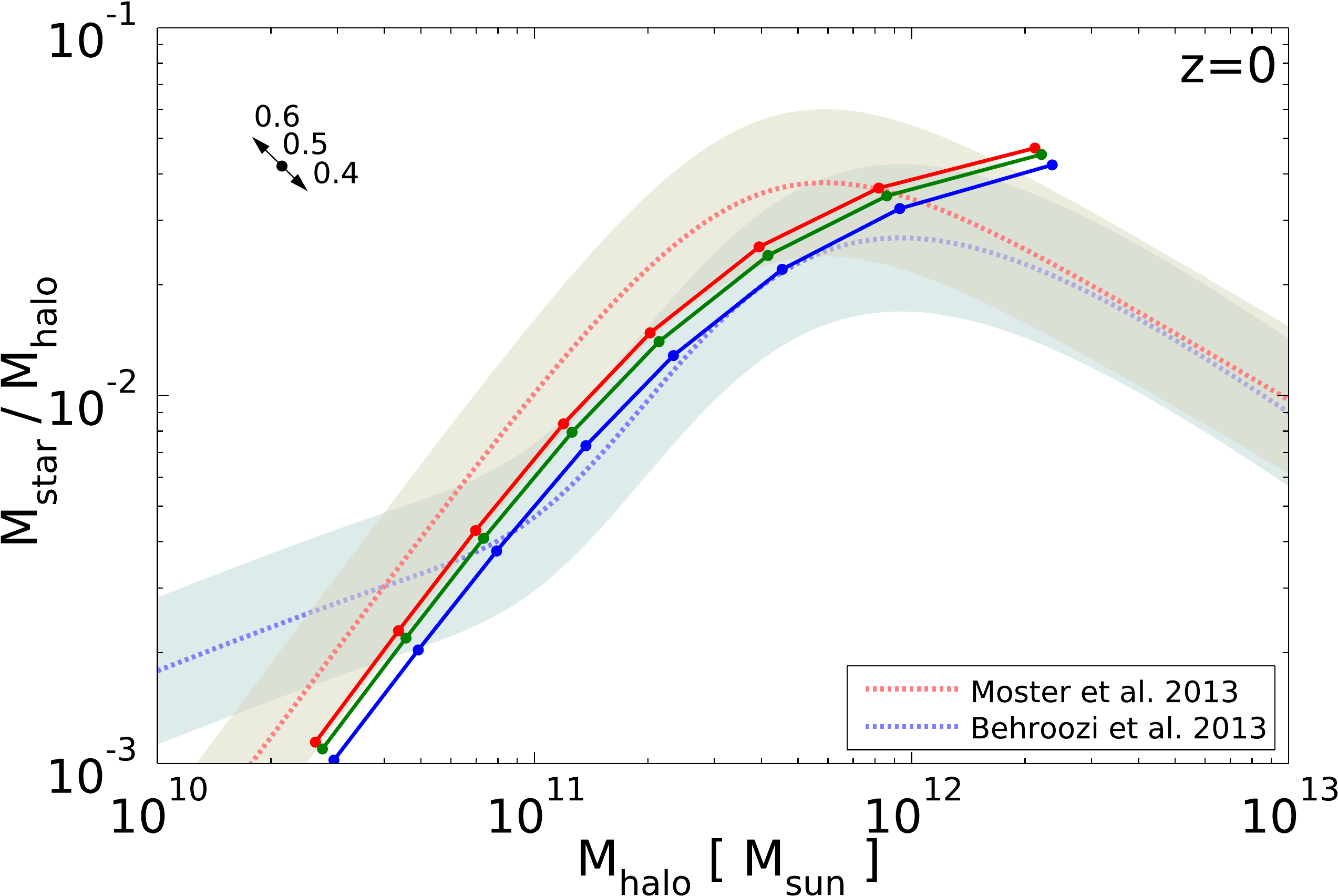} &
\includegraphics[width=85mm]{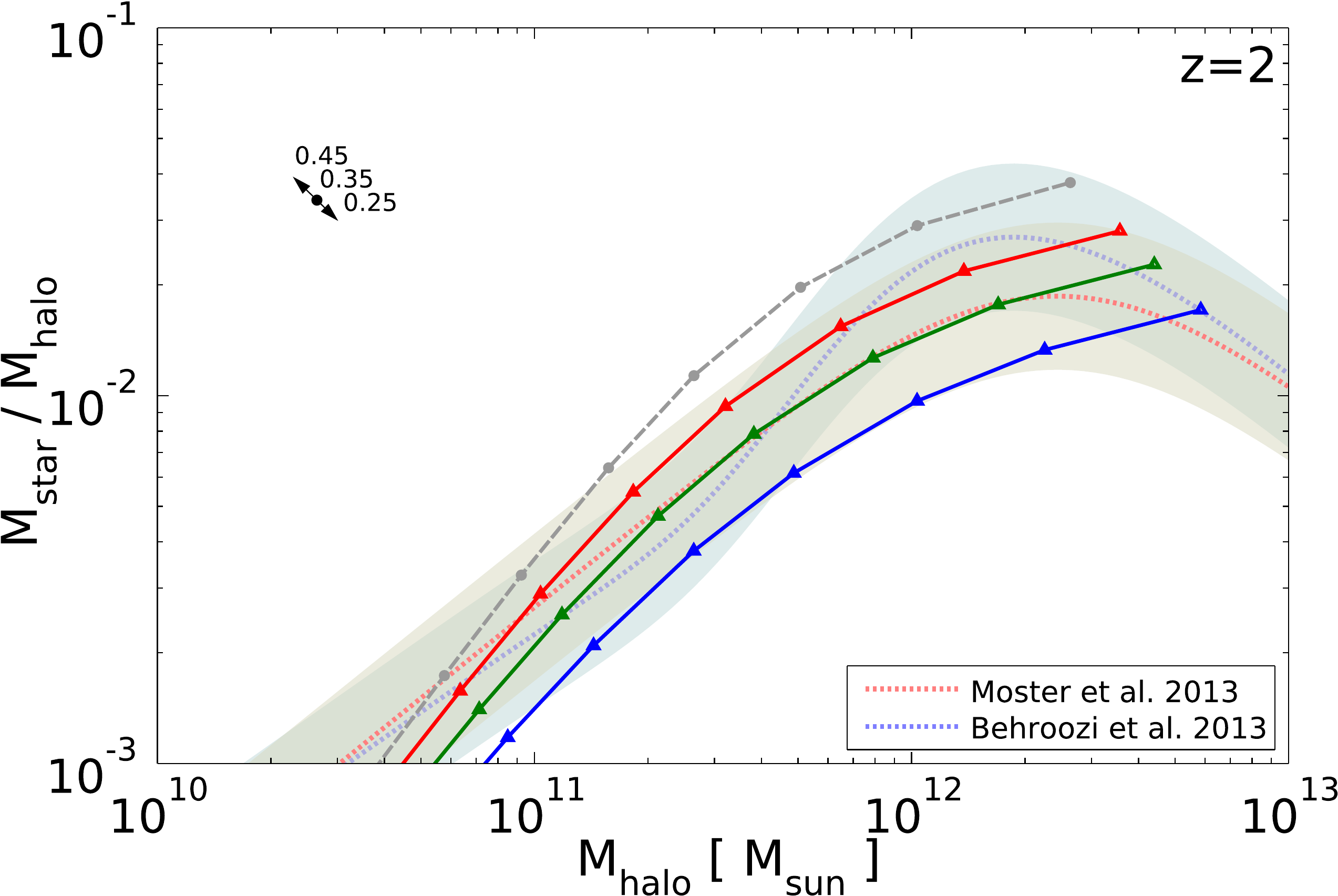} \\
\end{tabular}
\caption{Stellar mass to halo mass ratio (SHMR) as predicted by the dynamical model of section \ref{sect:DynModel} for our fiducial set of parameters. We show predictions for $z=0$ in the left-hand panel and for $z=2$ in the right-hand panel. Blue, green, and red solid lines in each plot correspond to galaxies above ($\mathcal{Q}_{\rm MS}=3$), on ($\mathcal{Q}_{\rm MS}=1$), and below ($\mathcal{Q}_{\rm MS}=1/3$) the main sequence of star formation, respectively. Circles denote stellar masses from $3\times{}10^7$ to $10^{11}$ $M_\odot$ in 0.5 dex steps (bottom left to top right). Dotted lines show empirical constraints by \protect\cite{2013MNRAS.428.3121M} and \protect\cite{2013ApJ...770...57B}, see legend. The shaded region around the dotted region indicates a scatter of 0.2 dex \protect\citep{2013ApJ...770...57B, 2013ApJ...771...30R}. The double-sided arrow at the top left shows how wind recycling (see the text) affects the predictions of the dynamical model. We adopt wind recycling fractions of 0.5 at $z=0$, and 0.35 at $z=2$. The gray dashed curve in the right-hand panel corresponds to the $z=0$ \& $\mathcal{Q}_{\rm MS}=1$ SHMR of the left-hand panel but with a wind recycling fraction of 0.35. The SHMR changes not only with $M_{\rm halo}$, but also with $\mathcal{Q}_{\rm MS}$ and with redshift. The dynamical model reproduces the empirically determined stellar mass -- halo mass relation for galaxies in halos with $M_{\rm halo}\lesssim{}10^{12}$ $M_\odot$.}
\label{fig:MstarMhalo}
\end{figure*}

Fig.~\ref{fig:MstarMhalo} shows the stellar mass -- halo mass relation as predicted by our dynamical model. The stellar mass to halo mass ratio (SHMR) changes with halo mass in agreement with empirical constraints \citep{2013MNRAS.428.3121M, 2013ApJ...770...57B} over a large range of stellar masses $M_*\sim{}3\times{}10^{7}$ -- $3\times{}10^{10}$ $M_\odot$. Our dynamical model does \emph{not} reproduce the observed drop in the SHMR at $M_{\rm halo}\gtrsim{}10^{12}$ $M_\odot$, however. This is expected as we do not model the quenched galaxy population that dominates the average value of the SHMR at the high mass end.

Star forming galaxies with larger $\mathcal{Q}_{\rm MS}$ reside in more massive halos. This trend is relatively inconspicuous at $z=0$ but gets more pronounced with redshift. The figure also shows that massive star forming galaxies (those with $M_*>3\times{}10^{10}$ $M_\odot$) have larger SHMRs than the average population of observed massive galaxies (most of them are quenched). Hence, massive quenched galaxies are predicted to have larger halo masses than star forming galaxies of the same stellar mass, see also \cite{2011MNRAS.410..210M} and \cite{2014arXiv1408.5407R}.

Fig.~\ref{fig:MstarMhalo} demonstrates further that the stellar mass -- halo mass relation evolves with redshift. A halo of a given mass hosts a lower mass galaxy at $z=2$ compared with $z=0$. Similarly, galaxies of a specified stellar mass live in more massive halos at higher redshift. Again this finding is in quantitative agreement with empirical measurements of the evolution of the stellar mass -- halo mass relation \citep{2013MNRAS.428.3121M, 2013ApJ...770...57B}.

The equilibrium framework explains most of these findings. Ignoring $\chi{}_{\rm rec}$, we see from equations (\ref{eq:dotMs}, \ref{eq:Mhalo}) that $\dot{M}_*/\dot{M}_{\rm halo}\sim{}f_{\rm bar}\,r$. Hence, galaxies with low $r$ will have low SHMRs, while galaxies with large $r$ will have large SHMRs. Low mass galaxies have low $r$ because their mass loading factors are large, see equation (\ref{eq:r}). In fact, the scaling of the mass loading factor with stellar mass imprints the corresponding scaling of the SHMR (see also \citealt{2013ApJ...772..119L}). Similarly, $r$ decreases as the sSFR increases for galaxies of a given stellar mass. Hence, at fixed stellar mass a positive offset from the main sequence of star formation implies larger halo masses. Finally, galaxies have larger sSFR and, hence, smaller $r$ at higher $z$, which explains the redshift evolution of the stellar mass -- halo mass relation.

We note that our model also reproduces (within systematic uncertainties) the observed stellar mass functions at $z=0$ and $z=2$  as a corollary of matching the stellar mass -- halo mass relation.

\subsection{Validating the equilibrium ansatz}
\label{sect:ValidateEquilibriumAnsatz}

The dynamical model provides us with a simple framework to test whether the equilibrium dust-to-gas ratio, equation (\ref{eq:Deq}), is indeed a good proxy for the actual dust-to-gas ratio of galaxies. In section \ref{sect:eqD}, we showed that the dust-to-gas ratio approaches the equilibrium value (on time scales of $\sim{}t_{\rm dep}$) for various simple functional forms of the gas inflow rate and a constant $\epsilon_{\rm out}$ and $t_{\rm dep}$. Here, we want to explore whether the equilibrium prediction (equation \ref{eq:Deq}) also holds under more realistic conditions, i.e., for mass loading factors, gas depletion times, and $\H2$ fractions etc. that evolve with the growing galaxy and for gas inflow rates that are derived from integrating the observed stellar mass -- SFR relation backward in time.

We compute the equilibrium predictions as follows. For a given $Z$, we compute $r$ via equation (\ref{eq:Zeq}) and insert $Z$ and $r$ into equation (\ref{eq:beta}) to compute $\langle{}D\rangle{}$ via equation (\ref{eq:Deq}). We use $\langle{}D\rangle{}/Z$ as our equilibrium estimate for $D/Z$. This procedure is slightly different from the alternative choice of computing $\langle{}Z\rangle{}$ and $\langle{}D\rangle{}$ based on the current value of $r$ given by the dynamical model. However, our adopted approach is advantageous for a number of reasons. First, we are interested in using the equilibrium ansatz to study the observed $Z$ -- $D$ relation. Hence, we want to be able to predict the dust-to-gas ratio based on the actual ISM metallicity $Z$ of a galaxy. Secondly, we find that computing $\langle{}D\rangle{}$ based on $Z$ results in a somewhat better agreement with the actual dust-to-gas ratios. Finally, $Z$ is observationally more accessible than $r$.

\begin{figure*}
\begin{tabular}{cc}
\includegraphics[width=85mm]{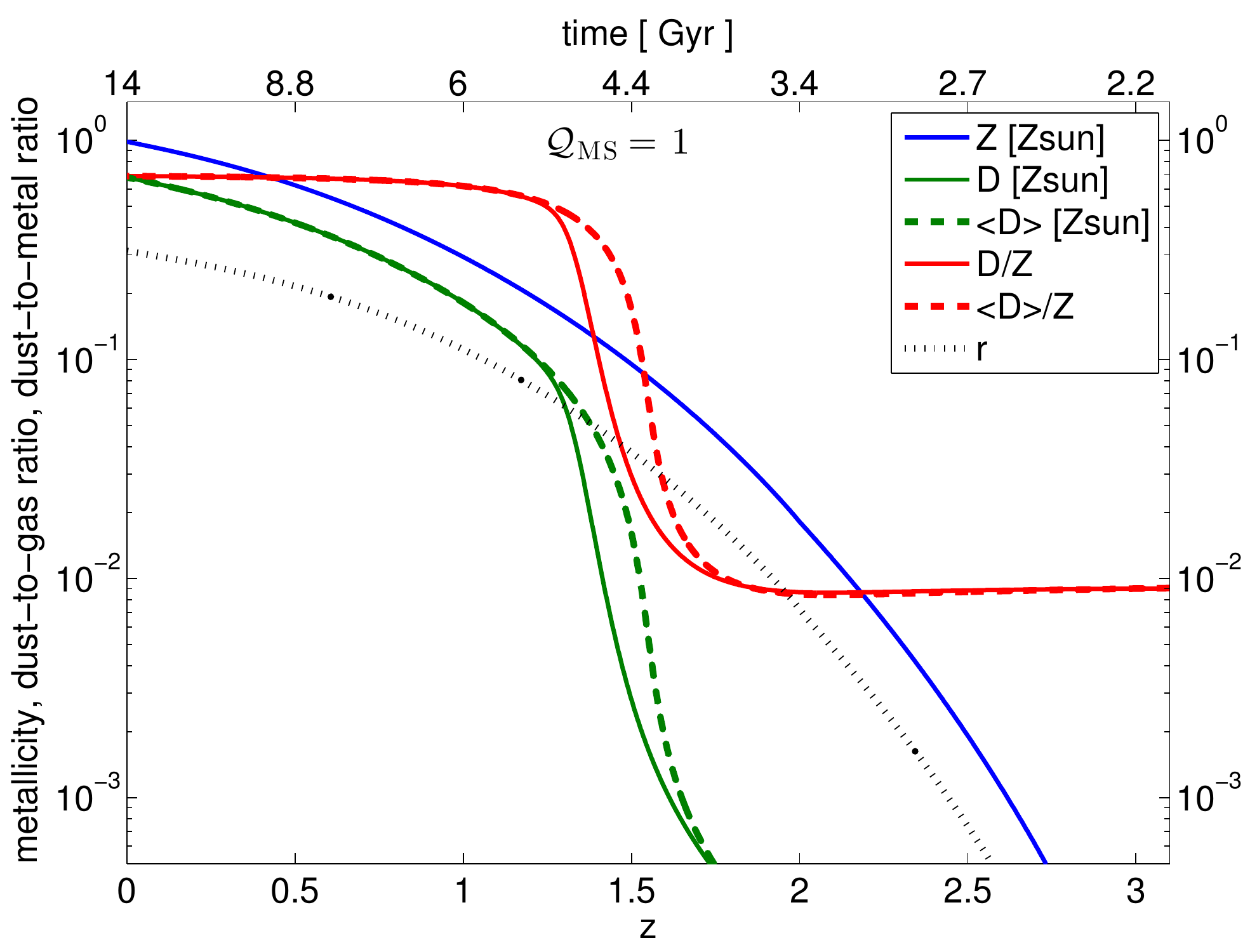} &
\includegraphics[width=85mm]{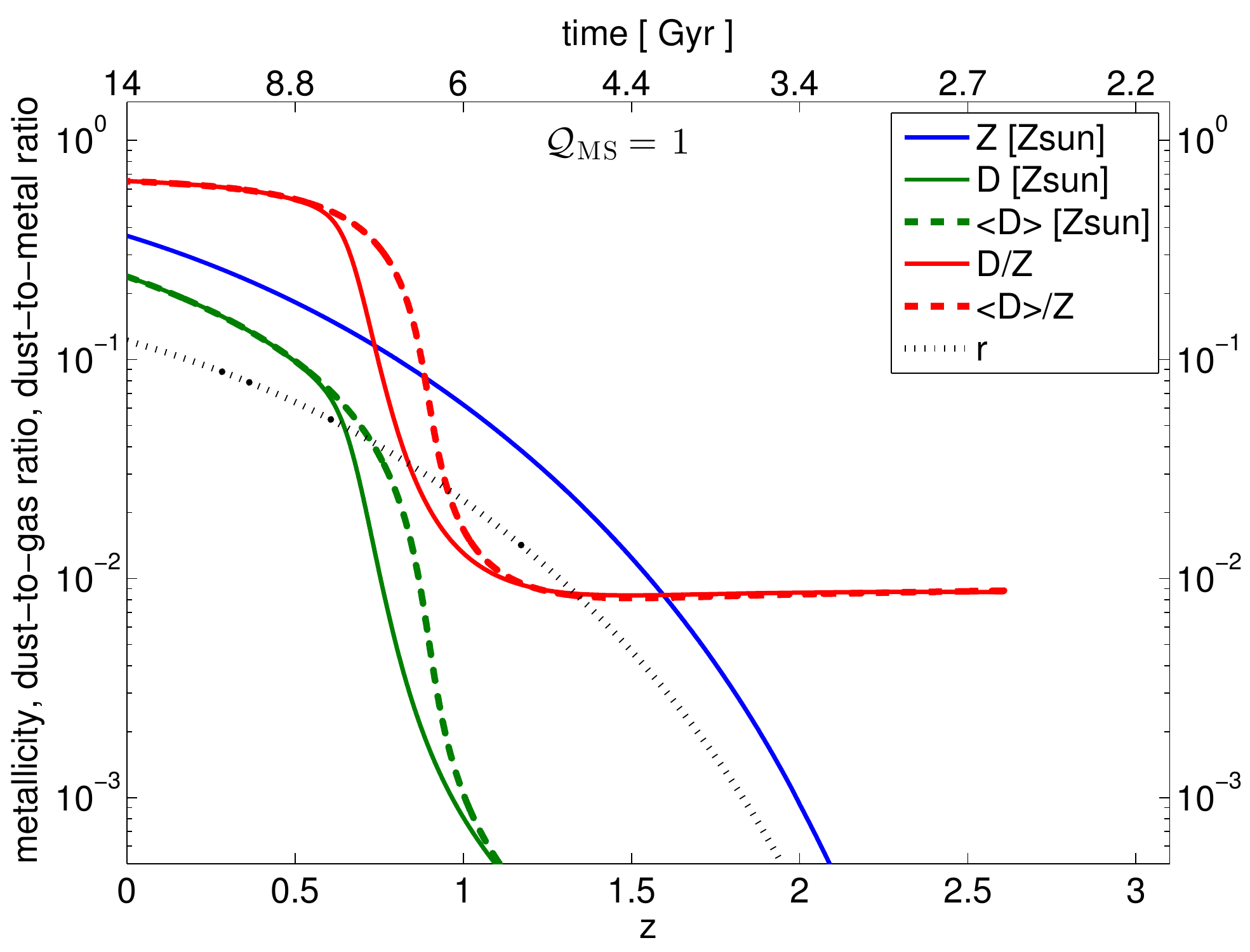} \\
\includegraphics[width=85mm]{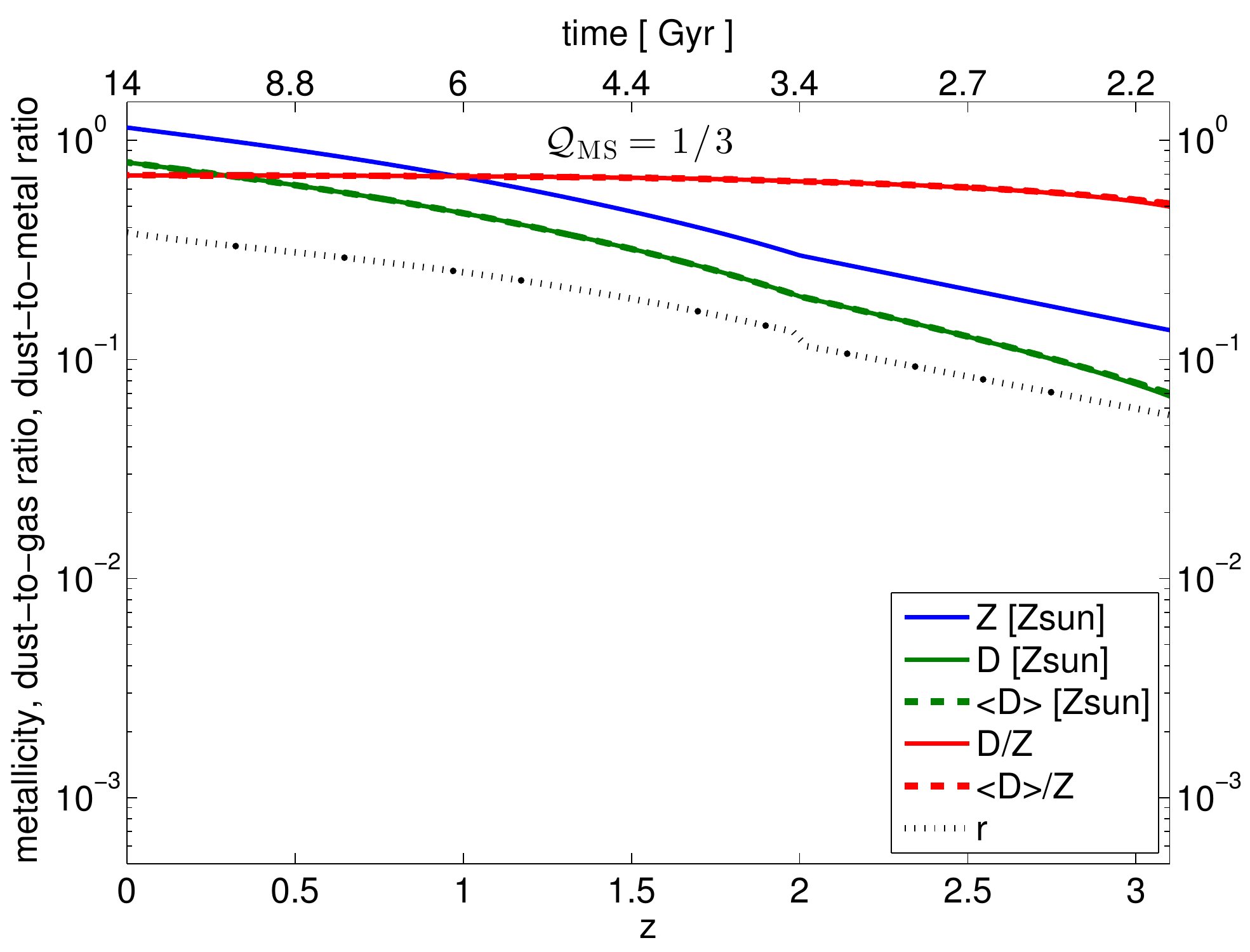} &
\includegraphics[width=85mm]{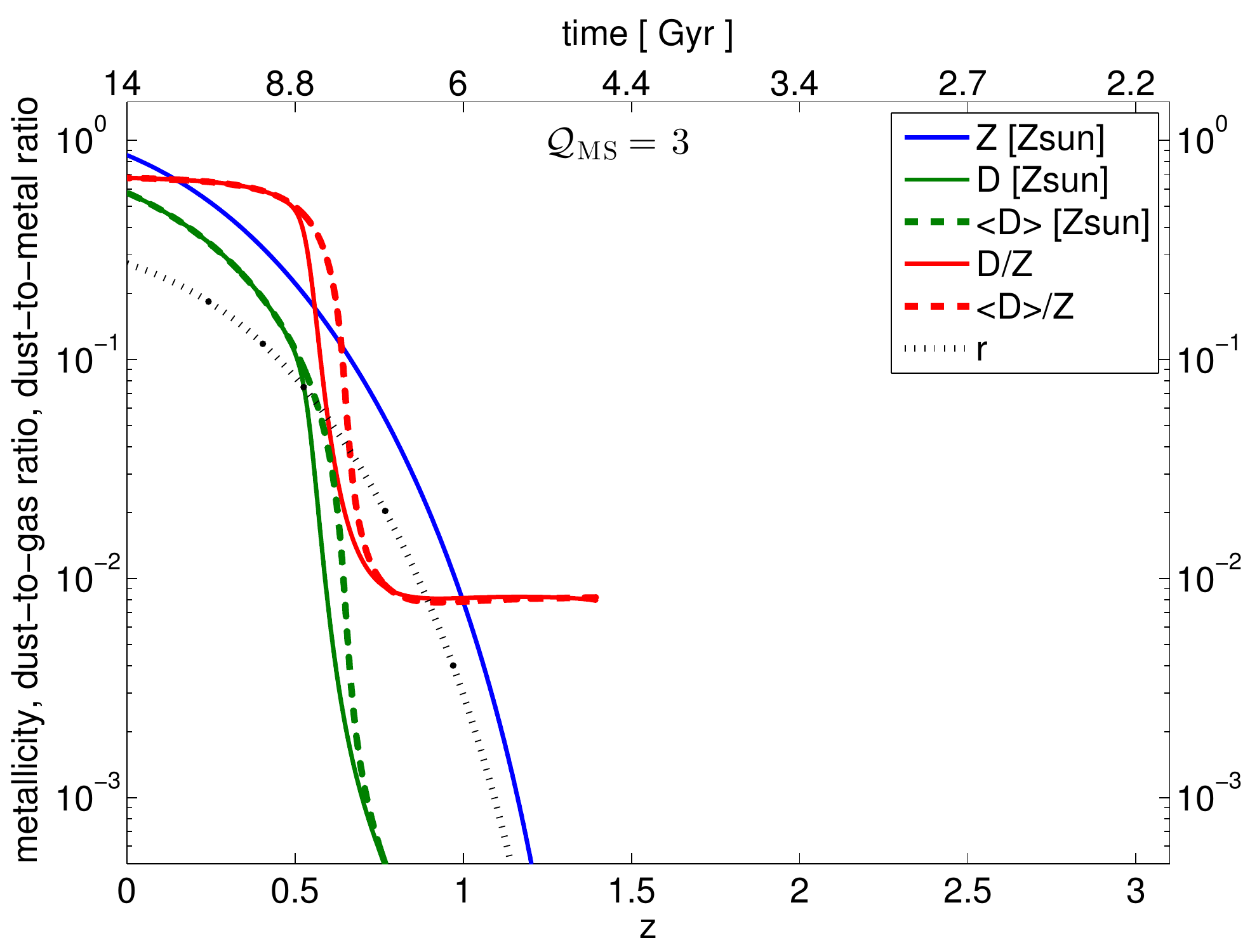} \\
\end{tabular}
\caption{Comparison between the predictions of the dynamical model and those based on the equilibrium ansatz. Each panel shows the ISM metallicity ($Z$, blue solid line), the dust-to-gas ratio ($D$, green solid line), the dust-to-metal ratio ($D/Z$, red solid line), and the ratio between SFR and gas inflow rate ($r$, black dotted line) as predicted by the dynamical model of section \ref{sect:DynModel} for our fiducial set of parameters.  The individual panels in this figure correspond to the panels in Fig.~\ref{fig:DynModelMassEvolution}. (Top left) for main-sequence galaxy with a stellar mass of $10^{10}$ $M_\odot$ at $z=0$. (Top right) same as top left, but for a galaxy with a stellar mass of $10^{9}$ $M_\odot$ at $z=0$. (Bottom left) Same as top left, but for a galaxy with $\mathcal{Q}_{\rm MS}=1/3$. The  redshift dependence of the sSFR changes at $z=2$, see equation (\ref{eq:SFRLilly}), resulting in a slight kink in $r$ at that time. (Bottom right) Same as bottom left, but for a galaxy with $\mathcal{Q}_{\rm MS}=1/3$. 
Dashed lines show the equilibrium predictions: the instantaneous equilibrium dust-to-gas ratio ($\langle{}D\rangle{}$, green dashed line) as well as the instantaneous dust-to-metal ratio ($\langle{}D\rangle{}/Z$, red dashed line). $\langle{}D\rangle{}$ follows from equation (\ref{eq:Deq}) upon inserting the current metallicity $Z$ and $r(Z)$ via equation (\ref{eq:Zeq}). 
The equilibrium ansatz slightly overestimates the redshift at which the transition from low to high dust-to-metal ratios takes place. However, outside this (brief) transition regime, $D$ and $D/Z$ of the dynamical model are in excellent agreement with the predictions based on the equilibrium approach. Lower mass galaxies and galaxies that live above the main sequence of star formation generally experience the transition from low to high dust-to-metal ratios at lower redshifts.}
\label{fig:DynModelDZeq}
\end{figure*}

In Fig.~\ref{fig:DynModelDZeq}, we compare the equilibrium predictions with the current values of $D$ and $D/Z$ for galaxies with various stellar masses and offsets from the main sequence of star formation. The equilibrium predictions match the actual time-dependent solutions well, especially outside the (brief) transitionary phase during which $D/Z$ increases by two orders of magnitude.

We can understand this behavior by analyzing the range of applicability of the equilibrium ansatz $\dot{D}\equiv\dot{Z}\equiv{}0$. In section \ref{sect:EquivZ} we showed that $\dot{Z}\sim{}0$ is a good ansatz as long as $t_{\rm dep}\dot{r}\ll{}1$, or, equivalently, as long as $t_{\rm dep}\dot{Z}\ll{}y(1-R)$. This inequality holds for most galaxies because (i) the metallicity increases rather steadily with time, i.e., $\dot{Z}\sim{}Z/t_{\rm H}$, (ii) $t_{\rm dep}/t_{\rm H}\lesssim{}1$, and (iii) $y$ is an upper limit to $Z$. Similarly, it is straightforward to show that $\dot{D}\sim0$ is a good approximation if $t_{\rm dep}\dot{D}\ll{}y_{\rm D}(1-R)$. The inequality is generally satisfied at low dust-to-metal ratios (here $D\sim{}Zy_{\rm D}/y$). At large dust-to-metal ratios, the ISM dust-to-gas ratio approaches $\beta/\gamma$, whether the above inequality is satisfied or not. Hence, the equilibrium dust-to-gas ratio is an accurate predictor of the actual ISM dust-to-gas ratio also in this case. The equilibrium ansatz is least accurate at intermediate dust-to-metal ratios when $\dot{D}$ is large, i.e., during a time of fast increase of the dust-to-gas ratio.

\begin{figure*}
\begin{tabular}{cc}
\includegraphics[width=85mm]{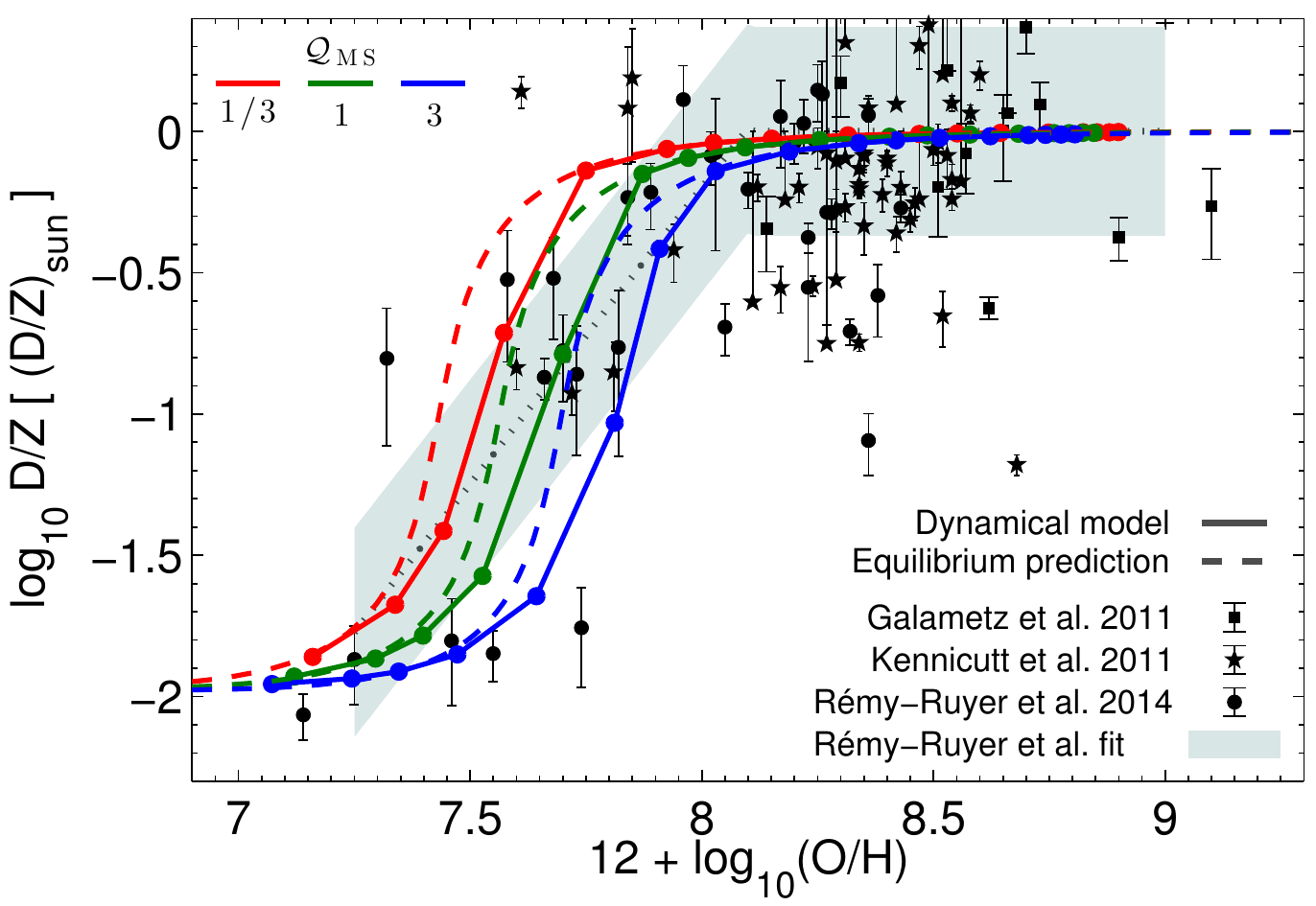} &
\includegraphics[width=85mm]{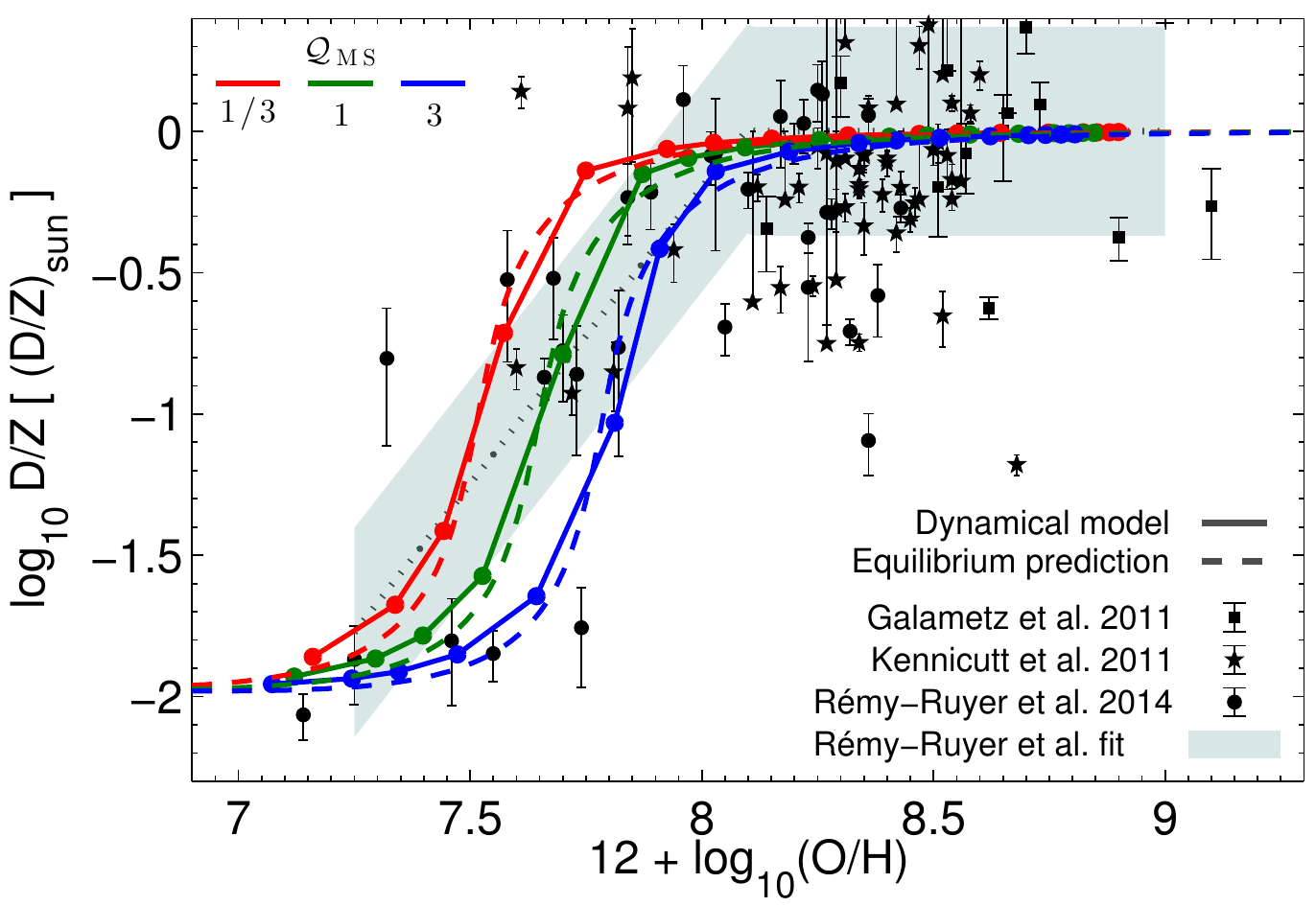} \\
\includegraphics[width=85mm]{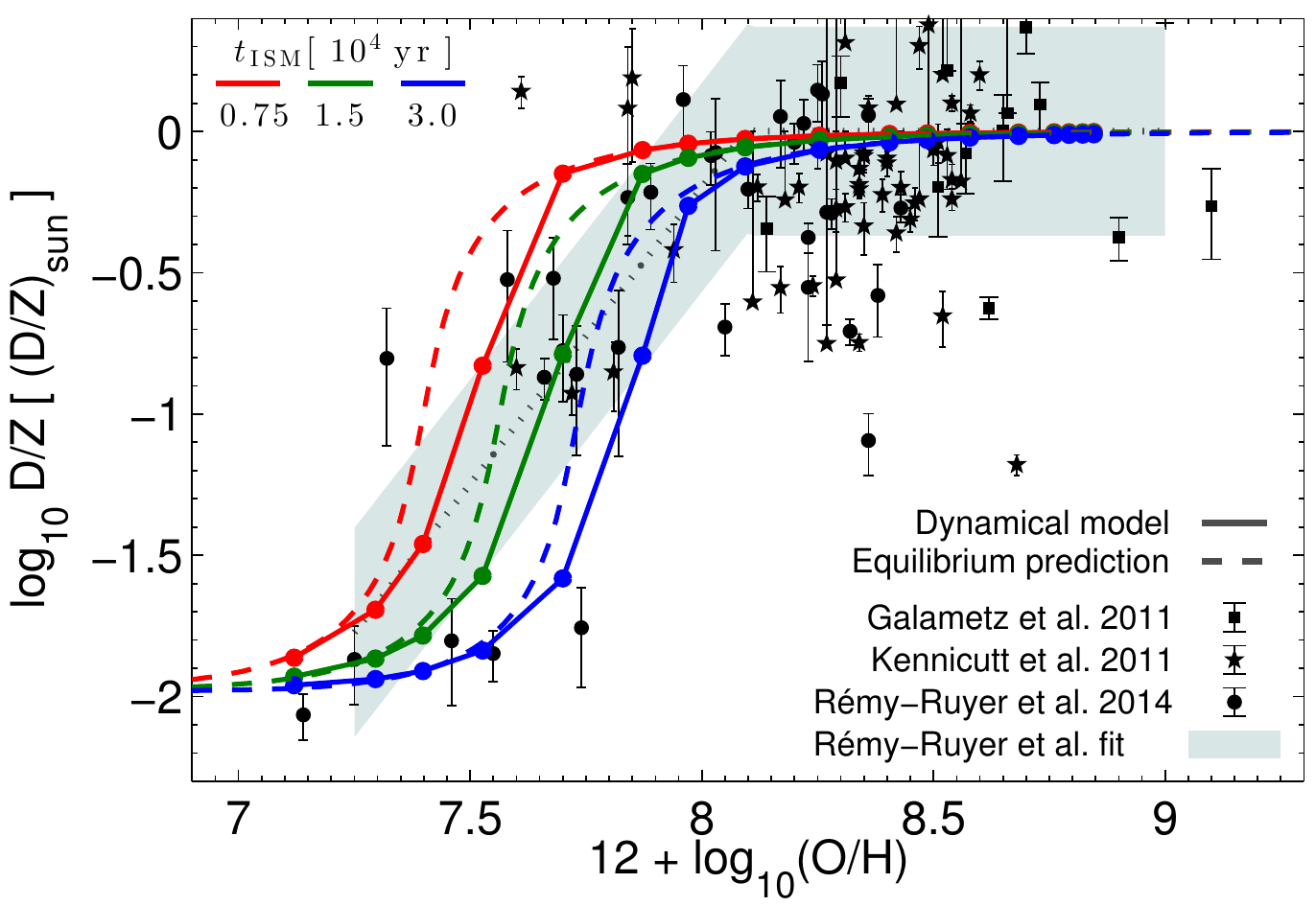} &
\includegraphics[width=85mm]{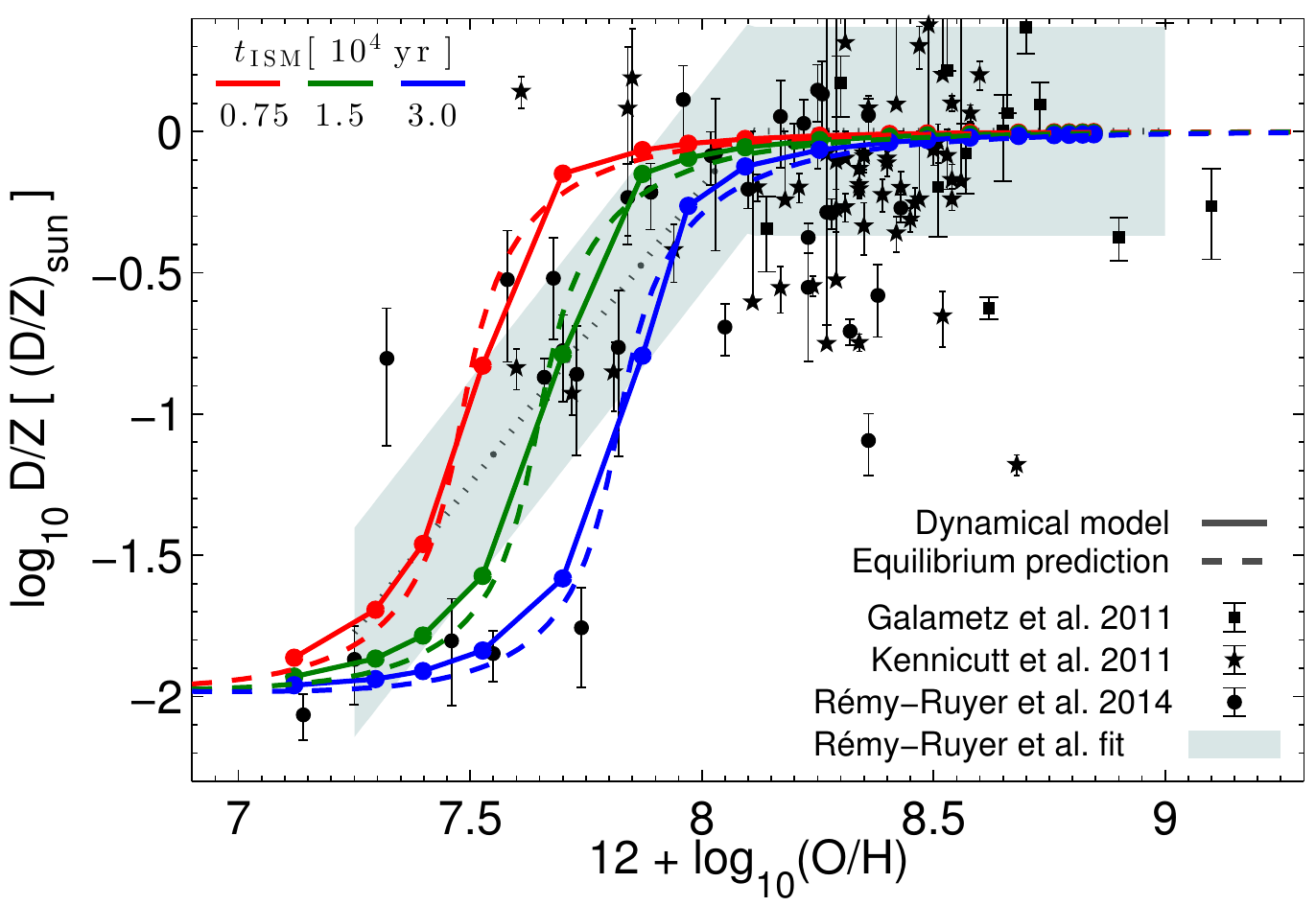}
\end{tabular}
\caption{Dust-to-metal ratio vs ISM metallicity (converted to oxygen-abundance) at $z=0$. Green solid lines show predictions of the dynamical model introduced in section \ref{sect:DynModel} for $t_{\rm ISM}=1.5\times{}10^4$ yr and $\mathcal{Q}_{\rm MS}=1$. The plotted dust-to-metal ratios are normalized to their value at solar metallicity $(D/Z)_\odot\sim{}f^{\rm dep}=0.7$.
(Top left) Predictions of the equilibrium approach, equations (\ref{eq:Zeq}, \ref{eq:Deq}), are shown by dashed lines. The $\H2$ depletion time at $z=0$, see equation (\ref{eq:tdepH2}), and the aforementioned value of $t_{\rm ISM}$ are used to calculate $\gamma$. Red and blue lines show results for galaxies offset from the main sequence of star formation, specifically for $\mathcal{Q}_{\rm MS}=1/3$ and $\mathcal{Q}_{\rm MS}=3$, respectively.
(Top right) same as left-hand panel, but equilibrium dust-to-gas ratios are computed after rescaling $\gamma$ by a factor $0.7$. This small adjustment improves the agreement of the equilibrium approach with the predictions of the dynamical model during the transition phase from low to high $D/Z$.
(Bottom left) Same as top left but red and blue lines now show the effect of varying $t_{\rm ISM}$ by a factor of 2.
(Bottom right) Same as bottom left but with $\gamma$ replaced by $0.7\gamma$ as in the top right panel.
Black symbols in each of the four panels are measurements of $Z$ and $D/Z$ in nearby galaxies \protect\citep{2011A&A...532A..56G, 2011PASP..123.1347K, 2014A&A...563A..31R}, see legend. These $D/Z$ values are normalized to the $D/Z$ value at solar metallicity adopted by \protect\cite{2014A&A...563A..31R}: $(D/Z)_\odot\sim{}0.44$.
Error bars indicate the uncertainties in the measurement of $D$. Metallicities are derived using the $R_{23}$ method with the \protect\cite{2005ApJ...631..231P} calibration and are uncertain at a level of  $\sim{}0.1$ dex (see \protect\citealt{2014A&A...563A..31R} for details). The shaded region (the same as in Fig.~\ref{fig:DZrelation}) is a fit to the observational data by \protect\cite{2014A&A...563A..31R}.
Both the dynamical model and the equilibrium approach reproduce the observed scaling of $D/Z$ with ISM metallicity. Modest variations of $t_{\rm ISM}$ and $\mathcal{Q}_{\rm MS}$ among galaxies can explain the scatter at low metallicities, but are probably not responsible for the scatter at $Z\gtrsim{}1/3\,Z_\odot$.
}
\label{fig:ZvsDZ}
\end{figure*}

We now turn to the main focus of this paper, the relation between ISM metallicity and the dust-to-metal (or dust-to-gas) ratio. Fig.~\ref{fig:ZvsDZ} shows the $Z$ -- $D/Z$ relation at $z=0$ as predicted by the dynamical model of section \ref{sect:DynModel} and compares it both with predictions of the equilibrium model and with observational data. 

In order to track the evolution of the dust-to-gas ratio with the help of our dynamical model, equation (\ref{eq:dotMd}), we need to specify an ISM growth time of the dust mass. We adopt as a fiducial value $t_{\rm ISM}=1.5\times{}10^4$ yr and explore the impact of modest variations (factor 2) in the bottom panels of Fig.~\ref{fig:ZvsDZ}. Increasing (Decreasing) $t_{\rm ISM}$ shifts the transition from low to high $D/Z$ towards higher (lower) metallicities as expected from equation (\ref{eq:Zcrit2}).  

Galaxies with elevated sSFRs ($\mathcal{Q}_{\rm MS}>1$) have shorter molecular gas depletion times, see equation (\ref{eq:tdepH2}), and thus lower $\gamma$. Hence, their critical metallicity will be higher, i.e., the transition from low to high $D/Z$ takes place at larger $Z$, as seen in Fig.~\ref{fig:ZvsDZ}. Galaxies with deficient sSFRs ($\mathcal{Q}_{\rm MS}<1$) have correspondingly a lower critical metallicity.

The transition from low to high $D/Z$ ratios occurs at low stellar masses. We estimate a transition stellar mass of $M_*\sim{}10^8$ $M_\odot$ for main sequence galaxies with $t_{\rm ISM}=1.5\times{}10^4$ yr. The transition mass is higher (lower) for galaxies above (below) the main sequence of star formation and for galaxies with longer (shorter) dust growth times $t_{\rm ISM}$.

Overall, we find a reasonable agreement between the predictions of the dynamical model and those of the equilibrium approach, especially before and after the transition from low to high $D/Z$. During the transition period, however, the equilibrium model somewhat overestimates the $D/Z$ ratio, see also Fig.~\ref{fig:DynModelDZeq}. We can mitigate this bias by increasing $t_{\rm ISM}$ by a factor of 1.43, i.e., by replacing $\gamma = t_{\rm dep, \H2}/t_{\rm ISM}$ with $0.7\gamma$. This small adjustment results in an overall improved match between the predictions of the dynamical model and the equilibrium approach, see the right-hand panels of Fig.~\ref{fig:ZvsDZ}.

We check our model predictions against the recent compilation of dust-to-gas ratios and metallicities by \cite{2014A&A...563A..31R}. Those authors combine observations of dwarf galaxies from the Dwarf Galaxy Survey (DGS; \citealt{2013PASP..125..600M}), the KINGFISH survey \citep{2011PASP..123.1347K}, and the sample by \cite{2011A&A...532A..56G}. The combined sample spans a large range of metallicities ($12+\log_{10}({\rm O}/{\rm H})\sim{}7$ -- $9$) and dust-to-gas ratios ($D\sim{}10^{-6}$ -- $10^{-2}$). Crucially, metallicities and dust-to-gas ratios of all galaxies in the sample are computed in a consistent fashion using the same methodology. Specifically, metallicities are derived based on the $R_{23}=([{\rm OII}]\lambda{}3727+{\rm OIII}]\lambda\lambda{}4959,5007)/{\rm H}\beta$ ratio and a \cite{2005ApJ...631..231P} calibration. Dust masses are estimated from a parametrized SED fit to the observed broad band IR emission (collected via various instruments including \emph{Spitzer} and \emph{Herschel}), following the approach by \cite{2011A&A...536A..88G}. $\HI$ and $\H2$ masses are obtained from a variety of previous works. A metallicity dependent ${\rm CO}-\H2$ conversion factor is used to account for the increased fraction of CO-dark molecular gas at low metallicities (e.g., \citealt{2010ApJ...716.1191W, 2012ApJ...747..124F, 2012ApJ...758..127F, 2012MNRAS.421.3127N, 2013ARA&A..51..207B, 2014A&A...564A.121C}). We refer the reader to \cite{2014A&A...563A..31R} and the references therein for further details on the observational data sets.

By comparing the prediction of the dynamical model with observations, we can narrowly constrain $t_{\rm ISM}$ to $1-3\times{}10^4$. However, the actual value of $t_{\rm ISM}$ is subject to systematic uncertainties of the metallicity estimates. According to \cite{2008ApJ...681.1183K}, the metallicity calibration used in the sample by \cite{2014A&A...563A..31R} typically results in metallicity estimates that are smaller than those of other methods. The shift can be up to 0.4 dex for low mass galaxies. We see from equation (\ref{eq:Zcrit2}) that $Z_{\rm crit}\sim{}q^{1/2}\propto{}t_{\rm ISM}^{1/2}$. Hence, larger ISM metallicities of observed galaxies imply larger $t_{\rm ISM}$. For instance, if the actual metallicities are higher by 0.4 dex, we will need to adopt $t_{\rm ISM}\sim{}0.6-1.9\times{}10^5$ yr. In this case, the time scale $t_{\rm grow}=t_{\rm ISM}/Z_\odot$ for dust to grow via the accumulation of metal atoms out of a solar metallicity ISM is $\sim{}0.4-1.4\times{}10^7$ yr.

The lower limit of $D/Z$ (normalized to its value at solar metallicity) is $\frac{y_{\rm D}}{y}\frac{1-r_{\rm Z}}{1-r_{\rm D}}\frac{1}{f^{\rm dep}}$, or $\sim{}1.1\times{}10^{-2}$ for our fiducial choice of model parameters. In other words, we can determine the ratio between dust yield and metal yield, $y_{\rm D}/y$, simply from measuring the floor value of $D/Z$ at low metallicities. This procedure offers a simple way to determine dust yields, provided the metal yields are known. We presented in section \ref{sect:Zcrit} an alternative method based on measuring the transition metallicity and dust-to-gas ratio, $Z_{\rm crit}$ and $D_{\rm crit}$. The largest systematic for both methods is the metallicity calibration which influences each method in a different way. The reason is that $D\propto{}Z$ (for $Z\ll{}Z_{\rm crit}$) but $D_{\rm crit}\propto{}Z_{\rm crit}^2$. Hence, a combination of both methods may provide a novel way to calibrate the metallicity scale in a low metallicity environment. 

Fig.~\ref{fig:ZvsDZ} shows that modest galaxy-to-galaxy variation of $t_{\rm ISM}$ and/or $\mathcal{Q}_{\rm MS}$ can explain most of the scatter displayed by $D/Z$ at low metallicities ($12+\log_{10}(\rm{O}/\rm{H})<8$). In contrast, even strong variations of these parameters cannot account for the observed scatter of $D/Z$ at $Z\gtrsim{}1/3$ $Z_\odot$. We speculate that processes not captured by the equilibrium approach, such as short time fluctuations of the gas inflow rate or a bursty star formation history, may instead be responsible (e.g., \citealt{2014A&A...562A..76Z}).

\section{What is driving metal and dust abundances in low mass galaxies?}
\label{sect:Driver}

\subsection{The Role of Outflows}
\label{sect:RoleOutflows}

Various physical processes affect the metallicity and dust-to-gas ratio of the ISM. Metal (dust) depleted inflows reduce $Z$ ($D$) as the inflowing material mixes with the ISM. Stars inject metals and dust into the ISM, but may also destroy dust (via supernova shockwaves). Dust can also grow in the cold phase of the ISM. In contrast, galactic outflows (unless they are metal/dust enriched or depleted compared with the ISM) do not change $Z$ or $D$ of a galaxy. It is thus perhaps counter-intuitive that outflows nonetheless drive metallicities and dust-to-gas ratios of low mass galaxies, as we will show in this section.

A fraction $\epsilon_{\rm out}\,r$ of the inflowing gas mass is expelled immediately via outflows. According to the mass conservation equation (\ref{eq:dotMg}), the remainder of the inflowing mass is either consumed in star formation ($(1-R)\SFR$) or added to the gas reservoir ($\dot{M}_{\rm g}$). Note that $(1-R)\SFR \ll{}\epsilon_{\rm out}\SFR$, i.e., outflows dominate the removal of the ISM in $M_*(z=0)\lesssim{}10^{10}$ $M_\odot$ galaxies. Hence, for the qualitative discussion in this section we will ignore the role of gas consumption via star formation.

We remind the reader that the (partly empirical) dynamical model introduced in the previous section reproduces many integral galaxy properties including the $Z$ -- $D/Z$ relation. We used an equilibrium approach to demonstrate that this relation arises from the scaling of $Z$ and $D$ with $r$, the ratio between SFR and gas inflow rate, see equations (\ref{eq:Zeq}, \ref{eq:Deq}). Low values of $r$ lead to low metallicities, low dust-to-gas ratios, and low dust-to-metal ratios, see section \ref{sect:Zcrit}. Hence, the process that sets $r$ is also the process that drives $Z$ and $D$ in galaxies. 

The value of $r$ is small if $\epsilon_{\rm out} + \dot{M}_{\rm g}/\SFR\gg{}1$, see equation (\ref{eq:r}). The mass loading factor is large ($\sim{}30$ for $M_*=10^8\,M_\odot$ at $z=0$) and all but ensures small $r$ for low mass galaxies unless $\dot{M}_{\rm g}/\SFR$ is negative and has a similar absolute value as $\epsilon_{\rm out}$. However, $r$ will also be small if the ISM mass increases quickly ($\dot{M}_{\rm g}/\SFR\gg{}1$). 

\begin{figure*}
\begin{tabular}{cc}
\includegraphics[width=85mm]{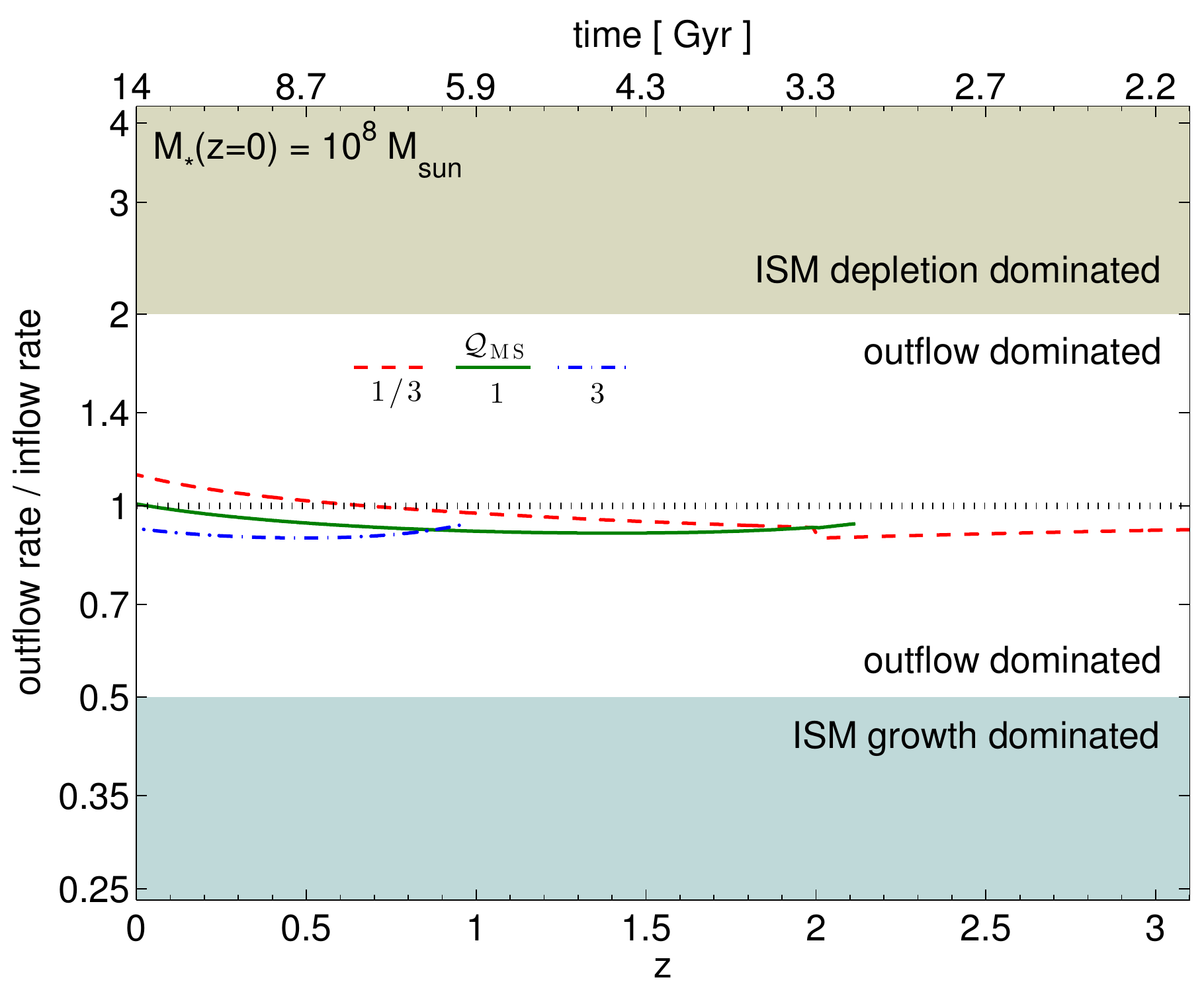} &
\includegraphics[width=85mm]{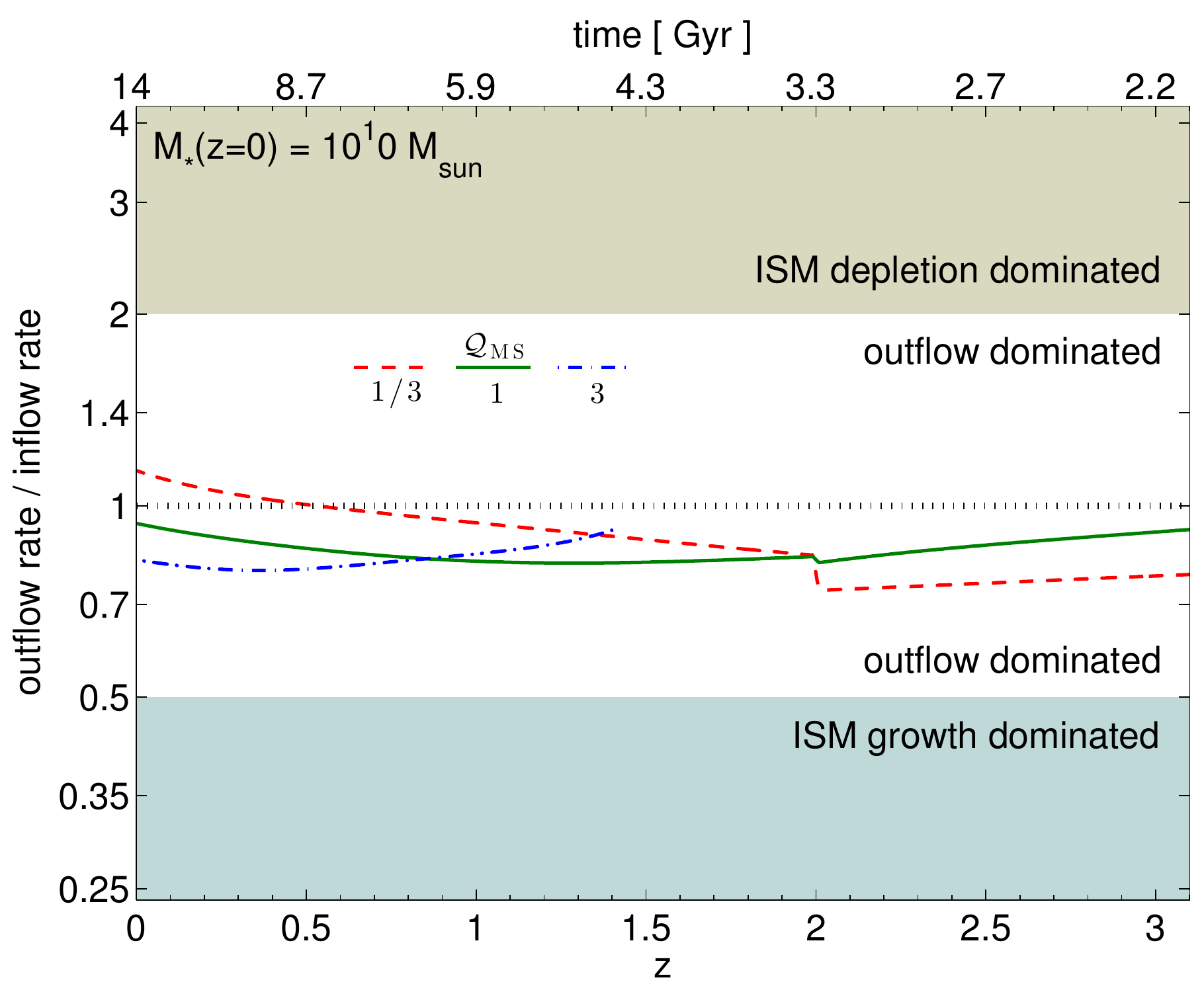} \\
\end{tabular}
\caption{Ratio between the outflow rate, $\epsilon_{\rm out}\,\SFR$, and the gas inflow rate, $\dot{M}_{\rm g, in}$, as predicted by the dynamical model introduced in section \ref{sect:DynModel}. A ratio close to unity implies that most of the inflowing material is removed from the galaxy via outflows with little impact on the total ISM mass (``outflow dominated''). A ratio below 0.5 indicates that a large fraction of the inflowing material is stored in a growing ISM mass (``ISM growth dominated''), while a ratio above 2 corresponds to small gas inflow rates and quickly decreasing ISM reservoirs (``ISM depletion dominated'').
(Left) for a galaxy with a $z=0$ stellar mass of $10^8$ $M_\odot$, i.e.., a typical mass for galaxies near the critical metallicity $Z_{\rm crit}$, see section \ref{sect:ValidateEquilibriumAnsatz}. (Right) same as left but for a galaxy with a stellar mass of $10^{10}$ $M_\odot$. In each panel we show the evolution of $\epsilon_{\rm out}\,r$ as function of redshift (bottom axis) and time (top axis) for 
three different offset from the main sequence of star formation. The green solid, red dashed, and blue dot-dashed lines correspond to a galaxy that lives on, below, and above the main sequence, respectively, see legend.
The redshift evolution of the sSFR changes slope at $z=2$, see equation (\ref{eq:SFRLilly}), resulting in a small jump of $r$ at that time.
Galaxies with $z=0$ stellar masses $\lesssim{}10^{10}$ $M_\odot$ reside in the ``outflow dominated'' regime from high redshifts until today. Hence, galactic outflows are primarily responsible for the small values of $r=\SFR/\dot{M}_{\rm g, in}$, and thus for the much lower metallicities, dust-to-gas ratios, and dust-to-metal ratios, of low mass galaxies.}
\label{fig:outflowVsgrowth}
\end{figure*}

To clarify the relative role of outflows and growing gas reservoirs we show in Fig.~\ref{fig:outflowVsgrowth} the ratio 
\[
\epsilon_{\rm out}\,r = \frac{\epsilon_{\rm out}\,\SFR}{\dot{M}_{\rm g,in}}
\]
between the outflow rate and the gas inflow rate. The results presented in Fig.~\ref{fig:outflowVsgrowth} are based on the dynamical model of section \ref{sect:DynModel} for our fiducial choice of parameters. Note that $\dot{M}_{\rm g,in} \sim{} \epsilon_{\rm out}\SFR + \dot{M}_{\rm g}$, i.e.,  $\epsilon_{\rm out}\,r$ measures how inflowing gas is distributed among the outflow and reservoir channels. If almost all of the inflowing gas is expelled (and the gas reservoir remains effectively constant), the value is about 1. Galaxies with negligible inflow rates have large ratios $\gg{}1$. In this case, outflows lead to a fast reduction of the ISM mass. In contrast, a ratio below 0.5 indicates that $\dot{M}_{\rm g}/\SFR>\epsilon_{\rm out}$, i.e., $r$ is driven by a growing ISM reservoir.

Fig.~\ref{fig:outflowVsgrowth} shows that low mass galaxies have $\epsilon_{\rm out}\gg{}\dot{M}_{\rm g}/\SFR$, i.e., they are outflow dominated. Gas inflows and outflows in low mass galaxies are tightly coupled since $z\sim{}3$. The physical picture is thus as follows:
\begin{itemize}
\item The expulsion of a large fraction of the inflowing gas limits the mass of the ISM and, thus, the SFR in low mass galaxies. In other words, stronger outflows allow for a larger inflow rate at a fixed SFR (and at a fixed stellar mass for a given $\mathcal{Q}_{\rm MS}$). Outflows are thus directly responsible for the low values of $r$ in low mass galaxies.
\item The inflowing material is dust- and metal-poor compared with the ISM. Hence, a large gas inflow rate results in a strong dilution of any existing amount of metals and dust in the ISM. Outflows reduce metallicities and dust-to-gas ratios in galaxies because they boost the dilution of the ISM and not because they remove any excess metals or dust.
\item The scaling of the mass loading factor with stellar mass is responsible for the reduction of $Z$, $D$, and $D/Z$ with decreasing $M_*$. If this scaling were absent, outflow dominated galaxies of any mass would have similar values of $r$ and, hence, similar metallicities, dust-to-gas ratios, and dust-to-metal ratios.
\end{itemize}

Hence, outflows drive the evolution of metals and dust in galaxies by regulating gas inflows (at fixed SFR), i.e., by regulating the dilution of the ISM. In contrast to previous works (e.g., \citealt{2007ApJ...658..941D, 2010A&A...514A..73S}), outflows and dilution of gas are thus not considered independent processes. Instead, galaxies naturally reach an equilibrium state between inflows, star formation, and outflows. The properties of the equilibrium state (e.g., $Z$ and $D$) depend on the mass loading factor of the galactic winds.

\subsection{Potential Caveats}
\label{sect:Caveats}

The physical picture outlined in section \ref{sect:RoleOutflows} connects outflows, inflows, and star formation in galaxies. We based some of our conclusions, e.g., that $r$ is set by outflows and not by a growing ISM mass, on the dynamical model of section \ref{sect:DynModel}. In this section, we discuss potential caveats of our approach and highlight future research opportunities.

The dynamical model of section \ref{sect:DynModel} is a one zone model, i.e., it does not capture spatial variations of any of the ISM components. For instance, inflowing material is completely mixed with the ISM because the model only traces the total metal, dust, and gas mass in the zone. However, a one-zone model is a reasonable (and commonly employed) choice as long as we focus on galaxy-integrated properties and on average scaling relation. Clearly, such an approach is not well suited to study the intrinsic scatter or the spatial variability of scaling relations. For these tasks, numerical simulations that incorporate a local dust formation and destruction model will be more suited.

A one zone model is also convenient as it limits the number of tunable parameters. We eliminate most of the free parameters of our model by incorporating galaxy-averaged scaling relations derived from observations and numerical simulations. Most of the parameters that remain, e.g., the stellar metal yield $y$ or the stellar return fraction $R$, are strongly constrained by observations or theory. A variation of these parameters within reasonable bounds will not change our results on a qualitative level. Indeed, we can quantify such changes with the equations provided by the equilibrium ansatz, section \ref{sect:equilibriumDust}. Other parameters, e.g., the dust yield $y_{\rm D}$ or the dust depletion of gas inflows $r_{\rm D}$, have limited impact on the predictions of the model. In fact, the only unconstrained, yet crucial, parameter of our model is $\gamma$. However, this parameter does not affect the conclusions reached in section \ref{sect:RoleOutflows}.

The dynamical model of section \ref{sect:DynModel} infers the evolution of stellar masses and gas inflow rates from the \emph{average} main sequence lives of galaxies. The model thus averages over fluctuations of the gas accretion rates that likely result in frequent star formation bursts in low mass galaxies (e.g., \citealt{2009ARA&A..47..371T, 2011A&A...531A.136Y}). The typical timescale of these fluctuations is set by the dynamical time of the central part of the galaxy (a few 100 Myr for dwarf galaxies), see \cite{2007ApJ...667..170S}. The timescale of star formation variability is thus an order of magnitude shorter than the gas depletion time ($\gtrsim{}$Gyr) that determines the long term, equilibrium evolution. It is thus likely justified to average over these fluctuations, as we do in our model, if all one cares about is the mean evolution of galaxy properties. However, this time variability will introduce significant scatter among galaxy properties, such as between $Z$ and $D$, see e.g., \citealt{2014A&A...562A..76Z}. In future work it may be worthwhile to add a stochastic component of gas accretion to our dynamical model and study the implications, see \cite{2014MNRAS.443..168F}.

We showed that outflows drive $Z$ and $D$ via the regulation of gas inflows (by setting $r$). Interestingly, outflows affect $Z$ and $D$ \emph{only} via $r$. Indeed, neither the equilibrium metallicity nor the equilibrium dust-to-gas ratio depend explicitly on $\epsilon_{\rm out}$ at fixed $r$, see equations (\ref{eq:Zeq}, \ref{eq:Deq}). This property of the equilibrium solution disappears, however, if metal-enriched/-depleted or dust-enriched/-depleted outflows are considered (see below). An important question to ask is thus whether metal and/or dust enriched outflows affect the conclusions of section \ref{sect:RoleOutflows} in a significant way.

Let the metallicity and dust-to-gas ratio of outflows be $Z\,r_{\rm Z}^{\rm out}$ and $D\,r_{\rm D}^{\rm out}$, respectively. After inserting these changes into equations (\ref{eq:dotMZ}) and (\ref{eq:dotMd}), we can re-derive the equations for the equilibrium metallicity and dust-to-gas ratio.
Equation (\ref{eq:Zeq}) becomes
\begin{equation}
\langle{}Z\rangle{} = \frac{y\,(1-R)\,r}{1-r_{\rm Z} + (r_{\rm Z}^{\rm out}-1)\,\epsilon_{\rm out}\,r},
\label{eq:ZeqNew}
\end{equation}
while equation (\ref{eq:beta}) becomes
\begin{equation}
\beta = \gamma{}f^{\rm dep}\langle{}Z\rangle{} - \left[ \epsilon_{\rm SN} + R + \frac{1-r_{\rm D} + (r_D^{\rm out}-1)\,\epsilon_{\rm out}\,r}{r}\right].
\end{equation}
Equations (\ref{eq:alpha}, \ref{eq:gamma}, \ref{eq:Deq}) remain unchanged.

The revised equations contain the mass loading factor only as part of the product $\epsilon_{\rm out}\,r$. Fig.~\ref{fig:outflowVsgrowth} demonstrates that $\epsilon_{\rm out}\,r\sim{}1$. Hence, modest metal or dust loadings ($1<r_{\rm Z,D}^{\rm out}<2$) have only a minor impact on the equilibrium metallicity and dust-to-gas ratio. A stronger metal loading results in lower $\langle{}Z\rangle{}$ and lower $Z_{\rm crit}$, while a stronger dust loading raises the value of $Z_{\rm crit}$. 

We note that accounting for metal and dust loadings amounts to replacing $r_{\rm Z}$ with $r_{\rm Z} + (1-r_{\rm Z}^{\rm out})\epsilon_{\rm out}\,r$ and $r_{\rm D}$ with $r_{\rm D} + (1-r_D^{\rm out})\epsilon_{\rm out}\,r$, i.e., metal and dust loadings can be absorbed into the definition of the relative dust and metal enrichments of inflowing material, $r_{\rm Z}$ and $r_{\rm D}$. Note however, that large metal (dust) loading factors result in a negative effective metal (dust) enrichment of the inflowing gas.

A sometimes proposed explanation for the stellar mass -- metallicity relation and its evolution is that metal enriched outflows reduce $Z$ in low mass galaxies (e.g., \citealt{2010A&A...514A..73S}). However, this mechanism requires that the metal loading depends on stellar mass \emph{and} redshift in a particular way. In contrast, we argued (cf. \citealt{2012MNRAS.421...98D}) that the dependence of the mass loading factor on stellar mass naturally introduces a scaling of $Z$ with stellar mass. Given that our model reproduces the observed stellar mass -- metallicity relation for $r_{Z}^{\rm out}=1$, the metal loading factor of outflows is probably not very large.

\section{Summary \& Conclusions}
\label{sect:Conclusions}

In this paper, we analyzed the origin and the functional form of the relation between the ISM metallicity $Z$ and the dust-to-gas ratio $D$ of low mass galaxies. To this end, we combined various empirical data into a simple one-zone model and added the crucial aspects of metal and dust chemistry including the production of metals and dust by stars, dust growth in the ISM, dust destruction via SNe shockwaves, as well as inflows and outflows. We tested this model against observations and against the predictions of an even simpler model derived from an equilibrium ansatz. Our main results are as follows:
\begin{itemize}

\item The dynamical, one-zone model matches a large number of core observations, including the stellar mass -- metallicity relation and its evolution out to $z=2$ (Fig.~\ref{fig:MZ}), the stellar mass -- halo mass relation and its evolution out to $z=2$ (Fig.~\ref{fig:MstarMhalo}), and the $Z$ -- $D$ relation at $z=0$ (Fig.~\ref{fig:ZvsDZ}).

\item Galaxies with different offsets from the main sequence of star formation, but with similar stellar masses, differ in their star formation and gas accretion histories (Fig.~\ref{fig:DynModelMassEvolution}). In particular, galaxies that live above (below) the main sequence form a larger (smaller) fraction of their stars at late times and have larger (smaller) gas fractions.

\item Galaxies switch from low dust-to-metal ratios at $Z<Z_{\rm crit}$ to large dust-to-metal ratios at $Z>Z_{\rm crit}$. The critical metallicity $Z_{\rm crit}$ is set by the competition between dust growth in the ISM and dilution via dust-poor gas inflows. Grain destruction in SN shockwaves plays a sub-dominant role. The concept of a critical metallicity can be replaced with the concept of a critical ratio between the SFR and the gas inflow rate into a galaxy, see section~\ref{sect:Zcrit}.

\item A model based on the equilibrium ansatz $\dot{D}=\dot{Z}\equiv{}0$ matches the $D$ -- $Z$ relation predicted by the dynamical, one-zone model. Hence, the equilibrium approach captures most of the physics of the dynamical model and can thus be used to study the origin of the $D$ -- $Z$ relation. 

\item The equilibrium ansatz provides simple analytic formulae for the metallicity, dust-to-gas ratio, and dust-to-metal ratio of galaxies, as well as for the critical metallicity, see equations (\ref{eq:Zeq}, \ref{eq:Deq}, \ref{eq:Zcrit2}). Metal and/or dust-loaded outflows can be easily incorporated into the equilibrium formalism, see section \ref{sect:Caveats}.

\item The ratio $r$ between the SFR and the gas inflow rate of a galaxy is the pivotal parameter that determines the ISM metallicity and dust-to-gas ratio. Small values of $r$ imply that (at a given SFR) the gas inflow rate is large. High inflow rates of depleted gas lead to a dilution of pre-existing amounts of metals and dust in the ISM and, hence, to low $Z$, $D$, and $D/Z$ values.

\item Galactic outflows suppress the available amount of gas of a galaxy (and thus the SFR) while enabling large gas inflow rates. Hence, outflows regulate the value of $r$ for a given galaxy. Using the one-zone, dynamical model we infer that outflows are indeed the main drivers behind the evolution of dust and metals in galaxies, see section \ref{sect:RoleOutflows}.
\end{itemize}

\begin{figure}
\begin{tabular}{c}
\includegraphics[width=85mm]{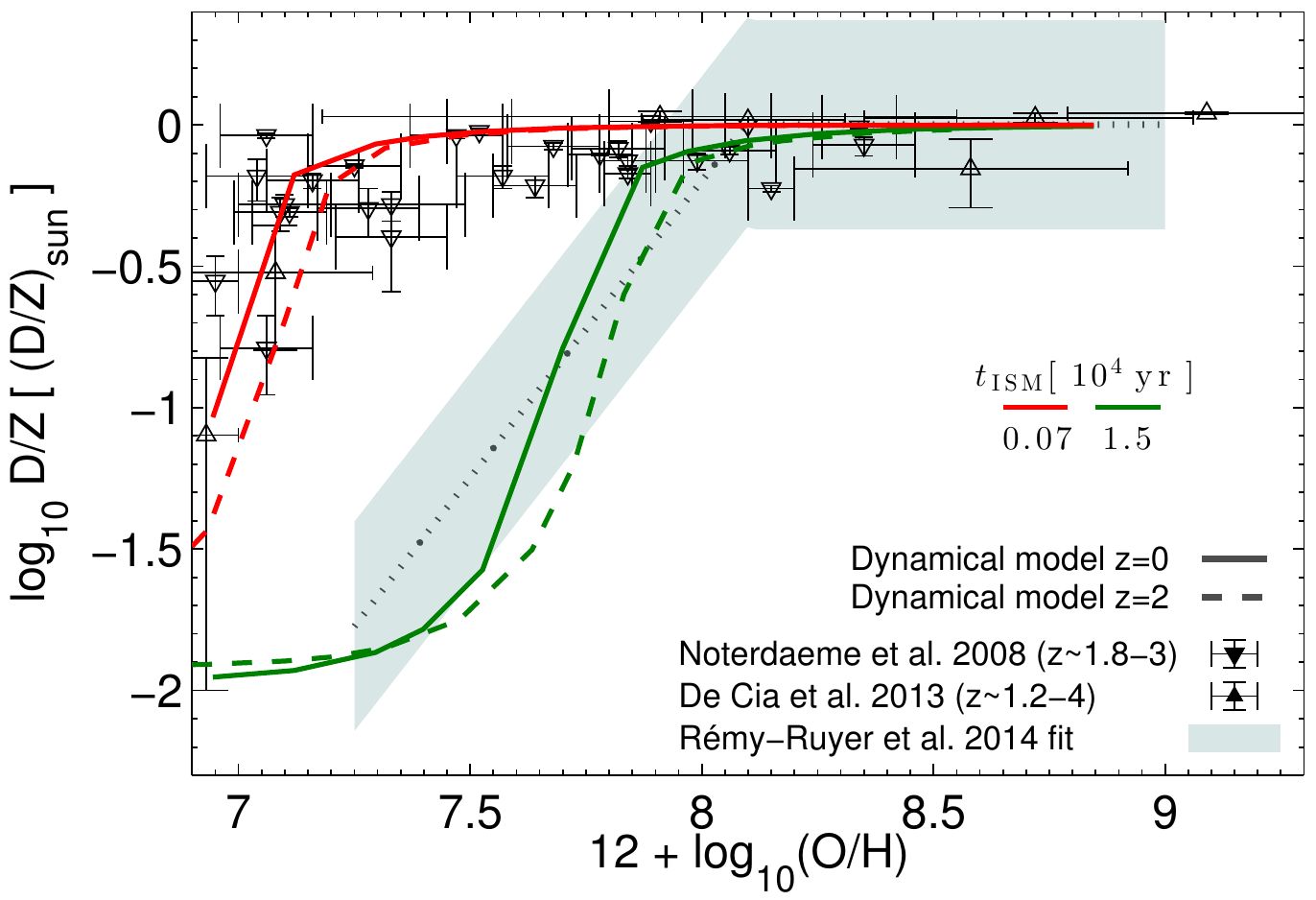}
\end{tabular}
\caption{Dust-to-metal ratio vs ISM metallicity (converted to oxygen-abundance) at $z=0$ (solid lines) and at $z=2$ (dashed lines) as predicted by the dynamical model introduced in section \ref{sect:DynModel}. We show the model results for $t_{\rm ISM}=1.5\times{}10^4$ yr (green lines; our fiducial value to match $z=0$ observations, see Fig.~\ref{fig:ZvsDZ}) and $t_{\rm ISM}=700$ yr (red lines). The dust-to-metal ratios are normalized to their value at solar metallicity. Triangles with error bars show the dust-to-metal ratio of $z\sim{}1-4$ host galaxies of gamma-ray bursts \protect\citep{2013A&A...560A..88D} and QSO absorbers \protect\citep{2008A&A...481..327N}. The dust-to-metal ratios are derived from the relative depletion of Zn and Fe using optical/UV absorption line spectroscopy, while metallicities are based on the Zn/H ratio, see \protect\cite{2013A&A...560A..88D} for details. The shaded region is the broken power-law fit by \protect\cite{2014A&A...563A..31R} to the $Z$ -- $D/Z$ relation of $z=0$ galaxies, see Fig.~\ref{fig:ZvsDZ}. The dynamical model predicts little evolution of the $Z$ -- $D/Z$ relation between $z=0$ and $z=2$ \emph{as long as the ratio between $t_{\rm ISM}$ and $t_{\rm dep,\H2}$ is approximately redshift independent.} The predictions of the dynamical model can be reconciled with the observations of high redshift damped Lyman-$\alpha$ absorbers if $t_{\rm ISM}$ is lower and/or $t_{\rm dep,\H2}$ is larger for such systems.}
\label{fig:DZhighZ}
\end{figure}

A precise observational characterization of the $Z$ -- $D$ relation offers promising possibilities to extract information about dust chemistry and the baryonic cycle of gas accretion, star formation, and outflows. For instance, as discussed in section \ref{sect:ValidateEquilibriumAnsatz} a lower bound on the $D/Z$ value constrains the stellar dust yield $y_{\rm D}$ (relative to the stellar metal yield $y$). In addition, we showed in section \ref{sect:Zcrit} how the stellar dust yield can be extracted from a precise measurement of the critical metallicity and the critical dust-to-gas ratio. We also showed, based on the equilibrium approach, that the critical metallicity scales with the ratio between the $\H2$ depletion time and the dust growth time in the ISM. Hence, the functional form of the $Z$ -- $D$ relation encodes important information about the physics of intra-cloud dust growth in galaxies.

To illustrate this latter point, we show in Fig.~\ref{fig:DZhighZ} the predicted $Z$ -- $D$ relation of main sequence galaxies at both $z=0$ and $z=2$ based on our dynamical model. Clearly, the $Z$ -- $D$ relation evolves little with redshift as long as $\gamma=t_{\rm dep,\H2}/t_{\rm ISM}$ at $z=2$ is close to its value at $z=0$. In particular, we predict that the dust-to-metal ratio of $z\sim{}2-3$ galaxies is similar to the dust-to-metal ratio of nearby galaxies of the same ISM metallicity. In contrast, \cite{2003PASJ...55..901I} argues that the $Z$ -- $D$ relation can be understood as a sequence of galaxies of constant stellar age. He suggests that galaxies at $z\sim{}3$ should have low dust-to-metal ratios because of their young age.

UV/optical absorption line spectroscopy of $z\sim{}1-4$ host galaxies of gamma-ray bursts and QSO absorbers show that dust-to-metal ratios are generally high, similar to those of metal-rich galaxies in the local Universe. Surprisingly, these large $D/Z$ values are even found in galaxies with $Z\sim{}1/10$ $Z_\odot$ \citep{2013A&A...560A..88D}. \cite{2013A&A...560A..26Z} obtain a similar result based on optical/UV extinction measurements.

A variety of explanations have been proposed to solve this conundrum. The most likely explanation is that the star formation efficiency is lower and/or the dust growth time in the ISM is shorter in these galaxies compared with nearby galaxies as suggested by \cite{2013MNRAS.436.1238K}. In other words, $\gamma=t_{\rm dep,\H2}/t_{\rm ISM}$ is potentially larger in these galaxies for some unknown reason. To see why this could work, we remind the reader that the critical metallicity scales as $Z_{\rm crit}\approx{}q^{1/2}\propto{}\left[y/\gamma\right]^{1/2}$, see section \ref{sect:Zcrit}, i.e., a large $\gamma$ implies that even moderately metal-poor galaxies have $Z>Z_{\rm crit}$ and, hence, large dust-to-metal ratios. Given the slow evolution of the $\H2$ depletion time with redshift for star forming galaxies, equation (\ref{eq:tdepH2}), these observations likely identify high $z$ galaxies / sight lines with suppressed star formation activity or with reduced dust growth times in the ISM.

A different possibility is that the stellar population integrated metal yield $y$ is lower in high redshift galaxies. Indeed, the ejected metal mass fractions of AGB stars \citep{2010MNRAS.403.1413K} and SNe \citep{2006NuPhA.777..424N} depend on the initial metallicity of the stellar population. However, Fig.~\ref{fig:DZhighZ} demonstrates that  $Z_{\rm crit}$ has to be lowered by a factor of $\sim{}5$, which would thus require an unrealistic decrease of $y$ by a factor 25. We note that lowering the dust destruction efficiency of SN, $\epsilon_{\rm SN}$, would not help as $Z_{\rm crit}$ is set by the balance of dust growth in the ISM and dust dilution from inflows, see section \ref{sect:Zcrit}.

A further possibility is that the stellar dust yield is enhanced in these galaxies compared with local galaxies (e.g., \citealt{2013MNRAS.436.1238K, 2014MNRAS.440.1562M}). Such a change does not affect the value of $Z_{\rm crit}$ but it raises the dust-to-metal ratio for $Z<Z_{\rm crit}$, see sections \ref{sect:eqD} and \ref{sect:Zcrit}. However, this approach would require $y_{\rm D}/y\sim{}0.5$ which is inconsistent with models for AGN and SN dust injection \citep{2013MNRAS.436.1238K}. Furthermore, it does not explain the drop in the dust-to-metal ratio at very low metallicities, see Fig.~\ref{fig:DZhighZ}.

High sensitivity radio arrays, such as ALMA or NOEMA, have the capacity to trace gas and dust based on their CO, C$_{\rm II}$, and infrared emission out to high redshift. Data taken with such observatories, supplemented with rest-frame UV/optical spectra, should be able to pin down the functional form of the $Z$ -- $D$ relation as function of redshift, galaxy type, and local environment. Such observations will thus provide important insights into the interplay between star formation and ISM chemistry in various galactic environments.

\acknowledgements

RF thanks Eliot Quataert and Chris McKee for valuable input on the draft of this paper. The author also thanks Ryan Sanders and Aur\'{e}lie R\'{e}my-Ruyer for providing the data presented in \cite{2014arXiv1408.2521S} and \cite{2014A&A...563A..31R}, respectively, in form of an easily accessible ASCII table. The author further thanks the referee for constructive comments that helped improving the paper. RF acknowledges support for this work by NASA through Hubble Fellowship grant HF-51304.01-A awarded by the Space Telescope Science Institute, which is operated by the Association of Universities for Research in Astronomy, Inc., for NASA, under contract NAS 5-26555. This work made extensive use of the NASA Astrophysics Data System and arXiv.org preprint server.\\

\bibliography{ms.bib}

\begin{thebibliography}{146}
\expandafter\ifx\csname natexlab\endcsname\relax\def\natexlab#1{#1}\fi

\bibitem[{{Allende Prieto}, {Lambert} \& {Asplund}(2001){Allende Prieto},
  {Lambert}, \& {Asplund}}]{2001ApJ...556L..63A}
{Allende Prieto} C., {Lambert} D.~L., {Asplund} M., 2001, \apjl, 556, L63

\bibitem[{{Andrews} \& {Martini}(2013)}]{2013ApJ...765..140A}
{Andrews} B.~H., {Martini} P., 2013, \apj, 765, 140

\bibitem[{{Asano} {et~al}\mbox{.}(2013){Asano}, {Takeuchi}, {Hirashita}, \&
  {Inoue}}]{2013EP&S...65..213A}
{Asano} R.~S., {Takeuchi} T.~T., {Hirashita} H., {Inoue} A.~K., 2013, Earth,
  Planets, and Space, 65, 213

\bibitem[{{Barlow}(1978)}]{1978MNRAS.183..367B}
{Barlow} M.~J., 1978, \mnras, 183, 367

\bibitem[{{Behroozi}, {Wechsler} \& {Conroy}(2013){Behroozi}, {Wechsler}, \&
  {Conroy}}]{2013ApJ...770...57B}
{Behroozi} P.~S., {Wechsler} R.~H., {Conroy} C., 2013, \apj, 770, 57

\bibitem[{{Bekki}(2013)}]{2013MNRAS.432.2298B}
{Bekki} K., 2013, \mnras, 432, 2298

\bibitem[{{Bigiel} {et~al}\mbox{.}(2008){Bigiel}, {Leroy}, {Walter}, {Brinks},
  {de Blok}, {Madore}, \& {Thornley}}]{2008AJ....136.2846B}
{Bigiel} F., {Leroy} A., {Walter} F., {Brinks} E., {de Blok} W.~J.~G., {Madore}
  B., {Thornley} M.~D., 2008, \aj, 136, 2846

\bibitem[{{Bolatto}, {Wolfire} \& {Leroy}(2013){Bolatto}, {Wolfire}, \&
  {Leroy}}]{2013ARA&A..51..207B}
{Bolatto} A.~D., {Wolfire} M., {Leroy} A.~K., 2013, \araa, 51, 207

\bibitem[{{Bothwell} {et~al}\mbox{.}(2013){Bothwell}, {Maiolino}, {Kennicutt},
  {Cresci}, {Mannucci}, {Marconi}, \& {Cicone}}]{2013MNRAS.433.1425B}
{Bothwell} M.~S., {Maiolino} R., {Kennicutt} R., {Cresci} G., {Mannucci} F.,
  {Marconi} A., {Cicone} C., 2013, \mnras, 433, 1425

\bibitem[{{Bouch{\'e}} {et~al}\mbox{.}(2010){Bouch{\'e}}, {Dekel}, {Genzel},
  {Genel}, {Cresci}, {F{\"o}rster Schreiber}, {Shapiro}, {Davies}, \&
  {Tacconi}}]{2010ApJ...718.1001B}
{Bouch{\'e}} N. {et~al.}, 2010, \apj, 718, 1001

\bibitem[{{Calzetti} {et~al}\mbox{.}(2010){Calzetti}, {Wu}, {Hong},
  {Kennicutt}, {Lee}, {Dale}, {Engelbracht}, {van Zee}, {Draine}, {Hao},
  {Gordon}, {Moustakas}, {Murphy}, {Regan}, {Begum}, {Block}, {Dalcanton},
  {Funes}, {Gil de Paz}, {Johnson}, {Sakai}, {Skillman}, {Walter}, {Weisz},
  {Williams}, \& {Wu}}]{2010ApJ...714.1256C}
{Calzetti} D. {et~al.}, 2010, \apj, 714, 1256

\bibitem[{{Carlberg} {et~al}\mbox{.}(1996){Carlberg}, {Yee}, {Ellingson},
  {Abraham}, {Gravel}, {Morris}, \& {Pritchet}}]{1996ApJ...462...32C}
{Carlberg} R.~G., {Yee} H.~K.~C., {Ellingson} E., {Abraham} R., {Gravel} P.,
  {Morris} S., {Pritchet} C.~J., 1996, \apj, 462, 32

\bibitem[{{Catinella} {et~al}\mbox{.}(2010){Catinella}, {Schiminovich},
  {Kauffmann}, {Fabello}, {Wang}, {Hummels}, {Lemonias}, {Moran}, {Wu},
  {Giovanelli}, {Haynes}, {Heckman}, {Basu-Zych}, {Blanton}, {Brinchmann},
  {Budav{\'a}ri}, {Gon{\c c}alves}, {Johnson}, {Kennicutt}, {Madore}, {Martin},
  {Rich}, {Tacconi}, {Thilker}, {Wild}, \& {Wyder}}]{2010MNRAS.403..683C}
{Catinella} B. {et~al.}, 2010, \mnras, 403, 683

\bibitem[{{Cazaux} \& {Tielens}(2004)}]{2004ApJ...604..222C}
{Cazaux} S., {Tielens} A.~G.~G.~M., 2004, \apj, 604, 222

\bibitem[{{Conroy}, {Wechsler} \& {Kravtsov}(2006){Conroy}, {Wechsler}, \&
  {Kravtsov}}]{2006ApJ...647..201C}
{Conroy} C., {Wechsler} R.~H., {Kravtsov} A.~V., 2006, \apj, 647, 201

\bibitem[{{Cormier} {et~al}\mbox{.}(2014){Cormier}, {Madden}, {Lebouteiller},
  {Hony}, {Aalto}, {Costagliola}, {Hughes}, {R{\'e}my-Ruyer}, {Abel}, {Bayet},
  {Bigiel}, {Cannon}, {Cumming}, {Galametz}, {Galliano}, {Viti}, \&
  {Wu}}]{2014A&A...564A.121C}
{Cormier} D. {et~al.}, 2014, \aap, 564, A121

\bibitem[{{Dalcanton}(2007)}]{2007ApJ...658..941D}
{Dalcanton} J.~J., 2007, \apj, 658, 941

\bibitem[{{Dav{\'e}}, {Finlator} \& {Oppenheimer}(2012){Dav{\'e}}, {Finlator},
  \& {Oppenheimer}}]{2012MNRAS.421...98D}
{Dav{\'e}} R., {Finlator} K., {Oppenheimer} B.~D., 2012, \mnras, 421, 98

\bibitem[{{De Cia} {et~al}\mbox{.}(2013){De Cia}, {Ledoux}, {Savaglio},
  {Schady}, \& {Vreeswijk}}]{2013A&A...560A..88D}
{De Cia} A., {Ledoux} C., {Savaglio} S., {Schady} P., {Vreeswijk} P.~M., 2013,
  \aap, 560, A88

\bibitem[{{Dekel} \& {Mandelker}(2014)}]{2014MNRAS.444.2071D}
{Dekel} A., {Mandelker} N., 2014, \mnras, 444, 2071

\bibitem[{{Draine}(1990)}]{1990ASPC...12..193D}
{Draine} B.~T., 1990, in Astronomical Society of the Pacific Conference Series,
  Vol.~12, The Evolution of the Interstellar Medium, {Blitz} L., ed., pp.
  193--205

\bibitem[{{Draine}(2003)}]{2003ARA&A..41..241D}
{Draine} B.~T., 2003, \araa, 41, 241

\bibitem[{{Draine}(2011)}]{2011piim.book.....D}
{Draine} B.~T., 2011, {Physics of the Interstellar and Intergalactic Medium}

\bibitem[{{Draine} \& {Salpeter}(1979)}]{1979ApJ...231..438D}
{Draine} B.~T., {Salpeter} E.~E., 1979, \apj, 231, 438

\bibitem[{{Dutton} \& {van den Bosch}(2009)}]{2009MNRAS.396..141D}
{Dutton} A.~A., {van den Bosch} F.~C., 2009, \mnras, 396, 141

\bibitem[{{Dwek}(1998)}]{1998ApJ...501..643D}
{Dwek} E., 1998, \apj, 501, 643

\bibitem[{{Dwek} \& {Scalo}(1980)}]{1980ApJ...239..193D}
{Dwek} E., {Scalo} J.~M., 1980, \apj, 239, 193

\bibitem[{{Eales} {et~al}\mbox{.}(2012){Eales}, {Smith}, {Auld}, {Baes},
  {Bendo}, {Bianchi}, {Boselli}, {Ciesla}, {Clements}, {Cooray}, {Cortese},
  {Davies}, {De Looze}, {Galametz}, {Gear}, {Gentile}, {Gomez}, {Fritz},
  {Hughes}, {Madden}, {Magrini}, {Pohlen}, {Spinoglio}, {Verstappen},
  {Vlahakis}, \& {Wilson}}]{2012ApJ...761..168E}
{Eales} S. {et~al.}, 2012, \apj, 761, 168

\bibitem[{{Erb} {et~al}\mbox{.}(2006){Erb}, {Shapley}, {Pettini}, {Steidel},
  {Reddy}, \& {Adelberger}}]{2006ApJ...644..813E}
{Erb} D.~K., {Shapley} A.~E., {Pettini} M., {Steidel} C.~C., {Reddy} N.~A.,
  {Adelberger} K.~L., 2006, \apj, 644, 813

\bibitem[{{Feldmann}(2013)}]{2013MNRAS.433.1910F}
{Feldmann} R., 2013, \mnras, 433, 1910

\bibitem[{{Feldmann}, {Gnedin} \& {Kravtsov}(2012{\natexlab{a}}){Feldmann},
  {Gnedin}, \& {Kravtsov}}]{2012ApJ...747..124F}
{Feldmann} R., {Gnedin} N.~Y., {Kravtsov} A.~V., 2012{\natexlab{a}}, \apj, 747,
  124

\bibitem[{{Feldmann}, {Gnedin} \& {Kravtsov}(2012{\natexlab{b}}){Feldmann},
  {Gnedin}, \& {Kravtsov}}]{2012ApJ...758..127F}
{Feldmann} R., {Gnedin} N.~Y., {Kravtsov} A.~V., 2012{\natexlab{b}}, \apj, 758,
  127

\bibitem[{{Ferrarotti} \& {Gail}(2006)}]{2006A&A...447..553F}
{Ferrarotti} A.~S., {Gail} H.-P., 2006, \aap, 447, 553

\bibitem[{{Forbes} {et~al}\mbox{.}(2014){Forbes}, {Krumholz}, {Burkert}, \&
  {Dekel}}]{2014MNRAS.443..168F}
{Forbes} J.~C., {Krumholz} M.~R., {Burkert} A., {Dekel} A., 2014, \mnras, 443,
  168

\bibitem[{{Fu} {et~al}\mbox{.}(2012){Fu}, {Kauffmann}, {Li}, \&
  {Guo}}]{2012MNRAS.424.2701F}
{Fu} J., {Kauffmann} G., {Li} C., {Guo} Q., 2012, \mnras, 424, 2701

\bibitem[{{Gail} \& {Sedlmayr}(1999)}]{1999A&A...347..594G}
{Gail} H.-P., {Sedlmayr} E., 1999, \aap, 347, 594

\bibitem[{{Galametz} {et~al}\mbox{.}(2011){Galametz}, {Madden}, {Galliano},
  {Hony}, {Bendo}, \& {Sauvage}}]{2011A&A...532A..56G}
{Galametz} M., {Madden} S.~C., {Galliano} F., {Hony} S., {Bendo} G.~J.,
  {Sauvage} M., 2011, \aap, 532, A56

\bibitem[{{Gall}, {Andersen} \& {Hjorth}(2011){Gall}, {Andersen}, \&
  {Hjorth}}]{2011A&A...528A..13G}
{Gall} C., {Andersen} A.~C., {Hjorth} J., 2011, \aap, 528, A13

\bibitem[{{Galliano}, {Dwek} \& {Chanial}(2008){Galliano}, {Dwek}, \&
  {Chanial}}]{2008ApJ...672..214G}
{Galliano} F., {Dwek} E., {Chanial} P., 2008, \apj, 672, 214

\bibitem[{{Galliano} {et~al}\mbox{.}(2011){Galliano}, {Hony}, {Bernard}, {Bot},
  {Madden}, {Roman-Duval}, {Galametz}, {Li}, {Meixner}, {Engelbracht},
  {Lebouteiller}, {Misselt}, {Montiel}, {Panuzzo}, {Reach}, \&
  {Skibba}}]{2011A&A...536A..88G}
{Galliano} F. {et~al.}, 2011, \aap, 536, A88

\bibitem[{{Genzel} {et~al}\mbox{.}(2010){Genzel}, {Tacconi}, {Gracia-Carpio},
  {Sternberg}, {Cooper}, {Shapiro}, {Bolatto}, {Bouch{\'e}}, {Bournaud},
  {Burkert}, {Combes}, {Comerford}, {Cox}, {Davis}, {Schreiber},
  {Garcia-Burillo}, {Lutz}, {Naab}, {Neri}, {Omont}, {Shapley}, \&
  {Weiner}}]{2010MNRAS.407.2091G}
{Genzel} R. {et~al.}, 2010, \mnras, 407, 2091

\bibitem[{{Genzel} {et~al}\mbox{.}(2015){Genzel}, {Tacconi}, {Lutz},
  {Saintonge}, {Berta}, {Magnelli}, {Combes}, {Garc{\'{\i}}a-Burillo}, {Neri},
  {Bolatto}, {Contini}, {Lilly}, {Boissier}, {Boone}, {Bouch{\'e}}, {Bournaud},
  {Burkert}, {Carollo}, {Colina}, {Cooper}, {Cox}, {Feruglio}, {F{\"o}rster
  Schreiber}, {Freundlich}, {Gracia-Carpio}, {Juneau}, {Kovac}, {Lippa},
  {Naab}, {Salome}, {Renzini}, {Sternberg}, {Walter}, {Weiner}, {Weiss}, \&
  {Wuyts}}]{2014arXiv1409.1171G}
{Genzel} R. {et~al.}, 2015, \apj, 800, 20

\bibitem[{{Glover} \& {Mac Low}(2007)}]{2007ApJS..169..239G}
{Glover} S.~C.~O., {Mac Low} M.-M., 2007, \apjs, 169, 239

\bibitem[{{Gnedin}, {Tassis} \& {Kravtsov}(2009){Gnedin}, {Tassis}, \&
  {Kravtsov}}]{2009ApJ...697...55G}
{Gnedin} N.~Y., {Tassis} K., {Kravtsov} A.~V., 2009, \apj, 697, 55

\bibitem[{{Gordon} {et~al}\mbox{.}(2014){Gordon}, {Roman-Duval}, {Bot},
  {Meixner}, {Babler}, {Bernard}, {Bolatto}, {Boyer}, {Clayton}, {Engelbracht},
  {Fukui}, {Galametz}, {Galliano}, {Hony}, {Hughes}, {Indebetouw}, {Israel},
  {Jameson}, {Kawamura}, {Lebouteiller}, {Li}, {Madden}, {Matsuura}, {Misselt},
  {Montiel}, {Okumura}, {Onishi}, {Panuzzo}, {Paradis}, {Rubio}, {Sandstrom},
  {Sauvage}, {Seale}, {Sewi{\l}o}, {Tchernyshyov}, \&
  {Skibba}}]{2014arXiv1406.6066G}
{Gordon} K.~D. {et~al.}, 2014, \apj, 797, 85

\bibitem[{{Groves} {et~al}\mbox{.}(2015){Groves}, {Schinnerer}, {Leroy},
  {Galametz}, {Walter}, {Bolatto}, {Hunt}, {Dale}, {Calzetti}, {Croxall}, \&
  {Kennicutt}}]{2014arXiv1411.2975G}
{Groves} B.~A. {et~al.}, 2015, \apj, 799, 96

\bibitem[{{Guo} {et~al}\mbox{.}(2011){Guo}, {White}, {Boylan-Kolchin}, {De
  Lucia}, {Kauffmann}, {Lemson}, {Li}, {Springel}, \&
  {Weinmann}}]{2011MNRAS.413..101G}
{Guo} Q. {et~al.}, 2011, \mnras, 413, 101

\bibitem[{{Guo} {et~al}\mbox{.}(2010){Guo}, {White}, {Li}, \&
  {Boylan-Kolchin}}]{2010MNRAS.404.1111G}
{Guo} Q., {White} S., {Li} C., {Boylan-Kolchin} M., 2010, \mnras, 404, 1111

\bibitem[{{Henriques} {et~al}\mbox{.}(2013){Henriques}, {White}, {Thomas},
  {Angulo}, {Guo}, {Lemson}, \& {Springel}}]{2013MNRAS.431.3373H}
{Henriques} B.~M.~B., {White} S.~D.~M., {Thomas} P.~A., {Angulo} R.~E., {Guo}
  Q., {Lemson} G., {Springel} V., 2013, \mnras, 431, 3373

\bibitem[{{Hinshaw} {et~al}\mbox{.}(2013){Hinshaw}, {Larson}, {Komatsu},
  {Spergel}, {Bennett}, {Dunkley}, {Nolta}, {Halpern}, {Hill}, {Odegard},
  {Page}, {Smith}, {Weiland}, {Gold}, {Jarosik}, {Kogut}, {Limon}, {Meyer},
  {Tucker}, {Wollack}, \& {Wright}}]{2013ApJS..208...19H}
{Hinshaw} G. {et~al.}, 2013, \apjs, 208, 19

\bibitem[{{Hirashita}(2000)}]{2000PASJ...52..585H}
{Hirashita} H., 2000, \pasj, 52, 585

\bibitem[{{Hirashita} \& {Ferrara}(2002)}]{2002MNRAS.337..921H}
{Hirashita} H., {Ferrara} A., 2002, \mnras, 337, 921

\bibitem[{{Hirashita} \& {Kuo}(2011)}]{2011MNRAS.416.1340H}
{Hirashita} H., {Kuo} T.-M., 2011, \mnras, 416, 1340

\bibitem[{{H{\"o}fner}(2008)}]{2008A&A...491L...1H}
{H{\"o}fner} S., 2008, \aap, 491, L1

\bibitem[{{Hollenbach} \& {Salpeter}(1971)}]{1971ApJ...163..155H}
{Hollenbach} D., {Salpeter} E.~E., 1971, \apj, 163, 155

\bibitem[{{Hopkins} {et~al}\mbox{.}(2014){Hopkins}, {Kere{\v s}}, {O{\~n}orbe},
  {Faucher-Gigu{\`e}re}, {Quataert}, {Murray}, \&
  {Bullock}}]{2014MNRAS.445..581H}
{Hopkins} P.~F., {Kere{\v s}} D., {O{\~n}orbe} J., {Faucher-Gigu{\`e}re} C.-A.,
  {Quataert} E., {Murray} N., {Bullock} J.~S., 2014, \mnras, 445, 581

\bibitem[{{Hopkins}, {Quataert} \& {Murray}(2012){Hopkins}, {Quataert}, \&
  {Murray}}]{2012MNRAS.421.3522H}
{Hopkins} P.~F., {Quataert} E., {Murray} N., 2012, \mnras, 421, 3522

\bibitem[{{Inoue}(2003)}]{2003PASJ...55..901I}
{Inoue} A.~K., 2003, \pasj, 55, 901

\bibitem[{{Inoue}(2011)}]{2011EP&S...63.1027I}
{Inoue} A.~K., 2011, Earth, Planets, and Space, 63, 1027

\bibitem[{{Jenkins}(2004)}]{2004oee..symp..336J}
{Jenkins} E.~B., 2004, Origin and Evolution of the Elements, 336

\bibitem[{{Jones}, {Tielens} \& {Hollenbach}(1996){Jones}, {Tielens}, \&
  {Hollenbach}}]{1996ApJ...469..740J}
{Jones} A.~P., {Tielens} A.~G.~G.~M., {Hollenbach} D.~J., 1996, \apj, 469, 740

\bibitem[{{Jones} {et~al}\mbox{.}(1994){Jones}, {Tielens}, {Hollenbach}, \&
  {McKee}}]{1994ApJ...433..797J}
{Jones} A.~P., {Tielens} A.~G.~G.~M., {Hollenbach} D.~J., {McKee} C.~F., 1994,
  \apj, 433, 797

\bibitem[{{Karakas}(2010)}]{2010MNRAS.403.1413K}
{Karakas} A.~I., 2010, \mnras, 403, 1413

\bibitem[{{Kennicutt} {et~al}\mbox{.}(2011){Kennicutt}, {Calzetti}, {Aniano},
  {Appleton}, {Armus}, {Beir{\~a}o}, {Bolatto}, {Brandl}, {Crocker}, {Croxall},
  {Dale}, {Meyer}, {Draine}, {Engelbracht}, {Galametz}, {Gordon}, {Groves},
  {Hao}, {Helou}, {Hinz}, {Hunt}, {Johnson}, {Koda}, {Krause}, {Leroy}, {Li},
  {Meidt}, {Montiel}, {Murphy}, {Rahman}, {Rix}, {Roussel}, {Sandstrom},
  {Sauvage}, {Schinnerer}, {Skibba}, {Smith}, {Srinivasan}, {Vigroux},
  {Walter}, {Wilson}, {Wolfire}, \& {Zibetti}}]{2011PASP..123.1347K}
{Kennicutt} R.~C. {et~al.}, 2011, \pasp, 123, 1347

\bibitem[{{Kennicutt}(1998)}]{1998ARA&A..36..189K}
{Kennicutt}, Jr. R.~C., 1998, \araa, 36, 189

\bibitem[{{Kewley} \& {Ellison}(2008)}]{2008ApJ...681.1183K}
{Kewley} L.~J., {Ellison} S.~L., 2008, \apj, 681, 1183

\bibitem[{{Kozasa}, {Hasegawa} \& {Nomoto}(1991){Kozasa}, {Hasegawa}, \&
  {Nomoto}}]{1991A&A...249..474K}
{Kozasa} T., {Hasegawa} H., {Nomoto} K., 1991, \aap, 249, 474

\bibitem[{{Kravtsov} \& {Klypin}(1999)}]{1999ApJ...520..437K}
{Kravtsov} A.~V., {Klypin} A.~A., 1999, \apj, 520, 437

\bibitem[{Krumholz(2014)}]{2014arXiv1402.0867K}
Krumholz M.~R., 2014, Physics Reports, 539, 49 , the Big Problems in Star
  Formation: the Star Formation Rate, Stellar Clustering, and the Initial Mass
  Function

\bibitem[{{Krumholz} \& {Dekel}(2012)}]{2012ApJ...753...16K}
{Krumholz} M.~R., {Dekel} A., 2012, \apj, 753, 16

\bibitem[{{Krumholz}, {McKee} \& {Tumlinson}(2008){Krumholz}, {McKee}, \&
  {Tumlinson}}]{2008ApJ...689..865K}
{Krumholz} M.~R., {McKee} C.~F., {Tumlinson} J., 2008, \apj, 689, 865

\bibitem[{{Kuo}, {Hirashita} \& {Zafar}(2013){Kuo}, {Hirashita}, \&
  {Zafar}}]{2013MNRAS.436.1238K}
{Kuo} T.-M., {Hirashita} H., {Zafar} T., 2013, \mnras, 436, 1238

\bibitem[{{Lara-L{\'o}pez} {et~al}\mbox{.}(2013){Lara-L{\'o}pez}, {Hopkins},
  {L{\'o}pez-S{\'a}nchez}, {Brough}, {Gunawardhana}, {Colless}, {Robotham},
  {Bauer}, {Bland-Hawthorn}, {Cluver}, {Driver}, {Foster}, {Kelvin}, {Liske},
  {Loveday}, {Owers}, {Ponman}, {Sharp}, {Steele}, {Taylor}, \&
  {Thomas}}]{2013MNRAS.434..451L}
{Lara-L{\'o}pez} M.~A. {et~al.}, 2013, \mnras, 434, 451

\bibitem[{{Larson}(1972)}]{1972NPhS..236....7L}
{Larson} R.~B., 1972, Nature Physical Science, 236, 7

\bibitem[{{Lattimer}, {Schramm} \& {Grossman}(1978){Lattimer}, {Schramm}, \&
  {Grossman}}]{1978ApJ...219..230L}
{Lattimer} J.~M., {Schramm} D.~N., {Grossman} L., 1978, \apj, 219, 230

\bibitem[{{Leitner}(2012)}]{2012ApJ...745..149L}
{Leitner} S.~N., 2012, \apj, 745, 149

\bibitem[{{Leitner} \& {Kravtsov}(2011)}]{2011ApJ...734...48L}
{Leitner} S.~N., {Kravtsov} A.~V., 2011, \apj, 734, 48

\bibitem[{{Lequeux} {et~al}\mbox{.}(1979){Lequeux}, {Peimbert}, {Rayo},
  {Serrano}, \& {Torres-Peimbert}}]{1979A&A....80..155L}
{Lequeux} J., {Peimbert} M., {Rayo} J.~F., {Serrano} A., {Torres-Peimbert} S.,
  1979, \aap, 80, 155

\bibitem[{{Leroy} {et~al}\mbox{.}(2011){Leroy}, {Bolatto}, {Gordon},
  {Sandstrom}, {Gratier}, {Rosolowsky}, {Engelbracht}, {Mizuno}, {Corbelli},
  {Fukui}, \& {Kawamura}}]{2011ApJ...737...12L}
{Leroy} A.~K. {et~al.}, 2011, \apj, 737, 12

\bibitem[{{Leroy} {et~al}\mbox{.}(2008){Leroy}, {Walter}, {Brinks}, {Bigiel},
  {de Blok}, {Madore}, \& {Thornley}}]{2008AJ....136.2782L}
{Leroy} A.~K., {Walter} F., {Brinks} E., {Bigiel} F., {de Blok} W.~J.~G.,
  {Madore} B., {Thornley} M.~D., 2008, \aj, 136, 2782

\bibitem[{{Lilly} {et~al}\mbox{.}(2013){Lilly}, {Carollo}, {Pipino}, {Renzini},
  \& {Peng}}]{2013ApJ...772..119L}
{Lilly} S.~J., {Carollo} C.~M., {Pipino} A., {Renzini} A., {Peng} Y., 2013,
  \apj, 772, 119

\bibitem[{{Lu} {et~al}\mbox{.}(2014){Lu}, {Wechsler}, {Somerville}, {Croton},
  {Porter}, {Primack}, {Behroozi}, {Ferguson}, {Koo}, {Guo}, {Safarzadeh},
  {Finlator}, {Castellano}, {White}, {Sommariva}, \&
  {Moody}}]{2014ApJ...795..123L}
{Lu} Y. {et~al.}, 2014, \apj, 795, 123

\bibitem[{{Madden} {et~al}\mbox{.}(2013){Madden}, {R{\'e}my-Ruyer}, {Galametz},
  {Cormier}, {Lebouteiller}, {Galliano}, {Hony}, {Bendo}, {Smith}, {Pohlen},
  {Roussel}, {Sauvage}, {Wu}, {Sturm}, {Poglitsch}, {Contursi}, {Doublier},
  {Baes}, {Barlow}, {Boselli}, {Boquien}, {Carlson}, {Ciesla}, {Cooray},
  {Cortese}, {de Looze}, {Irwin}, {Isaak}, {Kamenetzky}, {Karczewski}, {Lu},
  {MacHattie}, {O''Halloran}, {Parkin}, {Rangwala}, {Schirm}, {Schulz},
  {Spinoglio}, {Vaccari}, {Wilson}, \& {Wozniak}}]{2013PASP..125..600M}
{Madden} S.~C. {et~al.}, 2013, \pasp, 125, 600

\bibitem[{{Magdis} {et~al}\mbox{.}(2012){Magdis}, {Daddi}, {B{\'e}thermin},
  {Sargent}, {Elbaz}, {Pannella}, {Dickinson}, {Dannerbauer}, {da Cunha},
  {Walter}, {Rigopoulou}, {Charmandaris}, {Hwang}, \&
  {Kartaltepe}}]{2012ApJ...760....6M}
{Magdis} G.~E. {et~al.}, 2012, \apj, 760, 6

\bibitem[{{Magdis} {et~al}\mbox{.}(2011){Magdis}, {Daddi}, {Elbaz}, {Sargent},
  {Dickinson}, {Dannerbauer}, {Aussel}, {Walter}, {Hwang}, {Charmandaris},
  {Hodge}, {Riechers}, {Rigopoulou}, {Carilli}, {Pannella}, {Mullaney},
  {Leiton}, \& {Scott}}]{2011ApJ...740L..15M}
{Magdis} G.~E. {et~al.}, 2011, \apjl, 740, L15

\bibitem[{{Maier} {et~al}\mbox{.}(2005){Maier}, {Lilly}, {Carollo}, {Stockton},
  \& {Brodwin}}]{2005ApJ...634..849M}
{Maier} C., {Lilly} S.~J., {Carollo} C.~M., {Stockton} A., {Brodwin} M., 2005,
  \apj, 634, 849

\bibitem[{{Maiolino} {et~al}\mbox{.}(2008){Maiolino}, {Nagao}, {Grazian},
  {Cocchia}, {Marconi}, {Mannucci}, {Cimatti}, {Pipino}, {Ballero}, {Calura},
  {Chiappini}, {Fontana}, {Granato}, {Matteucci}, {Pastorini}, {Pentericci},
  {Risaliti}, {Salvati}, \& {Silva}}]{2008A&A...488..463M}
{Maiolino} R. {et~al.}, 2008, \aap, 488, 463

\bibitem[{{Mandelbaum} {et~al}\mbox{.}(2006){Mandelbaum}, {Seljak},
  {Kauffmann}, {Hirata}, \& {Brinkmann}}]{2006MNRAS.368..715M}
{Mandelbaum} R., {Seljak} U., {Kauffmann} G., {Hirata} C.~M., {Brinkmann} J.,
  2006, \mnras, 368, 715

\bibitem[{{Mannucci} {et~al}\mbox{.}(2010){Mannucci}, {Cresci}, {Maiolino},
  {Marconi}, \& {Gnerucci}}]{2010MNRAS.408.2115M}
{Mannucci} F., {Cresci} G., {Maiolino} R., {Marconi} A., {Gnerucci} A., 2010,
  \mnras, 408, 2115

\bibitem[{{Mattsson} {et~al}\mbox{.}(2014){Mattsson}, {De Cia}, {Andersen}, \&
  {Zafar}}]{2014MNRAS.440.1562M}
{Mattsson} L., {De Cia} A., {Andersen} A.~C., {Zafar} T., 2014, \mnras, 440,
  1562

\bibitem[{{McKee}(1989)}]{1989IAUS..135..431M}
{McKee} C., 1989, in IAU Symposium, Vol. 135, Interstellar Dust, {Allamandola}
  L.~J., {Tielens} A.~G.~G.~M., eds., p. 431

\bibitem[{{McKee} {et~al}\mbox{.}(1987){McKee}, {Hollenbach}, {Seab}, \&
  {Tielens}}]{1987ApJ...318..674M}
{McKee} C.~F., {Hollenbach} D.~J., {Seab} G.~C., {Tielens} A.~G.~G.~M., 1987,
  \apj, 318, 674

\bibitem[{{More} {et~al}\mbox{.}(2011){More}, {van den Bosch}, {Cacciato},
  {Skibba}, {Mo}, \& {Yang}}]{2011MNRAS.410..210M}
{More} S., {van den Bosch} F.~C., {Cacciato} M., {Skibba} R., {Mo} H.~J.,
  {Yang} X., 2011, \mnras, 410, 210

\bibitem[{{Moster}, {Naab} \& {White}(2013){Moster}, {Naab}, \&
  {White}}]{2013MNRAS.428.3121M}
{Moster} B.~P., {Naab} T., {White} S.~D.~M., 2013, \mnras, 428, 3121

\bibitem[{{Munshi} {et~al}\mbox{.}(2013){Munshi}, {Governato}, {Brooks},
  {Christensen}, {Shen}, {Loebman}, {Moster}, {Quinn}, \&
  {Wadsley}}]{2013ApJ...766...56M}
{Munshi} F. {et~al.}, 2013, \apj, 766, 56

\bibitem[{{Murray}, {Quataert} \& {Thompson}(2005){Murray}, {Quataert}, \&
  {Thompson}}]{2005ApJ...618..569M}
{Murray} N., {Quataert} E., {Thompson} T.~A., 2005, \apj, 618, 569

\bibitem[{{Nanni} {et~al}\mbox{.}(2013){Nanni}, {Bressan}, {Marigo}, \&
  {Girardi}}]{2013MNRAS.434.2390N}
{Nanni} A., {Bressan} A., {Marigo} P., {Girardi} L., 2013, \mnras, 434, 2390

\bibitem[{{Narayanan} {et~al}\mbox{.}(2012){Narayanan}, {Krumholz}, {Ostriker},
  \& {Hernquist}}]{2012MNRAS.421.3127N}
{Narayanan} D., {Krumholz} M.~R., {Ostriker} E.~C., {Hernquist} L., 2012,
  \mnras, 421, 3127

\bibitem[{{Nomoto} {et~al}\mbox{.}(2006){Nomoto}, {Tominaga}, {Umeda},
  {Kobayashi}, \& {Maeda}}]{2006NuPhA.777..424N}
{Nomoto} K., {Tominaga} N., {Umeda} H., {Kobayashi} C., {Maeda} K., 2006,
  Nuclear Physics A, 777, 424

\bibitem[{{Noterdaeme} {et~al}\mbox{.}(2008){Noterdaeme}, {Ledoux},
  {Petitjean}, \& {Srianand}}]{2008A&A...481..327N}
{Noterdaeme} P., {Ledoux} C., {Petitjean} P., {Srianand} R., 2008, \aap, 481,
  327

\bibitem[{{Nozawa} {et~al}\mbox{.}(2007){Nozawa}, {Kozasa}, {Habe}, {Dwek},
  {Umeda}, {Tominaga}, {Maeda}, \& {Nomoto}}]{2007ApJ...666..955N}
{Nozawa} T., {Kozasa} T., {Habe} A., {Dwek} E., {Umeda} H., {Tominaga} N.,
  {Maeda} K., {Nomoto} K., 2007, \apj, 666, 955

\bibitem[{{Nozawa} {et~al}\mbox{.}(2003){Nozawa}, {Kozasa}, {Umeda}, {Maeda},
  \& {Nomoto}}]{2003ApJ...598..785N}
{Nozawa} T., {Kozasa} T., {Umeda} H., {Maeda} K., {Nomoto} K., 2003, \apj, 598,
  785

\bibitem[{{Oppenheimer} \& {Dav{\'e}}(2006)}]{2006MNRAS.373.1265O}
{Oppenheimer} B.~D., {Dav{\'e}} R., 2006, \mnras, 373, 1265

\bibitem[{{Oppenheimer} \& {Dav{\'e}}(2008)}]{2008MNRAS.387..577O}
{Oppenheimer} B.~D., {Dav{\'e}} R., 2008, \mnras, 387, 577

\bibitem[{{Oppenheimer} {et~al}\mbox{.}(2010){Oppenheimer}, {Dav{\'e}},
  {Kere{\v s}}, {Fardal}, {Katz}, {Kollmeier}, \&
  {Weinberg}}]{2010MNRAS.406.2325O}
{Oppenheimer} B.~D., {Dav{\'e}} R., {Kere{\v s}} D., {Fardal} M., {Katz} N.,
  {Kollmeier} J.~A., {Weinberg} D.~H., 2010, \mnras, 406, 2325

\bibitem[{{Peeples} \& {Shankar}(2011)}]{2011MNRAS.417.2962P}
{Peeples} M.~S., {Shankar} F., 2011, \mnras, 417, 2962

\bibitem[{{Pilyugin} \& {Thuan}(2005)}]{2005ApJ...631..231P}
{Pilyugin} L.~S., {Thuan} T.~X., 2005, \apj, 631, 231

\bibitem[{{Reddick} {et~al}\mbox{.}(2013){Reddick}, {Wechsler}, {Tinker}, \&
  {Behroozi}}]{2013ApJ...771...30R}
{Reddick} R.~M., {Wechsler} R.~H., {Tinker} J.~L., {Behroozi} P.~S., 2013,
  \apj, 771, 30

\bibitem[{{R{\'e}my-Ruyer} {et~al}\mbox{.}(2014){R{\'e}my-Ruyer}, {Madden},
  {Galliano}, {Galametz}, {Takeuchi}, {Asano}, {Zhukovska}, {Lebouteiller},
  {Cormier}, {Jones}, {Bocchio}, {Baes}, {Bendo}, {Boquien}, {Boselli},
  {DeLooze}, {Doublier-Pritchard}, {Hughes}, {Karczewski}, \&
  {Spinoglio}}]{2014A&A...563A..31R}
{R{\'e}my-Ruyer} A. {et~al.}, 2014, \aap, 563, A31

\bibitem[{{Roche} \& {Aitken}(1984)}]{1984MNRAS.208..481R}
{Roche} P.~F., {Aitken} D.~K., 1984, \mnras, 208, 481

\bibitem[{{Rodr{\'{\i}}guez-Puebla}
  {et~al}\mbox{.}(2015){Rodr{\'{\i}}guez-Puebla}, {Avila-Reese}, {Yang},
  {Foucaud}, {Drory}, \& {Jing}}]{2014arXiv1408.5407R}
{Rodr{\'{\i}}guez-Puebla} A., {Avila-Reese} V., {Yang} X., {Foucaud} S.,
  {Drory} N., {Jing} Y.~P., 2015, \apj, 799, 130

\bibitem[{{Rowlands} {et~al}\mbox{.}(2014){Rowlands}, {Dunne}, {Dye},
  {Arag{\'o}n-Salamanca}, {Maddox}, {da Cunha}, {Smith}, {Bourne}, {Eales},
  {Gomez}, {Smail}, {Alpaslan}, {Clark}, {Driver}, {Ibar}, {Ivison},
  {Robotham}, {Smith}, \& {Valiante}}]{2014MNRAS.441.1017R}
{Rowlands} K. {et~al.}, 2014, \mnras, 441, 1017

\bibitem[{{Saintonge} {et~al}\mbox{.}(2011{\natexlab{a}}){Saintonge},
  {Kauffmann}, {Kramer}, {Tacconi}, {Buchbender}, {Catinella}, {Fabello},
  {Graci{\'a}-Carpio}, {Wang}, {Cortese}, {Fu}, {Genzel}, {Giovanelli}, {Guo},
  {Haynes}, {Heckman}, {Krumholz}, {Lemonias}, {Li}, {Moran},
  {Rodriguez-Fernandez}, {Schiminovich}, {Schuster}, \&
  {Sievers}}]{2011MNRAS.415...32S}
{Saintonge} A. {et~al.}, 2011{\natexlab{a}}, \mnras, 415, 32

\bibitem[{{Saintonge} {et~al}\mbox{.}(2011{\natexlab{b}}){Saintonge},
  {Kauffmann}, {Wang}, {Kramer}, {Tacconi}, {Buchbender}, {Catinella},
  {Graci{\'a}-Carpio}, {Cortese}, {Fabello}, {Fu}, {Genzel}, {Giovanelli},
  {Guo}, {Haynes}, {Heckman}, {Krumholz}, {Lemonias}, {Li}, {Moran},
  {Rodriguez-Fernandez}, {Schiminovich}, {Schuster}, \&
  {Sievers}}]{2011MNRAS.415...61S}
{Saintonge} A. {et~al.}, 2011{\natexlab{b}}, \mnras, 415, 61

\bibitem[{{Saintonge} {et~al}\mbox{.}(2013){Saintonge}, {Lutz}, {Genzel},
  {Magnelli}, {Nordon}, {Tacconi}, {Baker}, {Bandara}, {Berta}, {F{\"o}rster
  Schreiber}, {Poglitsch}, {Sturm}, {Wuyts}, \& {Wuyts}}]{2013ApJ...778....2S}
{Saintonge} A. {et~al.}, 2013, \apj, 778, 2

\bibitem[{{Saintonge} {et~al}\mbox{.}(2012){Saintonge}, {Tacconi}, {Fabello},
  {Wang}, {Catinella}, {Genzel}, {Graci{\'a}-Carpio}, {Kramer}, {Moran},
  {Heckman}, {Schiminovich}, {Schuster}, \& {Wuyts}}]{2012ApJ...758...73S}
{Saintonge} A. {et~al.}, 2012, \apj, 758, 73

\bibitem[{{Sanders} {et~al}\mbox{.}(2015){Sanders}, {Shapley}, {Kriek},
  {Reddy}, {Freeman}, {Coil}, {Siana}, {Mobasher}, {Shivaei}, {Price}, \& {de
  Groot}}]{2014arXiv1408.2521S}
{Sanders} R.~L. {et~al.}, 2015, \apj, 799, 138

\bibitem[{{Santini} {et~al}\mbox{.}(2014){Santini}, {Maiolino}, {Magnelli},
  {Lutz}, {Lamastra}, {Li Causi}, {Eales}, {Andreani}, {Berta}, {Buat},
  {Cooray}, {Cresci}, {Daddi}, {Farrah}, {Fontana}, {Franceschini}, {Genzel},
  {Granato}, {Grazian}, {Le Floc'h}, {Magdis}, {Magliocchetti}, {Mannucci},
  {Menci}, {Nordon}, {Oliver}, {Popesso}, {Pozzi}, {Riguccini}, {Rodighiero},
  {Rosario}, {Salvato}, {Scott}, {Silva}, {Tacconi}, {Viero}, {Wang}, {Wuyts},
  \& {Xu}}]{2014A&A...562A..30S}
{Santini} P. {et~al.}, 2014, \aap, 562, A30

\bibitem[{{Savage} \& {Sembach}(1996)}]{1996ARA&A..34..279S}
{Savage} B.~D., {Sembach} K.~R., 1996, \araa, 34, 279

\bibitem[{{Schaye} {et~al}\mbox{.}(2015){Schaye}, {Crain}, {Bower}, {Furlong},
  {Schaller}, {Theuns}, {Dalla Vecchia}, {Frenk}, {McCarthy}, {Helly},
  {Jenkins}, {Rosas-Guevara}, {White}, {Baes}, {Booth}, {Camps}, {Navarro},
  {Qu}, {Rahmati}, {Sawala}, {Thomas}, \& {Trayford}}]{2015MNRAS.446..521S}
{Schaye} J. {et~al.}, 2015, \mnras, 446, 521

\bibitem[{{Scoville}(2003)}]{2003JKAS...36..167S}
{Scoville} N., 2003, Journal of Korean Astronomical Society, 36, 167

\bibitem[{{Scoville} {et~al}\mbox{.}(2014){Scoville}, {Aussel}, {Sheth},
  {Scott}, {Sanders}, {Ivison}, {Pope}, {Capak}, {Vanden Bout}, {Manohar},
  {Kartaltepe}, {Robertson}, \& {Lilly}}]{2014ApJ...783...84S}
{Scoville} N. {et~al.}, 2014, \apj, 783, 84

\bibitem[{{Shi} {et~al}\mbox{.}(2014){Shi}, {Armus}, {Helou}, {Stierwalt},
  {Gao}, {Wang}, {Zhang}, \& {Gu}}]{2014Natur.514..335S}
{Shi} Y., {Armus} L., {Helou} G., {Stierwalt} S., {Gao} Y., {Wang} J., {Zhang}
  Z.-Y., {Gu} Q., 2014, \nat, 514, 335

\bibitem[{{Skibba} \& {Sheth}(2009)}]{2009MNRAS.392.1080S}
{Skibba} R.~A., {Sheth} R.~K., 2009, \mnras, 392, 1080

\bibitem[{{Somerville} {et~al}\mbox{.}(2012){Somerville}, {Gilmore}, {Primack},
  \& {Dom{\'{\i}}nguez}}]{2012MNRAS.423.1992S}
{Somerville} R.~S., {Gilmore} R.~C., {Primack} J.~R., {Dom{\'{\i}}nguez} A.,
  2012, \mnras, 423, 1992

\bibitem[{{Spitoni} {et~al}\mbox{.}(2010){Spitoni}, {Calura}, {Matteucci}, \&
  {Recchi}}]{2010A&A...514A..73S}
{Spitoni} E., {Calura} F., {Matteucci} F., {Recchi} S., 2010, \aap, 514, A73

\bibitem[{{Stinson} {et~al}\mbox{.}(2013){Stinson}, {Brook}, {Macci{\`o}},
  {Wadsley}, {Quinn}, \& {Couchman}}]{2013MNRAS.428..129S}
{Stinson} G.~S., {Brook} C., {Macci{\`o}} A.~V., {Wadsley} J., {Quinn} T.~R.,
  {Couchman} H.~M.~P., 2013, \mnras, 428, 129

\bibitem[{{Stinson} {et~al}\mbox{.}(2007){Stinson}, {Dalcanton}, {Quinn},
  {Kaufmann}, \& {Wadsley}}]{2007ApJ...667..170S}
{Stinson} G.~S., {Dalcanton} J.~J., {Quinn} T., {Kaufmann} T., {Wadsley} J.,
  2007, \apj, 667, 170

\bibitem[{{Thompson} {et~al}\mbox{.}(2015){Thompson}, {Fabian}, {Quataert}, \&
  {Murray}}]{2014arXiv1406.5206T}
{Thompson} T.~A., {Fabian} A.~C., {Quataert} E., {Murray} N., 2015, \mnras,
  449, 147

\bibitem[{{Thompson}, {Quataert} \& {Murray}(2005){Thompson}, {Quataert}, \&
  {Murray}}]{2005ApJ...630..167T}
{Thompson} T.~A., {Quataert} E., {Murray} N., 2005, \apj, 630, 167

\bibitem[{{Tolstoy}, {Hill} \& {Tosi}(2009){Tolstoy}, {Hill}, \&
  {Tosi}}]{2009ARA&A..47..371T}
{Tolstoy} E., {Hill} V., {Tosi} M., 2009, \araa, 47, 371

\bibitem[{{Tremonti} {et~al}\mbox{.}(2004){Tremonti}, {Heckman}, {Kauffmann},
  {Brinchmann}, {Charlot}, {White}, {Seibert}, {Peng}, {Schlegel}, {Uomoto},
  {Fukugita}, \& {Brinkmann}}]{2004ApJ...613..898T}
{Tremonti} C.~A. {et~al.}, 2004, \apj, 613, 898

\bibitem[{{Vogelsberger} {et~al}\mbox{.}(2014){Vogelsberger}, {Genel},
  {Springel}, {Torrey}, {Sijacki}, {Xu}, {Snyder}, {Nelson}, \&
  {Hernquist}}]{2014MNRAS.444.1518V}
{Vogelsberger} M. {et~al.}, 2014, \mnras, 444, 1518

\bibitem[{{Wang}(1991)}]{1991ApJ...374..456W}
{Wang} B., 1991, \apj, 374, 456

\bibitem[{{Weingartner} \& {Draine}(1999)}]{1999ApJ...517..292W}
{Weingartner} J.~C., {Draine} B.~T., 1999, \apj, 517, 292

\bibitem[{{Whitaker} {et~al}\mbox{.}(2012){Whitaker}, {van Dokkum}, {Brammer},
  \& {Franx}}]{2012ApJ...754L..29W}
{Whitaker} K.~E., {van Dokkum} P.~G., {Brammer} G., {Franx} M., 2012, \apjl,
  754, L29

\bibitem[{{Winters} {et~al}\mbox{.}(1997){Winters}, {Fleischer}, {Le Bertre},
  \& {Sedlmayr}}]{1997A&A...326..305W}
{Winters} J.~M., {Fleischer} A.~J., {Le Bertre} T., {Sedlmayr} E., 1997, \aap,
  326, 305

\bibitem[{{Wolfire}, {Hollenbach} \& {McKee}(2010){Wolfire}, {Hollenbach}, \&
  {McKee}}]{2010ApJ...716.1191W}
{Wolfire} M.~G., {Hollenbach} D., {McKee} C.~F., 2010, \apj, 716, 1191

\bibitem[{{Yamasawa} {et~al}\mbox{.}(2011){Yamasawa}, {Habe}, {Kozasa},
  {Nozawa}, {Hirashita}, {Umeda}, \& {Nomoto}}]{2011ApJ...735...44Y}
{Yamasawa} D., {Habe} A., {Kozasa} T., {Nozawa} T., {Hirashita} H., {Umeda} H.,
  {Nomoto} K., 2011, \apj, 735, 44

\bibitem[{{Yang}, {Mo} \& {van den Bosch}(2003){Yang}, {Mo}, \& {van den
  Bosch}}]{2003MNRAS.339.1057Y}
{Yang} X., {Mo} H.~J., {van den Bosch} F.~C., 2003, \mnras, 339, 1057

\bibitem[{{Yin}, {Matteucci} \& {Vladilo}(2011){Yin}, {Matteucci}, \&
  {Vladilo}}]{2011A&A...531A.136Y}
{Yin} J., {Matteucci} F., {Vladilo} G., 2011, \aap, 531, A136

\bibitem[{{Zafar} \& {Watson}(2013)}]{2013A&A...560A..26Z}
{Zafar} T., {Watson} D., 2013, \aap, 560, A26

\bibitem[{{Zehavi} {et~al}\mbox{.}(2004){Zehavi}, {Weinberg}, {Zheng},
  {Berlind}, {Frieman}, {Scoccimarro}, {Sheth}, {Blanton}, {Tegmark}, {Mo},
  {Bahcall}, {Brinkmann}, {Burles}, {Csabai}, {Fukugita}, {Gunn}, {Lamb},
  {Loveday}, {Lupton}, {Meiksin}, {Munn}, {Nichol}, {Schlegel}, {Schneider},
  {SubbaRao}, {Szalay}, {Uomoto}, {York}, \& {SDSS
  Collaboration}}]{2004ApJ...608...16Z}
{Zehavi} I. {et~al.}, 2004, \apj, 608, 16

\bibitem[{{Zhukovska}(2014)}]{2014A&A...562A..76Z}
{Zhukovska} S., 2014, \aap, 562, A76

\bibitem[{{Zhukovska}, {Gail} \& {Trieloff}(2008){Zhukovska}, {Gail}, \&
  {Trieloff}}]{2008A&A...479..453Z}
{Zhukovska} S., {Gail} H.-P., {Trieloff} M., 2008, \aap, 479, 453

\bibitem[{{Zubko}, {Dwek} \& {Arendt}(2004){Zubko}, {Dwek}, \&
  {Arendt}}]{2004ApJS..152..211Z}
{Zubko} V., {Dwek} E., {Arendt} R.~G., 2004, \apjs, 152, 211

\end{thebibliography}

\end{document}